\documentclass[twocolumn,tighten]{aastex61}

\received{October 20, 2016}
\revised{\today}
\accepted{May 30, 2016}
\published{???}
\submitjournal{the Astronomical Journal}

\newcounter{column_number}
\newcommand{\numberthecolumn}{\colhead{(\arabic{column_number})}\stepcounter{column_number}}
\newcommand{\anchorparen}[2]{\anchor{#1}{#2} (\url{#1})}
\newcommand{\tnm} {\tablenotemark}

\newcommand{\hr}{$^\mathrm{h}$}
\newcommand{\mn}{$^\mathrm{m}$}
\newcommand{\se}{$^\mathrm{s}$}
\newcommand{\de}{$^\circ$}
\newcommand{\am}{$^\prime$}
\newcommand{\as}{$^{\prime\prime}$}

\shorttitle{NGC~6231: Stellar Population}
\shortauthors{Kuhn et al.}

\begin{document}

\title{The Structure of the Young Star Cluster NGC 6231. I. Stellar Population}

\correspondingauthor{Michael A. Kuhn}
\email{mkuhn1@gmail.com}

\author[0000-0002-0631-7514]{Michael A. Kuhn}
\affil{Millennium Institute of Astrophysics, Vicu\~na Mackenna 4860, 7820436 Macul, Santiago, Chile}
\affiliation{Instituto de Fisica y Astronom\'{i}a, Universidad de Valpara\'{i}so, Gran Breta\~{n}a 1111, Playa Ancha, Valpara\'{i}so, Chile}

\author{Nicol\'as Medina}
\affil{Millennium Institute of Astrophysics, Vicu\~na Mackenna 4860, 7820436 Macul, Santiago, Chile}
\affiliation{Instituto de Fisica y Astronom\'{i}a, Universidad de Valpara\'{i}so, Gran Breta\~{n}a 1111, Playa Ancha, Valpara\'{i}so, Chile}

\author{Konstantin V. Getman}
\affiliation{Department of Astronomy \& Astrophysics, 525 Davey Laboratory, Pennsylvania State University, University Park, PA 16802, USA}

\author{Eric D. Feigelson}
\affiliation{Department of Astronomy \& Astrophysics, 525 Davey Laboratory, Pennsylvania State University, University Park, PA 16802, USA}
\affiliation{Millennium Institute of Astrophysics, Vicu\~na Mackenna 4860, 7820436 Macul, Santiago, Chile}

\author{Mariusz Gromadzki}
\affiliation{Warsaw University Astronomical Observatory, Al. Ujazdowskie 4, 00-478 Warszawa, Poland}
\affil{Millennium Institute of Astrophysics, Vicu\~na Mackenna 4860, 7820436 Macul, Santiago, Chile}
\affiliation{Instituto de Fisica y Astronom\'{i}a, Universidad de Valpara\'{i}so, Gran Breta\~{n}a 1111, Playa Ancha, Valpara\'{i}so, Chile}

\author{Jordanka Borissova}
\affiliation{Millennium Institute of Astrophysics, Vicu\~na Mackenna 4860, 7820436 Macul, Santiago, Chile}
\affiliation{Instituto de Fisica y Astronom\'{i}a, Universidad de Valpara\'{i}so, Gran Breta\~{n}a 1111, Playa Ancha, Valpara\'{i}so, Chile}

\author{Radostin Kurtev}
\affiliation{Millennium Institute of Astrophysics, Vicu\~na Mackenna 4860, 7820436 Macul, Santiago, Chile}
\affiliation{Instituto de Fisica y Astronom\'{i}a, Universidad de Valpara\'{i}so, Gran Breta\~{n}a 1111, Playa Ancha, Valpara\'{i}so, Chile}



\begin{abstract}

NGC~6231 is a young cluster (age $\sim$2--7~Myr) dominating the Sco~OB1 association (distance $\sim$1.59~kpc) with $\sim$100 O and B stars and a large pre--main-sequence stellar population. We combine a reanalysis of archival {\it Chandra} X-ray data with multi-epoch NIR photometry from the VVV survey and published optical catalogs to obtain a catalog of 2148 probable cluster members. This catalog is 70\% larger than previous censuses of probable cluster members in NGC~6231, and it includes many low-mass stars detected in the NIR but not in the optical and some B-stars without previously noted X-ray counterparts. In addition, we identify 295 NIR variables, about half of which are expected to be pre--main-sequence stars.
With the more-complete sample, we estimate a total population in the {\it Chandra} field of 5700--7500 cluster members down to 0.08~$M_\odot$ (assuming a universal initial mass function) with a completeness limit at 0.5~$M_\odot$.
A decrease in stellar X-ray luminosities is noted relative to other younger clusters. However, within the cluster, there is little variation in the distribution of X-ray luminosities for ages less than 5~Myr. X-ray spectral hardness for B stars may be useful for distinguishing between early-B stars with X-rays generated in stellar winds and B-star systems with X-rays from a pre--main-sequence companions ($>$35\% of B stars). A small fraction of catalog members have unusually high X-ray median energies or reddened near-infrared colors, which might be explained by absorption from thick or edge-on disks or being background field stars.  

\end{abstract}




\section{Introduction}

\objectname[NGC 6231]{NGC~6231} is the dominant young stellar cluster ($\sim$2--7~Myr old) at the center of the \objectname[Sco OB 1]{Sco~OB1} association \citep[$d\approx1.59$~kpc;][]{1998AJ....115..734S}. This cluster is notable in having a large population of pre--main-sequence stars and early-type stars \citep{1998AJ....115..734S,2008hsf2.book..401R,2007MNRAS.377..945S,2016arXiv160708860D}. The natal molecular cloud has already been dispersed, leading to a cessation of star-formation activity, so the cluster represents the final product of the star-formation process. Nevertheless, Sco~OB1 also contains other regions, some of which are forming stars, including the Large Elephant Trunk to the north-west and \objectname[IC 4628]{IC~4628} to the north-east, as well as the young stellar cluster \objectname[Trumpler 21]{Tr~21} to the north-west. NGC~6231 itself is probably substantially larger than the $\sim$0.1~square degrees {\it Chandra}/ACIS-I field, which only includes the central region of the cluster. 

Clusters containing OB stars can be important laboratories for understanding star formation because most low-mass stars, like our Sun, likely formed in regions containing massive stars \citep{2012A&A...545A...4G,2012ApJ...754...56D}. NGC~6231 is host to a number of massive stars, including a Wolf-Rayet (WR) star, 15 O-type stellar systems, and at least 91 B-type stellar systems. In addition to the WR star (\objectname[HD 152270]{HD~152270}; WC7+O5-8 binary), several other massive stars are evolved, including $>$20 OB stars classified as giants or supergiants by \citet{1978ApJS...38..309H}. Among them is the earliest-type star in the cluster, \objectname[HD 152233]{HD~152233} (O6III(f)+O9? binary; $L_\mathrm{bol}=10^{5.7}$~$L_\odot$), and the most luminous system in the cluster, \objectname[HD 152248]{HD~152248} (O7.5III(f)+O7III(f) binary; $L_\mathrm{bol}=10^{5.8}$~$L_\odot$). The massive star \objectname[HD 153919]{HD~153919} (O6.5Iaf$^{+}$), with an angular separation of $\sim$4$^\circ$ from the cluster, is probably a runaway cluster member \citep{1974PASP...86..284F,2001A&A...370..170A} that was discovered to be a high-mass X-ray binary (HMXB) with a neutron-star or black-hole companion \citep{1973ApJ...181L..43J,2002A&A...392..909C}. Several other WR stars lie outside the cluster center, including \objectname[HD 151932]{HD~151932} (WN9ha) and \objectname[HD 152408]{HD~152408} (WN7h) \citep{2014AJ....147..115F}.

NGC~6231 is at a critical stage in its evolutionary history for either the birth of a bound open cluster or the dispersal of its stellar population into the Galactic field. Most clusters of newly formed stars disperse once the star formation activity in a region has ceased \citep{2003ARA&A..41...57L}. Clusters can become unbound due to either mass loss from the dispersal of molecular cloud mass \citep{1978A&A....70...57T,1980ApJ...235..986H} or through tidal interactions with other giant molecular clouds \citep{2012MNRAS.426.3008K}. Cloud dispersal has recently occurred in NGC~6231, possibly due to one or more supernovas, including a possible explosion 3 Myr ago that formed the HMXB. 

In these two papers, we aim to understand the cluster's evolution through investigation of the final stellar population produced by star formation and modeling of the cluster's structure and dynamics. For a study of structure, a large, representative sample of the cluster's low-mass stars is essential. 
In particular, care must be taken to construct a mass-complete sample of stars to avoid systematic biases due to selection effects from missing stars \citep[e.g.,][]{2009A&A...495..147A}.

In this article (Paper~I) we obtain a new, more complete census of probable cluster members using deeper near-infrared (NIR)  survey data and a reanalysis of the archival {\it Chandra X-ray Observatory} observations. Paper~II will model the spatial structure of the NGC~6231 cluster, including the cluster's density profile, subclusters of stars within the cluster, mass segregation, and age gradients. The results of these investigations will be used to comment on both the formation of NGC~6231 and its fate, based on theoretical understandings of cluster assembly and cluster dissolution.

\subsection{X-ray and Infrared Methods for Young Stellar Populations}

Low-mass stars in young stellar clusters and star-forming regions are difficult to disentangle from field stars \citep{2013ApJS..209...26F}, and this is particularly true for NGC~6231 in Quadrant~4 of the Galactic Plane, $(\ell,b)=(343.5,+1.2)$. Some previous studies have used optical methods to identify young stars, including H$\alpha$ emission \citep[e.g.,][]{1998AJ....115..734S} or placement of objects on optical color magnitude diagrams \citep[e.g.,][]{2016arXiv160708860D}. However, most cluster members have been discovered using X-ray observations, by both {\it XMM Newton} \citep{2004MNRAS.350..809S,2005A&A...441..213S,2006MNRAS.371...67S,2006A&A...454.1047S,2006MNRAS.372..661S,2007ApJ...659.1582S,2007MNRAS.377..945S,2008MNRAS.386..447S,2008NewA...13..202S} and the {\it Chandra X-ray Observatory} \citep[][henceforth DMS2016]{2016arXiv160708860D}.
\defcitealias{2016arXiv160708860D}{DMS2016}	

Our study of NGC~6231 adopts methods for studying young stellar populations developed by the Penn State group in a variety of X-ray/infrared projects, most directly from the Massive Young Star-Forming Complex Study in Infrared and X-ray \citep[MYStIX;][]{2013ApJS..209...26F}. MYStIX combined a reanalysis of archival {\it Chandra} data with NIR catalogs from UKIDSS+2MASS and mid-infrared (MIR) catalogs from {\it Spitzer}/IRAC to identify young stars at various evolutionary stages, ranging from protostars to disk-free pre--main-sequence stars. The MYStIX project surveyed 20 star-forming regions, each containing at least one O-type star, that range in distance from 0.4 to 3.6~kpc, and provide a catalog of $>$30,000 MYStIX Probable Complex Members \citep[MPCM;][]{2013ApJS..209...32B}. The MPCM catalog has served as a basis for investigations of cluster structure and evolution \citep{2014ApJ...787..107K,2015ApJ...802...60K,2015ApJ...812..131K,2015ApJ...798..126J}, star-formation history and spatial gradients in stellar ages \citep{2014ApJ...787..109G, 2014ApJ...787..108G}, circumstellar-disk evolution \citep{2015ApJ...811...10R}, pre--main-sequence evolution of X-ray luminosity \citep{2016MNRAS.457.3836G}, and previously unknown populations of protostars \citep{2016ApJ...833..193R} and OB stars \citep{2017ApJ...838...61P}. 

Other studies following similar strategies include the {\it Chandra} Orion Ultradeep Project \citep[COUP;][]{2005ApJS..160..319G}, the {\it Chandra} Carina Complex Project \citep[CCCP;][]{2011ApJS..194....1T}, and Star Formation in Nearby Clouds \citep[SFiNCs;][]{2017ApJS..229...28G}, and studies of many individual regions including 
\objectname[NGC 1333]{NGC~1333} \citep{2002ApJ...575..354G}, 
\objectname[RMC 136]{30~Doradus} \citep{2006AJ....131.2140T}, 
\objectname[Cep B]{Cep~B} \citep{2006ApJS..163..306G}, 
\objectname[IC 1396 North]{IC~1396N} \citep{2007ApJ...654..316G}, 
\objectname[NGC 6357]{NGC~6357} \citep{2007ApJS..168..100W}, 
\objectname[NGC 6618]{M17} \citep{2007ApJS..169..353B}, 
\objectname[GUM 29]{RCW~49} \citep{2007ApJ...665..719T}, 
\objectname[CG 12]{CG~12} \citep{2008ApJ...673..331G}, 
\objectname[W 3]{W3} \citep{2008ApJ...673..354F}, 
the \objectname[Rosette Nebula]{Rosette Nebula} \citep{2008ApJ...675..464W}, 
\objectname[NGC 6334]{NGC~6334} \citep{2009AJ....138..227F}, 
\objectname[LBN 028.77+03.43]{W40} \citep{2010ApJ...725.2485K}, and 
\objectname[Elephant Trunk Nebula]{IC~1396A} and \objectname[IC 1396]{Tr 37} \citep{2012MNRAS.426.2917G}. 

In most of these studies, NIR counterparts to X-ray sources are used as the primary indicator of cluster membership, with only a small fraction of these sources pruned as likely foreground or background sources. In the case of NGC~6231, more than twice as many X-ray sources have $K_s$-band counterparts than have $V$-band counterparts in the optical and infrared catalogs that are available (Section~\ref{nir.sec}). This is both an effect of the relative brightness of M stars (the typical spectral type of a 0.5-$M_\odot$ star at 3--7~Myr) in the infrared compared to the optical and an effect of the moderate absorption of the cluster. Thus, restricting classification to sources detected in the optical will omit nearly half the detected cluster members. Furthermore, neither the optical nor the infrared color-magnitude diagrams can provide a definitive indication of membership, so some level of contamination is inevitable whether a sample is defined in the optical or NIR.

Our X-ray source detection and extraction methodology is designed to make optimum use of {\it Chandra}'s sensitivity.  Source detection uses maximum-likelihood (ML) image reconstruction to identify sources in crowded regions, and point-source validation is performed using sophisticated source and background modeling \citep{2010ApJ...714.1582B,2014ApJS..213....1T}. The data analysis recipes and {\it ACIS Extract} software \citep{2010ApJ...714.1582B,2012ascl.soft03001B}, which implements these strategies, typically detect 1.5--2 times more X-ray sources (many of which have counterparts in other wavebands) than other leading software such as {\it wavdetect} \citep{2002ApJS..138..185F} and {\it pwdetect} \citep{1997ApJ...483..350D}. We have developed techniques for nonparametric estimation of X-ray luminosity and absorption for faint sources \citep{2010ApJ...708.1760G}.

Both the X-ray and NIR strategies are designed to produce large catalogs of probable cluster members, while allowing some contaminants. This allows us to take full advantage of the effective area of the {\it Chandra X-ray Observatory}, doubling our sample size. The contamination rate can be approximately estimated by simulations (Section~\ref{contam.sec}), and is found not to be significantly higher than in previous studies. In the statistical literature of the last 20 years, the bias towards minimizing False Positives has been revised in favor of less stringent controls on Type~I error that have greater statistical power \citep[cf.\ False Discovery Rate;][]{MR1325392}. For the scientific purpose of modeling cluster structure in Paper~II, the larger sample is a major advantage, while the contaminants have little effect and can be accounted for as a spatially uniformly distributed population of sources in the model \citep[e.g.,][]{2014ApJ...787..107K}.

\subsection{Outline of this Paper}

Section~\ref{previous.sec} provides basic cluster properties from the literature. Section~\ref{oanddr.sec} describes observations and data reduction. Section~\ref{members.sec} derives a catalog of probable cluster members. Section~\ref{prop.sec} derives stellar properties from infrared and X-ray data. Section~\ref{ob.sec} discusses the OB stellar population. Section~7 provides the conclusion. 

\section{Basic Cluster Properties\label{previous.sec}}

This work makes use of some basic cluster properties available from the literature. Summaries of older studies are provided by \citet{2006A&A...454.1047S} and \citet{2008hsf2.book..401R}. Since then, expanded catalogs of cluster members have been provided by \citet{2013AJ....145...37S} and \citetalias{2016arXiv160708860D}. 

\begin{description}
\item[Distance] \citet{2006A&A...454.1047S} summarize a variety of estimates of distance modulus and report a weighted mean of $DM=11.07\pm0.04$. Distance can also be estimated independently using parallax measurements of nine of the OB stars in the Tycho-Gaia Astrometric Solution (TGAS) catalog \citep{2016arXiv160904172G} yielding a distance estimate of $d=1.37\pm0.42$~kpc (Appendix~\ref{gaia.sec}). We follow \citetalias{2016arXiv160708860D} and adopt a distance modulus of $DM=11.0$, corresponding to a distance of $d=1.59$~kpc. 

\item[Age] Recent estimates of median stellar age have clustered around $\sim$3.5~Myr and suggest a significant age spread of $\Delta age\sim3$--7~Myr \citep{2007MNRAS.377..945S,2013AJ....145...37S, 2016arXiv160708860D}. 
The estimates of stellar ages in the literature come from both pre--main-sequence and post--main-sequence evolution on the Hertzsprung--Russell (HR) diagram and color--magnitude diagrams. For high-mass stars, \citet{2013AJ....145...37S} found the distribution to be bracketed by 3--4~Myr isochrones from \citet{2011A&A...530A.115B} or 4--7~Myr isochrones from \citet{2012A&A...537A.146E}. \citetalias[][]{2016arXiv160708860D} (their Figure~5) show that spectroscopically identified OB stars on a $V$ vs.\ $B-V$ diagram are scattered around the 3~Myr isochrone from \citet{2012A&A...537A.146E}. 
For low-mass, pre--main-sequence stars, \citet{2013AJ....145...37S} estimate stellar ages using the \citet{2000A&A...358..593S} models, finding ages ranging from 1 to 7~Myr, with a peak at $\sim$3~Myr. For pre--main-sequence stars \citetalias{2016arXiv160708860D} report a $V$-band magnitude distribution consistent with an age range of 1.5 to 7~Myr. They also note that on the $J$ vs. $J-H$ color-magnitude diagram the stars follow a 5-Myr isochrone well. 

The HD~153919 HMXB may provide an additional constraint on age, as was previously noted by \citet{2001A&A...370..170A}. The TGAS catalog reports a proper motion of $\Delta\mathrm{\alpha}=2.28\pm0.04$~mas~yr$^{-1}$ and $\Delta\mathrm{\delta}=4.95\pm0.03$~mas~yr$^{-1}$ for HD~153919.  This traceback vector passes through NGC~6231 at $\sim$2.9~Myr ago. 
According to \citet{2012A&A...537A.146E}, the minimum lifetime of a massive star is 3.54 Myr, so the system must have been formed at least $6.4$~Myr ago if it originated in NGC~6231. We note that this traceback vector also passes through other parts of the Sco~OB1 association, so its origin in NGC~6231, while likely, is not certain, and constraints on its age would depend on where in Sco~OB1 it originated. For example, the traceback vector also passes through the star-forming region IC~4628 1.9~Myr ago. 

To infer stellar properties, we use an age of 3.2~Myr (Section~\ref{age.sec}) and also an alternate age of 6.4~Myr. A younger median age of stars in the cluster is not necessarily inconsistent with the presence of an older star, given the considerable age spread indicated by previous studies.  

\item[Absorption] The natal molecular cloud of NGC~6231 appears completely dispersed.  \citet{2017MNRAS.465.1023M} note that stellar winds flow unimpeded for more than 2~pc in the center of the cluster, based on their observations of a C\,{\sc iv} absorption feature.

Studies suggest that most of the reddening of the cluster occurs in foreground clouds between 100 and 1300~pc in distance and in a possible shell of material surrounding Sco~OB1 \citep{2006A&A...454.1047S}. A map of reddening by \citet[][their Figure~4]{2013AJ....145...37S} shows variations $E(B-V)$ from 0.40 to 0.55~mag in a $40^\prime\times40^\prime$ region around the cluster. The cluster itself is in a local hole in the extinction with $E(B-V)\approx0.45$~mag corresponding to $A_V\approx1.4$~mag for $R=3.1$. \citetalias{2016arXiv160708860D} report uniform extinction of cluster members. 

\end{description}

\section{Observations and Data Reduction \label{oanddr.sec}}

\subsection{X-ray Data\label{xraydata.sec}}

\begin{figure*}[h]
\centering
\includegraphics[width=1.0\textwidth]{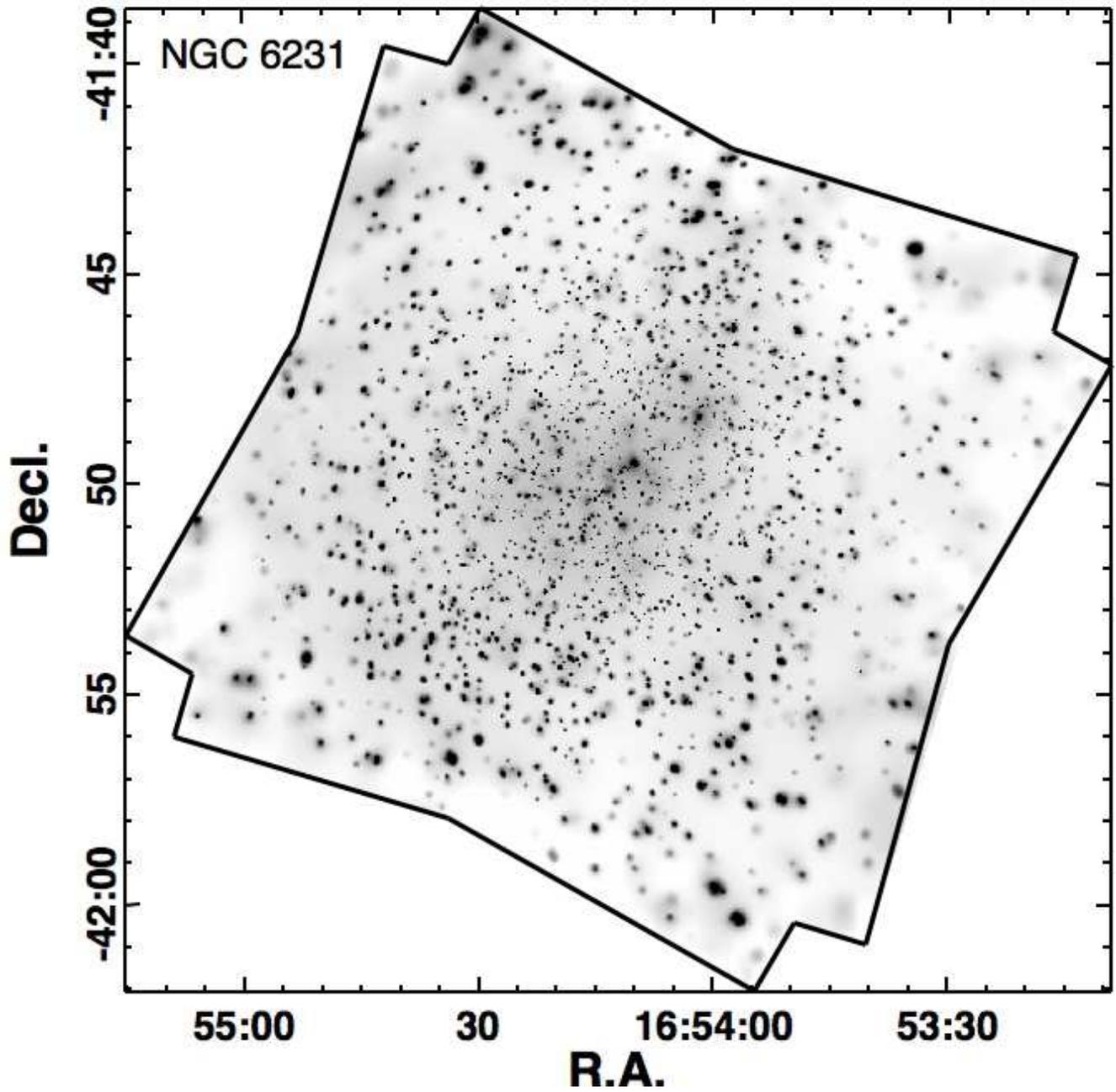} 
\caption{X-ray image of the NGC~6231 cluster from {\it Chandra}'s ACIS-I instrument, with brightness shown on a logarithmic grayscale. Smoothing was performed on the total 0.5--8.0~keV band using an algorithm by \citet{2006MNRAS.368...65E}.\label{ximg.fig}}
\end{figure*}

{\it Chandra} X-ray observations were made using the imaging array on the Advanced CCD Imaging Spectrometer \citep[ACIS-I;][]{2003SPIE.4851...28G}. This instrument is an array of four CCD detectors that subtends $17^\prime\times17^\prime$. (The ACIS-S2 and S3 chips were also active during the observation, but we exclude these data due to {\em Chandra}'s reduced angular resolution far off axis.)  The target was observed in July 2005 (Sequence\,200307; PI S.\ Murray) in two observations (ObsID\,5372 and 6291) and the data were retrieved from the Chandra Data Archive. These observations were both taken in faint mode, with exposure times 77,165\,s and 44,954\,s, roll angles 299$^\circ$ and 287$^\circ$, and an aimpoint at 16\hr54\mn12\se.092 $-$41\de50\am23\as.02 (J2000). The {\it Chandra} event files provide time, position, and energy of each photon-detection event on the CCD. The X-ray image of NGC\,6231 is shown in Figure~\ref{ximg.fig} produced by the \citet{2006MNRAS.368...65E} adaptive-smoothing algorithm.

\subsubsection{X-ray Source Extraction}
The X-ray data reduction for NGC~6231 follows the MYStIX data-reduction procedures given by \citet{2010ApJ...714.1582B}, \citet{2014ApJS..213....1T}, and \citet{2013ApJS..209...27K}. 
These procedures make use of ACIS Extract and TARA recipes available from the Astrophysics Source Code Library \citep{2012ascl.soft03001B}. This analysis requires additional software such as CIAO \citep{2006SPIE.6270E..1VF}, MARX \citep{2012SPIE.8443E..1AD}, HEASoft \citep{2014ascl.soft08004N}, and the Astronomy User's Library \citep[AstroLib;][]{1993ASPC...52..246L}; and the work was carried out using the IDL programming language. Briefly, data products are rebuilt from the ``Level~1'' satellite telemetry applying a variety of calibrations and corrections developed at Penn State. Source detection is performed on deconvolved images generated with the Lucy-Richardson algorithm \citep{1974AJ.....79..745L}. The source list is iteratively refined (improved positions, optimal backgrounds, etc.) using the ACIS Extract procedures, and a measure of source significance $ProbNoSrc\_min$ is calculated. This source significance is the $p$-value for the no-source hypothesis as defined by \citet[][\S4.3]{2010ApJ...714.1582B}, and only sources having $ProbNoSrc\_min<0.01$ are retained at the end of each iteration. Iterations are necessary because removal or shifting of a source will affect the background regions and source significance of other sources, requiring $ProbNoSrc\_min$ to be recalculated for those sources.

For the NGC\,6231 data, 14 iterations were necessary before convergence to the final list of 2411 {\it Chandra} X-ray sources that are reported here. While some sources were detected with thousands of counts, most sources are much fainter. The detected X-ray sources have a median of 10 net counts ($NetCounts$; the number of X-ray events detected in the source extraction region minus the expected number of background counts), while the mode of the histogram of $\log NetCounts$ is $\sim$30~counts.

\subsubsection{X-ray Catalog}

The {\it Chandra} X-ray point-source catalog is presented in Table~1. This table includes both ``X-ray photometry'' quantities obtained from ACIS Extract and ``X-ray Spectral Model'' quantities which are inferred using XPHOT software \citep{2010ApJ...708.1760G}; the definitions of these quantities are identical to those from \citet[][their Table~4]{2013ApJS..209...27K}. 
Photometric quantities are often calculated for several energy bands: a ``soft'' band including X-ray events with energies between 0.5 and 2.0~keV, a ``hard'' band including X-ray events between 2.0 and 8.0~keV, and a ``total'' band including X-ray events with energies between 0.5 and 8.0~keV. Several important quantities, including X-ray median energy, X-ray luminosity, and X-ray variability, are described below. 

X-ray median energy ($ME$) is a measure of spectral hardness, calculated by taking the median of the energies of a source's events on the CCD detector. Estimates of $ME$ are more robust than the estimates of the hardness ratio for sources with few counts, which comprise the majority of X-ray sources in this observation. For thermal X-ray sources, median energy is moderately sensitive to plasma temperature and strongly sensitive to absorption from a thick molecular cloud \citep{2010ApJ...708.1760G}. However, in the case of NGC~6231, where the obscuration of the cluster is $A_V\approx1.6$~mag (corresponding to $N_H\approx3.5\times10^{21}$~cm$^{-2}$; Section~\ref{nir2.sec}), attenuation of X-rays by interstellar gas will be on the order of 1\%, so $ME$ will be mostly affected by plasma temperature and local absorption.

X-ray luminosities and absorbing columns are estimated using the XPHOT algorithms from \citet{2010ApJ...708.1760G}. For the vast majority of sources, the number of counts in the low-resolution X-ray spectrum extracted from the event list is too low for parametric fitting with software like XSPEC \citep{1996ASPC..101...17A}. XPHOT provides a more robust method of estimating spectral properties, using photometric quantities like $NetCounts\_t$ and $ME$ along with empirical relations relating these properties to the X-ray spectroscopic properties of pre--main-sequence stars. These empirical relations were found by \citeauthor{2010ApJ...708.1760G} based on a sample of low-mass stars from COUP \citep{2005ApJS..160..319G,2005ApJS..160..379F}. The XPHOT spectral model properties reported in Table~\ref{xray_properties.tab} include both statistical errors, due to measurement uncertainties in source photometry, and systematic uncertainties, due to intrinsic scatter in the empirical relations. These relations are only valid for the spectral characteristics of pre--main-sequence stars, so XSPEC analysis is performed for early-type stars in Section~\ref{ob.sec}. In this paper, the variable $L_X$ is used to denote absorption-corrected X-ray luminosity in the total band (listed as {\it LOG\_LTC} in the table).

X-ray variable stars were identified by {\it ACIS Extract} by testing for constant count rates (both within one observation and for the two observations combined) using the Kolmogorov--Smirnov (K--S) test. Most $p$-values lie between 0.01 and 1 (no statistically significant deviation from a constant count rate), but 356 sources do have $p$-values ranging from $<$10$^{-5}$ to 0.01. The K--S test is more sensitive to sources with more counts, so more-luminous X-ray variables are more likely to be identified than less-luminous X-ray variables. 

Bright X-ray sources can be affected by pile-up, which occurs when multiple X-ray photons arrive at the same location on the detector during a single 3.2-s frame, causing them to be read as a single event with greater energy. The ACIS Extract photometry recipe notes that a flux of 0.075~counts~frame$^{-1}$ in a 3$\times$3-pixel island leads to a decrease in count rate by a factor of 1.1. For NGC~6231, the sources most affected by pile-up are CXOU~J165401.84-414823.0 and CXOU~J165410.06-414930.1, with photon fluxes of 0.07 and 0.06~counts~frame$^{-1}$, respectively. CCD pile-up has therefore been ignored in this study.

\startlongtable
\onecolumngrid
\begin{deluxetable}{lll}
\tablecaption{X-ray Sources and Properties \label{xray_properties.tab}}
\tablewidth{0in}
\tabletypesize{\scriptsize}

\tablehead{      
\colhead{Column Label} & \colhead{Units} & \colhead{Description} \\ 
\numberthecolumn & \numberthecolumn & \numberthecolumn  
\setcounter{column_number}{1}
}
\startdata
\\
  \multicolumn{3}{l}{\raggedleft {\bf X-ray Photometry}  \citep[][ACIS Extract]{2010ApJ...714.1582B}}\\
Name                       & \nodata              & \parbox[t]{3.5in}{X-ray source name; prefix is CXOU~J} \\
  & & ({\em Chandra X-ray Observatory})\\
Label\tnm{a}               & \nodata              & source name generated by ACIS Extract       \\
RAdeg                      & deg                  & right ascension (J2000)                                         \\
DEdeg                      & deg                  & declination (J2000)                                             \\
PosErr                     & arcsec               & 1-$\sigma$ error circle around (RAdeg,DEdeg)                               \\
PosType                    & \nodata              & algorithm used to estimate position \citep[][ \S7.1]{2010ApJ...714.1582B}  \\
ProbNoSrc\_min             & \nodata              & smallest of ProbNoSrc\_t, ProbNoSrc\_s, ProbNoSrc\_h      \\
ProbNoSrc\_t               & \nodata              & $p$-value\tnm{b} for no-source hypothesis \citep[][ \S4.3]{2010ApJ...714.1582B} \\
ProbNoSrc\_s               & \nodata              & $p$-value for no-source hypothesis                                \\
ProbNoSrc\_h               & \nodata              & $p$-value for no-source hypothesis                                \\
ProbKS\_single\tnm{c}      & \nodata              & \parbox[t]{3.5in}{smallest $p$-value for the one-sample Kolmogorov-Smirnov statistic under the no-variability null hypothesis within a single-observation}\\
ProbKS\_merge\tnm{c}       & \nodata              & \parbox[t]{3.5in}{smallest $p$-value for the one-sample Kolmogorov-Smirnov statistic under the no-variability null hypothesis over merged observations}      \\
ExposureTimeNominal        & s                    & total exposure time in merged observations                     \\
ExposureFraction\tnm{d}    & \nodata              & fraction of ExposureTimeNominal that source was observed\\
NumObservations            & \nodata              & total number of observations extracted                                \\
NumMerged                  & \nodata              & \parbox[t]{3.5in}{number of observations merged to estimate photometry properties}\\
MergeBias                  & \nodata              & fraction of exposure discarded in merge\\
Theta\_Lo                  & arcmin               & smallest off-axis angle for merged observations                 \\
Theta                      & arcmin               &  average off-axis angle for merged observations                 \\
Theta\_Hi                  & arcmin               &  largest off-axis angle for merged observations                 \\
PsfFraction                & \nodata              & average PSF fraction (at 1.5 keV) for merged observations \\
SrcArea                    & (0.492 arcsec)$^2$   & average aperture area for merged observations                   \\
AfterglowFraction\tnm{e}   & \nodata              & suspected afterglow fraction                   \\
SrcCounts\_t               & count                & observed counts in merged apertures                             \\
SrcCounts\_s               & count                & observed counts in merged apertures                             \\
SrcCounts\_h               & count                & observed counts in merged apertures                             \\
BkgScaling                 & \nodata              & scaling of the background extraction \citep[][ \S5.4]{2010ApJ...714.1582B}                 \\
BkgCounts\_t               & count                & observed counts in merged background regions                    \\
BkgCounts\_s               & count                & observed counts in merged background regions                    \\
BkgCounts\_h               & count                & observed counts in merged background regions                    \\
NetCounts\_t               & count                & net counts in merged apertures                                  \\
NetCounts\_s               & count                & net counts in merged apertures                                  \\
NetCounts\_h               & count                & net counts in merged apertures                                  \\
NetCounts\_Lo\_t\tnm{f}    & count                & 1-sigma lower bound on NetCounts\_t               \\
NetCounts\_Hi\_t           & count                & 1-sigma upper bound on NetCounts\_t               \\
NetCounts\_Lo\_s           & count                & 1-sigma lower bound on NetCounts\_s               \\
NetCounts\_Hi\_s           & count                & 1-sigma upper bound on NetCounts\_s               \\
NetCounts\_Lo\_h           & count                & 1-sigma lower bound on NetCounts\_h               \\
NetCounts\_Hi\_h           & count                & 1-sigma upper bound on NetCounts\_h             \\  
MeanEffectiveArea\_t\tnm{g}& cm$^2$~count~photon$^{-1}$ & mean ARF value                                                  \\
MeanEffectiveArea\_s       & cm$^2$~count~photon$^{-1}$ & mean ARF value                                                  \\
MeanEffectiveArea\_h       & cm$^2$~count~photon$^{-1}$ & mean ARF value                                                  \\
MedianEnergy\_t\tnm{h}     & keV                  & median energy, observed spectrum                                \\
MedianEnergy\_s            & keV                  & median energy, observed spectrum                                \\
MedianEnergy\_h            & keV                  & median energy, observed spectrum                                \\
PhotonFlux\_t\tnm{i}       & photon /cm$^2$ /s     & log incident photon flux \\
PhotonFlux\_s              & photon /cm$^2$ /s     & log incident photon flux \\
PhotonFlux\_h              & photon /cm$^2$ /s     & log incident photon flux \\
        & &                    \\
  \multicolumn{3}{l}{\raggedleft {\bf X-ray Spectral Model\tnm{j}}  \citep[][XPHOT]{2010ApJ...708.1760G}}\\
{F\_H}                      & erg /cm$^2$ /s               & X-ray flux, 2:8 keV                       \\
{F\_HC}                     & erg /cm$^2$ /s               & absorption-corrected X-ray flux, 2:8 keV  \\
{SF\_HC\_STAT}              & erg /cm$^2$ /s               & 1-sigma statistical uncertainty on F\_HC       \\
{SF\_HC\_SYST}              & erg /cm$^2$ /s               & 1-sigma systematic  uncertainty on F\_HC       \\
F\_T                      & erg /cm$^2$ /s               & X-ray flux, 0.5:8 keV                     \\
{F\_TC}                     & erg /cm$^2$ /s               & absorption-corrected X-ray flux, 0.5:8 kev\\
{SF\_TC\_STAT}              & erg /cm$^2$ /s               & 1-sigma statistical uncertainty on F\_TC       \\
{SF\_TC\_SYST}              & erg /cm$^2$ /s               & 1-sigma systematic  uncertainty on F\_TC       \\
{LOG\_NH}                 & /cm$^2$               & gas column density                              \\
{SLOG\_NH\_STAT}     & /cm$^2$               & 1-sigma statistical uncertainty on LOG\_NH   \\
{SLOG\_NH\_SYST}     & /cm$^2$               & 1-sigma systematic uncertainty on LOG\_NH   \\
{LOG\_LTC}                      & erg /s               & log X-ray luminosity, 0.5:8 keV                       \\
{LOG\_LHC}                      & erg /s               & log X-ray luminosity, 2:8 keV                       \\
{ERR\_LOG\_LTC}     & erg /s               & 1-sigma statistical uncertainty on LOG\_LTC   \\
{ERR\_LOG\_LHC}     & erg /s               & 1-sigma statistical uncertainty on LOG\_LHC   \\
\enddata
\tablecomments{X-ray source properties from ACIS Extract and XPHOT. Column definitions are identical to those in \citet{2013ApJS..209...27K}. Rows are sorted by R.A.
\\~\\
The suffixes ``\_t'', ``\_s'', and ``\_h'' on names of photometric quantities designate the {\em total} (0.5--8~keV), {\em soft} (0.5--2~keV), and {\em hard} (2--8~keV) energy bands. 
\\~\\
Source significance quantities (ProbNoSrc\_t, ProbNoSrc\_s, ProbNoSrc\_h, ProbNoSrc\_min) are computed using a subset of each source's extractions chosen to maximize significance \citep[][ \S6.2]{2010ApJ...714.1582B}.
Source position quantities (RAdeg, DEdeg, PosErr) are computed using a subset of each source's extractions chosen to minimize the position uncertainty \citep[][ \S6.2 and 7.1]{2010ApJ...714.1582B}.   
All other quantities are computed using a subset of each source's extractions chosen to balance the conflicting goals of minimizing photometric uncertainty and of avoiding photometric bias \citep[][ \S6.2 and 7]{2010ApJ...714.1582B}. 
}

\tablenotetext{a}{Source labels identify a {\it Chandra} pointing; they do not convey membership in astrophysical clusters.}

\tablenotetext{b}{In statistical hypothesis testing, the $p$-value is the probability of obtaining a test statistic at least as extreme as the one that was actually observed when the null hypothesis is true.}

\tablenotetext{c}{See \citet[][ \S7.6]{2010ApJ...714.1582B} for a description of the variability metrics, and caveats regarding possible spurious indications of variability using the ProbKS\_merge metric. }

\tablenotetext{d}{Due to dithering over inactive portions of the focal plane, a {\it Chandra} source is often not observed during some fraction of the nominal exposure time.  (See \url{http://cxc.harvard.edu/ciao/why/dither.html}.)  The reported quantity is FRACEXPO produced by the CIAO tool {\em mkarf}.}

\tablenotetext{e}{Some background events arising from an effect known as ``afterglow'' (\url{http://cxc.harvard.edu/ciao/why/afterglow.html}) may contaminate source extractions, despite careful procedures to identify and remove them during data preparation \citep[][ \S3]{2010ApJ...714.1582B}.
After extraction, we attempt to identify afterglow events using the tool ae\_afterglow\_report, and report the fraction of extracted events attributed to afterglow; see the \anchorparen{http://www.astro.psu.edu/xray/acis/acis_analysis.html}{{\it ACIS Extract} manual}.} 

\tablenotetext{f}{Confidence intervals (68\%) for NetCounts quantities are estimated by the CIAO tool {\em aprates} (\url{http://asc.harvard.edu/ciao/ahelp/aprates.html}).}

\tablenotetext{g}{The ancillary response file (ARF) in ACIS data analysis represents both the effective area of the observatory and the fraction of the observation for which data were actually collected for the source (ExposureFraction).}

\tablenotetext{h}{MedianEnergy is the median energy of extracted events, corrected for background \citep[][ \S7.3]{2010ApJ...714.1582B}. } 

\tablenotetext{i}{PhotonFlux = (NetCounts / MeanEffectiveArea / ExposureTimeNominal) \citep[][ \S7.4]{2010ApJ...714.1582B} }

\tablenotetext{j}{XPHOT assumes X-ray spectral shapes of young, low-mass stars, which come from coronal X-ray emission. XPHOT quantities will therefore be unreliable for high-mass stars, for which X-ray emission is associated with the stellar wind   \citep{2010ApJ...708.1760G}. }

\end{deluxetable}

\subsubsection{Comparison to the DMS2016 X-ray Catalog}

\begin{figure*}[t]
\centering
\includegraphics[width=0.7\textwidth]{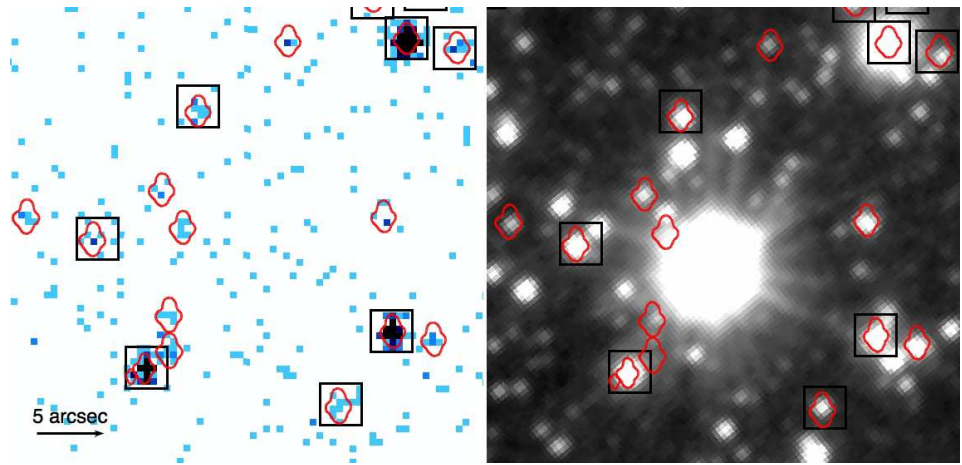} 
\caption{X-ray sources over-plotted on the X-ray image (left) and $J$-band image (right) in the vicinity of the B0.5V star CPD$-$41~7743. X-ray source extractions for our catalog are shown in red, and the X-ray source locations from \citetalias{2016arXiv160708860D} are indicated by black boxes.   
\label{j.fig}}
\end{figure*}

Our catalog of 2411 X-ray point sources includes $\sim$1.5 times more sources than the 1613 X-ray sources reported by \citetalias{2016arXiv160708860D}, which is similar to the increase in number of sources for other studies in which {\it Chandra} observations were reanalyzed using this methodology \citep{2013ApJS..209...27K}. In particular, many new faint X-ray sources are included. Figure~\ref{j.fig} shows sources from the two X-ray catalogs marked on an X-ray image (left) and a VVV $J$-band image (right) of part of the field of view near star \objectname[V* V1208 Sco]{CPD$-$41~7743}. In this region, 5 of the new X-ray sources have $J$-band counterparts, while 4 do not. Although the deeper catalogs may include additional spurious sources or extragalactic X-ray sources, most of these will be filtered out later in the analysis by matching with NIR counterparts.

\begin{figure*}
\centering
\includegraphics[width=0.8\textwidth]{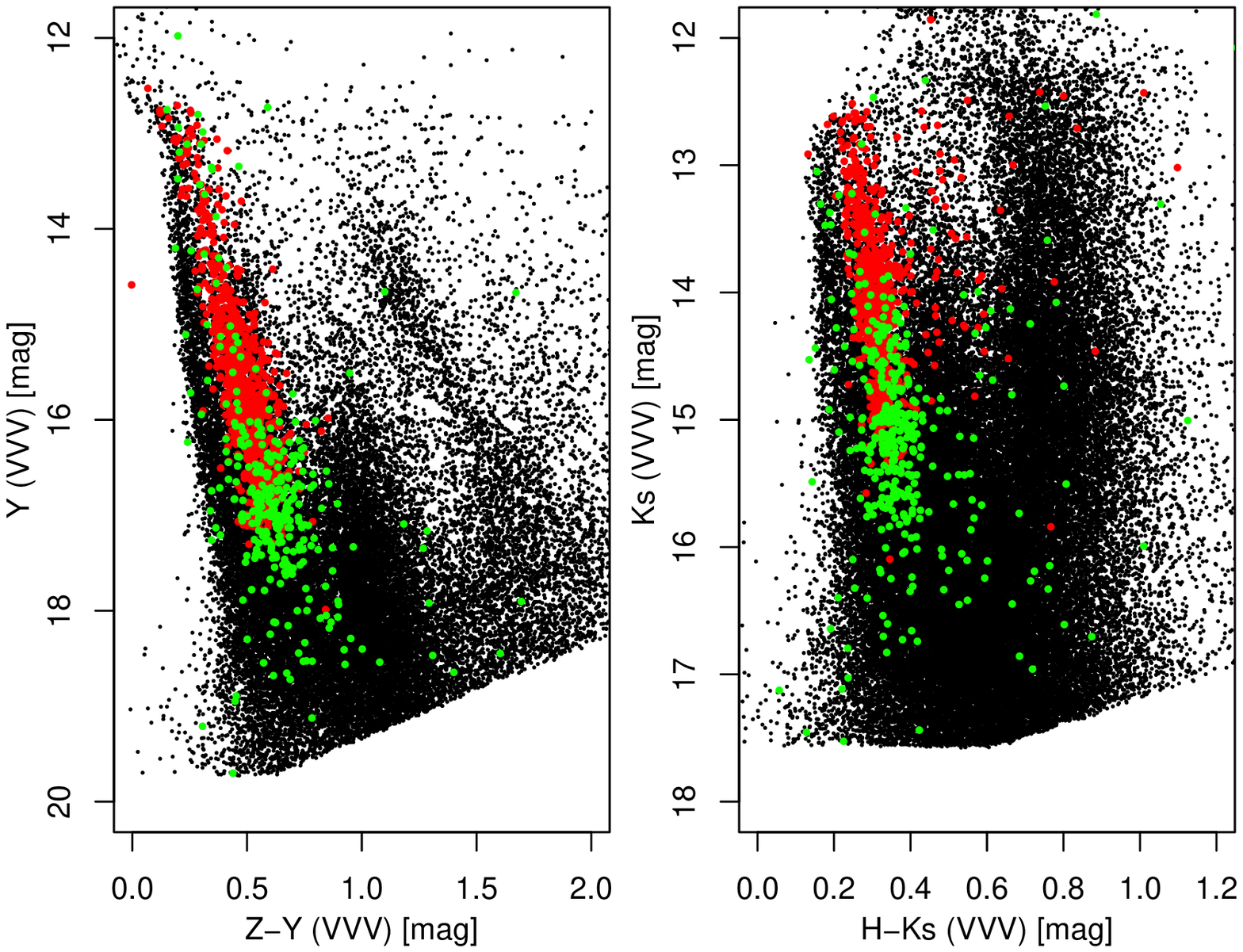} 
\caption{VVV color-magnitude diagrams showing all stars in the {\it Chandra} field of view. Left: $Y$ vs.\ $Z-Y$, and right: $H$ vs.\ $H-K_s$. Field stars are shown as black points, while probable cluster members are shown either as red points (also contained in the \citetalias{2016arXiv160708860D} catalog) or green points (new members).
\label{new_sources.fig}}
\end{figure*}

A comparison of reported X-ray source properties from the two catalogs shows that, for the majority of stars, X-ray net counts ($CT$ from \citetalias{2016arXiv160708860D} vs.\ $NetCounts_t$) are scattered around the $y=x$ line with a root-mean-squared deviation of 0.1~dex and a mean offset of 0.05~dex. (These differences are smaller than the typical Poisson $\sqrt(N)$ uncertainty on net counts.) Sources of scatter include differences in shape and size of the extraction regions, differences in filtering of events, and different algorithms for estimating source background. A small fraction of X-ray sources do have up to a factor of 2 times more counts in the \citetalias{2016arXiv160708860D} catalog, but these are all sources with close neighbors in the ACIS-Extract source list that are included as a single source by \citetalias{2016arXiv160708860D}.  

A comparison of X-ray luminosities ($LX$ from \citetalias{2016arXiv160708860D} vs.\ $LOG\_LTC$) for probable cluster members shows that X-ray luminosities derived here using XPHOT ($LOG\_LTC$) are on average 0.1~dex lower than those derived by \citetalias{2016arXiv160708860D} through spectral fitting, with 0.2~dex scatter. The magnitude of this shift is similar to the typical magnitude of the discrepancies found in the SFiNCs project between X-ray luminosities derived using the same methods we use here and from the published literature \citep[][their Figure~A2]{2017ApJS..229...28G}, albeit usually in the opposite direction. Typical uncertainty on $LOG\_LTC$ (statistical and systematic added in quadrature) is 0.5~dex.

\subsection{Infrared and Optical Photometry \label{nir.sec}}

The NIR $ZYJHK_s$ data were obtained from the VISTA Variables in the V\'ia Lact\'ea (VVV) survey \citep{2010NewA...15..433M,2012A&A...537A.107S}. VVV is a multi-epoch NIR survey that covers both the Galactic Bulge and an adjacent Galactic Disk region that was carried out using the 4.1-m VISTA telescope on Cerro Paranal. The VVV data were taken with VIRCAM \citep[VISTA Infrared CAMera;][]{2006SPIE.6269E..0XD}, a $4\times4$ array of Raytheon VIRGO $2048\times2048$ 20-$\mu$m-pixel detectors, with a pixel scale of $0.\!^{\prime\prime}34$. The individual detectors are separated by gaps, with a width that is 42.5\% of the detector width in the vertical direction and 90\% of the detector width in horizontal direction, forming the ``paw-print'' field of view that covers 0.59~square degrees. A series of 6 exposures with various horizontal and vertical shifts combine to form a single rectangular tile with an area of $1.5^\circ\times1.1^\circ=1.64$~square degrees \citep{2012A&A...537A.107S}. 

For color-color and color-magnitude diagram analysis, we used the first epoch $ZYJHK_s$ images from VVV tile ``d148'' in the disk region of the survey. These images all had top ``Quality Control'' grade and sub-arcsecond seeing. The observation log for these data is presented in Table~\ref{log.tab}. Aperture photometry was performed with v1.3 of the VISTA pipeline developed by the Cambridge Astronomical Survey Unit \citep[CASU;][]{2010ASPC..434...91L} and downloaded via the CASU webpage.\footnote{\url{http://casu.ast.cam.ac.uk/vistasp/}} Flags indicate morphological classification of sources, which are generated based on curve-of-growth analysis \citep{2004SPIE.5493..411I}. We make use of sources with flags ``-1'' stellar, ``-2'' borderline stellar, and ``-9'' saturated in the analysis. The flag ``0'' indicates noise, ``-7'' indicates bad pixels, and ``+1'' non stellar, and catalog entries with these flags are omitted. For sources that are saturated in the $JHK_s$ VVV images, photometry from the Two Micron All Sky Survey \citep[2MASS;][]{2006AJ....131.1163S} is substituted using the transformations from \citet{2013A&A...552A.101S} to convert the photometry in the 2MASS system to the VVV system. 

For variability analysis, photometry was obtained from $\sim$30 $K_s$-band epochs with net exposure times of 8 or 40~seconds\footnote{The first and last epochs had the longer exposure times.} in both tile ``d148'' and the adjacent tile ``d110.'' Only images with top ``Quality Control'' grade and arcsecond or subarcseond seeing were used. These images were observed between March 2010 and July 2015, with a mean cadence of 0.025~day$^{-1}$ (gaps between observations were distributed with first-quartile, median, and third-quartile time delays of 0.07, 1.0, and 20~days). Data reduction followed the photometry procedures described by \citet{2016arXiv160701795N}, who use PSF-fitting photometry on individual chip images to avoid spurious photometric variability caused by PSF variability. The photometry was extracted using the {\it DoPHOT} pipeline \citep{1993PASP..105.1342S}, and objects with {\it DoPHOT} flags indicating reliable photometry were kept. Photometry was calibrated using a set of isolated, non-variable stars in the images with VVV $K_s$-band magnitudes between 11 and 15~mag. 

In addition to the VVV photometry, public optical or infrared catalogs are available from surveys and publications. We have included  VPHAS+ photometry \citep{2014MNRAS.440.2036D}, $UBVRI$ (Johnson--Cousins system) and H$\alpha$ photometry from \citet{2013AJ....145...37S}, and {\it Spitzer}/IRAC photometry from the GLIMPSE survey \citep{2003PASP..115..953B}.\footnote{Note that the GLIMPSE survey region only overlaps the southern corner of the {\it Chandra}/ACIS-I field of view.} Matching between these catalogs and the VVV catalog is performed using TOPCAT \citep{2005ASPC..347...29T}.  Among the optical and NIR bands, there are significantly more counterparts for probable cluster members when going to longer wavelengths: e.g.\ more than 2 times as many $K_s$-band counterparts from VVV+2MASS than $V$-band counterparts from \citet{2013AJ....145...37S}. Reported right ascensions and declinations in the merged catalogs are taken from the VVV $K_s$-band catalog where available. 

\begin{deluxetable*}{llccccc}
\tablecaption{VISTA Observing Log \label{log.tab}}
\tabletypesize{\small}\tablewidth{0pt}
\tablehead{
\colhead{Band} & \colhead{FOV Center} &\colhead{Date} & \colhead{Exp.} & \colhead{Airmass} & \colhead{Seeing} & \colhead{Mag.\ Lim.}\\
\colhead{} & \colhead{(J2000)} &\colhead{} & \colhead{(s)} & \colhead{} & \colhead{(arcsec)} & \colhead{(mag)}
}
\startdata
$Z$ & 16:52:52.5 $-$41:34:10 & 2011-08-14T02:05:55&80&1.154&0.93&20.35\\
$Y$ &16:52:52.5 $-$41:34:11  &2011-08-14T01:58:51&80&1.140&0.84&19.73\\
$J$ &16:52:52.5 $-$41:34:10&  2010-03-26T07:38:12&80&1.124&0.88&19.05\\
$H$ &16:52:52.5 $-$41:34:11 & 2010-03-26T07:24:26&80&1.148&0.87&18.13\\
$Ks$ &16:52:52.5 $-$41:34:10  &2010-03-26T07:31:27&80&1.135&0.85&17.52\\
\enddata
\end{deluxetable*}

\subsection{NIR--X-ray Catalog Matching \label{match.sec}}

Matching between the X-ray sources and NIR sources was performed using the sky coordinate matching algorithm from the CRAN package {\it celestial} \citep{2016ascl.soft02011R}.
A preliminary match radius of 2$^{\prime\prime}$ was used to identify candidate matches, which were then pruned based on uncertainty in the source positions. Uncertainty on NIR source position was assumed to be $0.\!^{\prime\prime}$3, while uncertainty on X-ray source position was calculated by {\it ACIS Extract} and ranged from $0.\!^{\prime\prime}$1 to 1$^{\prime\prime}$ (median $0.\!^{\prime\prime}$2). Positions differing by less than 2 times the quadratically combined uncertainty on position are accepted. A total of 1735 matches were found between X-ray sources and VVV+2MASS sources, giving a match rate of 72\% for sources in the X-ray catalog.

In deep images of the Galactic Plane, the high numbers of sources in the NIR catalogs can lead to error in 1) whether an X-ray source has a NIR counterpart and, if so, 2) which NIR source is the correct counterpart \citep{2013ApJS..209...30N}. We investigate these two problems for the $K$-band--X-ray matching. For X-ray--$K$-band matching, 72\% of X-ray sources have a primary match (the match to the closest $K$-band source with the matching radius), while 1\% of sources have a secondary match (the second closest $K$-band source within the matching radius), and 0.04\% have a tertiary match (the match to the third closest $K$-band source within the matching radius). When multiple $K$-band matches are possible, in $\sim$70\% of the cases the primary match is the brightest source. This result agrees with the general expectation from \citeauthor{2013ApJS..209...30N} that when there are multiple possible matches, the correct match is usually brighter. 

The number of X-ray sources with spurious counterparts is estimated by artificially shifting the $\sim$700 X-ray sources without $K$-band matches and redoing the matching procedure. This simulation indicates that 90$\pm$10 (5\%) of counterparts are spurious for our catalog of 1735 X-ray/infrared matches. This contamination rate is significantly more optimistic than the contamination rate estimated by \citet{2006A&A...454.1047S} for matching between {\it XMM Newton} sources and NIR catalogs ($>$100 spurious counterparts for a catalog of 610 X-ray sources). The lower rate of such contaminants is an advantage of the higher spatial resolution of the {\it Chandra X-ray Observatory}. There is a strong correlation between $K$-band luminosity and X-ray net counts among actual matches, and this correlation is non-existent for the simulated spurious matches.

\subsection{Catalog Completeness Limits \label{completeness.sec}}

The detection sensitivity for X-ray point-sources is related to photon flux (photon~s$^{-1}$~cm$^{-2}$) in the bands used for source detection. This sensitivity varies across the ACIS-I field as a result of telescope vignetting and the degradation of the point-spread function with off-axis angle.\footnote{These effects are described in \S4 of the ``{\it Chandra} Proposers' Observatory Guide'' by the {\it Chandra} X-ray Center.}  As a result, a larger number of faint X-ray sources are detected near the center of the field, where sensitivity is greatest, which can produce an artificial clustering effect as noted by \citet{2011ApJS..194....2B}. For studies of cluster structure, the artificial clustering can be corrected by using only X-ray sources with photon fluxes greater than the completeness limit for the full X-ray catalog. Such a strategy has also been recommended by \citet{2009A&A...495..147A} to avoid interpretation of observational effects of varying sensitivity as astrophysical phenomenon, such as cluster mass segregation. \citet{2011ApJS..194....9F} and \citet{2014ApJ...787..107K,2015ApJ...802...60K,2015ApJ...812..131K} truncated X-ray selected samples of pre--main-sequence stars using photon-flux thresholds, allowing the identification of spatial structure masked in the full sample by artificial clustering. 

For NGC~6231, the completeness limit is estimated empirically using histograms of photon flux for X-ray point sources stratified by off-axis angle, using radial divisions at 0.$\!^\prime$3, 5.$\!^\prime$1, 6.$\!^\prime$3, 7.$\!^\prime$5, 8.$\!^\prime$3, and 12.$\!^\prime$3. The photon-flux limit increased with each larger-radius stratum, giving a completeness limit for the full sample of $\log F_\mathrm{photon}=-5.95$ in the 0.5--8.0~keV band. This limit is similar to the completeness limits of many of the MYStIX {\it Chandra} observations \citep{2013ApJS..209...27K}.

The completeness limits for the VVV catalogs reported in Table~\ref{log.tab} were calculated by artificial source insertion and extraction by the CASU v1.1 pipeline reduction, and these NIR catalogs are generally deeper than the X-ray catalogs. 

\section{Cluster Membership \label{members.sec}}

\subsection{Simulations of Contaminants \label{contam.sec}}

Classification of X-ray sources as cluster members and non-members is based on a decision tree, using X-ray, NIR, and optical properties. The sources of contaminants include extragalactic sources, foreground fields stars, and background field stars, with extragalactic sources being the more numerous \citep{2011ApJS..194....3G}. A variety of rules have been used to select probable members in X-ray studies of young stellar clusters, which include machine-learning strategies based on training sets \citep[e.g.,][]{2013ApJS..209...32B} and decision trees based on different criteria \citep[e.g.,][]{2017ApJS..229...28G}. However, unlike MYStIX and SFiNCs, mid-infrared photometry from the {\it Spitzer Space Telescope} is available only for a small region at the south edge of the field. Optical catalogs can be used to remove some likely foreground or background field stars, but optical photometry is only available for half as many stars as is VVV NIR photometry.

\begin{figure*}[t]
\centering
\includegraphics[width=1.0\textwidth]{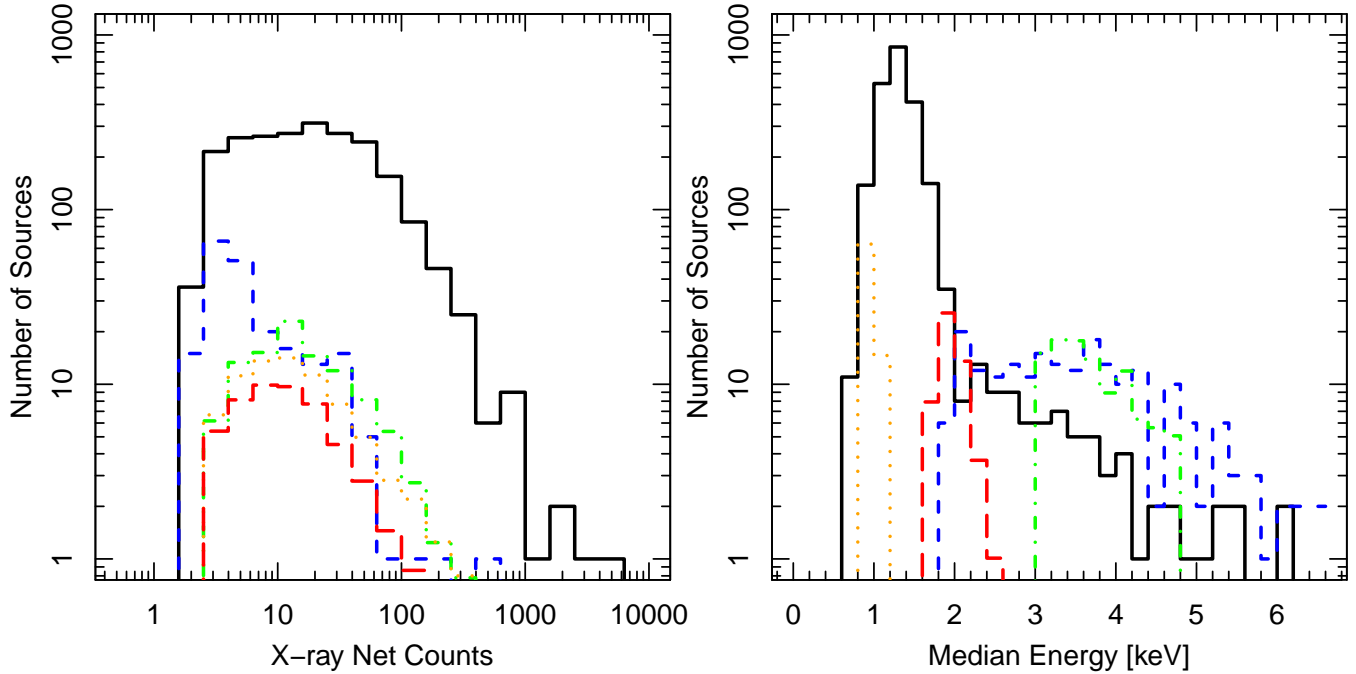} 
\caption{Left: Distributions of net counts ($NetCounts\_t$) for observed and simulated X-ray sources. Right: Distribution of median energies ($ME$) for observed and simulated sources. In both panels the distributions for ``probable cluster members'' are shown in black (solid line), distributions of ``unclassified'' sources are shown in blue (dashed line), distributions of simulated extragalactic sources are shown in green (dotdash line), distributions of simulated foreground sources are shown in orange (dotted line), and distributions of simulated background sources are shown in red (longdash line).
\label{sim.fig}}
\end{figure*}

We simulate X-ray contaminants (extragalactic sources, foreground field stars, and background field stars) to estimate the expected level of contaminants and see how their X-ray properties would compare with the properties of the observed stars. Extragalactic X-ray sources are mostly active galactic nuclei, with some starburst galaxies. These objects are seen through a Galactic neutral hydrogen column density $N_H=1.2\times10^{22}$~cm$^{-2}$ \citep{1990ARA&A..28..215D}, with little contribution to absorption from the region itself. Galactic field stars are typically not detected in the X-ray; however, the high number of field stars means that they are an important source of contaminants. 
We follow the simulation methods described by \citet{2011ApJS..194....3G}. For extragalactic sources, we use the extragalactic $\log N$--$\log S$ relationship from \citet{2003ApJ...588..696M} and assume power-law X-ray spectra based on relations from \citet{2001AJ....122.2810B}. Stars are simulated using the Besan\c{c}on models of the Galaxy \citep{2003A&A...409..523R}, and we assume thermal plasma models based on observed distributions of main-sequence and giant-star X-ray luminosities.  Artificial X-ray sources were simulated using the {\it fakeit} tool in the XSPEC software package \citep{1996ASPC..101...17A} and count rates were obtained using the Portable Interactive Multi-Mission Simulator \citep[PIMMS;][]{1993Legac...3...21M}. A cut based on the detection sensitivity of the NGC~6231 source list was then applied to the artificial sources to produce the final lists. These simulations produce 100$\pm$10 extragalactic sources, 80$\pm$10 foreground stars, and 50$\pm$8 background stars. 

The distributions of X-ray net counts ($NetCounts\_t$) and X-ray median energies ($ME$) for simulated contaminants and observed X-ray sources (with and without NIR counterparts) are shown in Figure~\ref{sim.fig}. 

Typical simulated extragalactic sources have $J>20$~mag, so these X-ray contaminants are not expected to have NIR counterparts in the VVV catalogs. In contrast, foreground and background field-star contaminants will have VVV counterparts, so they are harder to disentangle from cluster members, and the lack of strong absorption by a molecular cloud associated with NGC~6231 means that their NIR reddening and X-ray extinction will not be very different for these three classes. However, cluster members are expected to dominate foreground and background field stars by a ratio of 28:1 and 44:1, respectively, so, without additional evidence suggesting otherwise, an X-ray source with a NIR counterpart is more likely than not to be a cluster member. However, the 80+50=130 stellar contaminants may have $K_s$-band counterparts, and thus can constitute up to 7\% of the 1735 X-ray sources matched to NIR. We thus can conclude that 5\% (spurious matches) + 7\% (stellar contaminants) = 12\% of these X-ray sources with NIR counterparts are not valid members.  

 The Besan\c{c}on-model approach may be limited because it does not take the spiral structure of the Galaxy into account. \citet{2012A&A...538A.106R} note that spiral arms do not significantly contribute to NIR star counts in 2MASS. However, younger stellar populations in spiral arms are more likely to be detectable in X-ray emission, so background field-star contamination may be higher than estimated by the model.

\subsection{Probable Cluster Members \label{pcm.sec}}
Based on the arguments above, we classify any X-ray source with a NIR counterpart a ``probable cluster member'' unless further evidence suggests that it is either a foreground or background field star. Figure~\ref{new_sources.fig} guarantees that X-ray-derived field-star contamination is low.

Although optical photometry is not available for the faintest sources, these bands can help identify field stars when they are available. Foreground stars will lie above the cluster members on the $V$ vs.\ $V-I$ color-magnitude diagram, and we use the polygonal regions in the ($V$,$V-I$) color space from \citetalias{2016arXiv160708860D} (their Figure~4) to remove these sources. Background stars will lie below the zero-age main-sequence track on color-magnitude diagrams. We use both the $V$ and $I$ photometry from \citet{2013AJ....145...37S} and the deeper $r$ and $i$ photometry from VPHAS+ to identify these sources. Overall, 16 X-ray sources are identified as likely foreground field stars and 123 X-ray sources are identified as likely background field stars. The overall number of field stars is similar to the $\sim$130 field star predictions from the simulations, although the simulations anticipated more foreground contaminants and fewer background contaminants.

Variability of the X-ray light curve that is statistically significant ($p<0.01$ as measured by {\it ACIS Extract}) is also considered to be evidence of membership, and these sources are also included as ``probable cluster members.'' The X-ray contaminants are not expected to be as variable as pre--main-sequence stars, while pre--main-sequence stars typically have variability in the full (0.5--8.0~keV) X-ray band of 0.15--4.5~dex on time scales of 0.4--10 days due to stochastic flaring, with some contribution from variable absorption \citep{2012A&A...548A..85F}.  There are 356 variable X-ray sources that meet this criterion. 

Extragalactic sources are expected to have significantly higher X-ray median energies than cluster members, as seen in Figure~\ref{sim.fig} where their simulated $ME$ values range from 1.9~keV to 5~keV. Thus, sources with $ME<1.9$~keV are unlikely to be extragalactic, and are more likely to be either cluster members or field stars, so they are also included as ``probable cluster members.''  On visual inspection of the 459 X-ray sources with $ME<1.9$ that lack a match to a VVV source, many do appear to have VVV counterparts that are either just outside the $2\sigma$ match radius or are flagged as a non-point sources by the CASU pipeline, while others are located near bright NIR sources. However, some have no obvious explanation for the lack of a NIR counterpart. These sources typically have 4--20 X-ray net counts (median of 7 net counts).

All X-ray sources that are in catalogs of previously known Wolf-Rayet, O, and B-type cluster members \citep[compiled by][]{2006MNRAS.372..661S} are also included. The {\it Chandra} X-ray catalog provides improved detection of O and B-type stars, with 13 out of 13  (100\%) O-stars system detected and 41 out of 82 (50\%) B-star systems detected (see \S\ref{ob.sec}). However, the X-ray emission for many of systems containing later-type B stars (which are not expected to emit X-rays) may be produced by a T-Tauri companion. 

Late-B and A stars are not expected to produce X-ray emission, so stars with these spectral types are likely to be missing from the X-ray catalog \citep{2007MNRAS.381.1569H}. However, some AB stars may have pre--main-sequence companions that are detected in X-rays, and some $>$2.5-$M_\odot$ stars may have later spectral types of G or K if they are younger than $\sim$2.5~Myr. Overall, 114 of the X-ray selected probable cluster members lie in the region of the $V$ vs.\ $V-I$ diagram consistent with OB or AF stars (as defined by \citetalias{2016arXiv160708860D}).

Of the 2411 X-ray sources, 2148 are classified as ``probable cluster members,'' while sources that do not meet any of the above criteria are considered ``unclassified'' and likely include extragalactic sources (mostly quasars and Seyfert galaxies), field stars, cluster members with missing NIR photometry, and spurious sources. Spurious X-ray sources may be either fluctuations in the background level with significance $p<0.01$, or they may be X-ray events from the wings of bright sources that were erroneously deconvolved into distinct sources \citep[cf.][]{2013ApJS..209...27K}. Most spurious sources will lack NIR counterparts, and will therefore be listed as ``unclassified.'' Nevertheless, most of the faint (3--5 count) X-ray sources in our NGC~6231 catalog do have NIR counterparts (the black histogram in Figure~\ref{sim.fig}), which is evidence that these are bonafide X-ray sources. The ``unclassified'' sources making up the blue histogram can mostly be explained as being astrophysical contaminants rather than spurious sources.

We note that we do not distinguish between young stars in NGC~6231 and possible other members of the Sco~OB1 association that may be projected along the same line of sight. Superposition of clusters in the plane of the sky has been detected in other regions, like the Orion Complex, where \objectname[NGC 1980]{NGC~1980} and  \objectname[NGC 1981]{NGC~1981} lie in front of the \objectname[Orion Nebula Cluster]{Orion Nebula Cluster} \citep{2014A&A...564A..29B}. As in the case of Orion, source selection based on a simple X-ray/optical/infrared photometric analysis will not distinguish between these populations. 

Table~\ref{nirprop.tab} lists 2148 probable cluster members and their properties. Each cluster member is given an IAU designation: a prefix ``CXOVVV J'' followed by a sequence based on the truncated source coordinates. For each item, the table provides X-ray source, VVV, GLIMPSE, and optical photometry, spectral types for OB stars, and cross-correlation with cluster members from \citetalias{2016arXiv160708860D}. In addition, inferred stellar properties are presented (derived in the next sections) including an indicator of $K_s$ variability, absorption ($A_K$), estimated stellar age, stellar mass (estimated using both 3.2 and 6.4~Myr age assumptions), bolometric luminosity (based on the mass estimates), an indicator of $K_s$-band excess, and classifications of GLIMPSE infrared excess sources. 

\startlongtable
\onecolumngrid
\begin{deluxetable*}{lll}
\tablecaption{NGC 6231 CXOVVV Probable Cluster Member Catalog \label{nirprop.tab}}
\tablewidth{0in}
\tabletypesize{\scriptsize}

\tablehead{      
\colhead{Column Label} & \colhead{Units} & \colhead{Description} \\ 
\numberthecolumn & \numberthecolumn & \numberthecolumn  
\setcounter{column_number}{1}
}
\startdata
\\
Name                       &                         & IAU source name; the prefix is CXOVVV\\
Catalog\_RAdeg                      & deg                  & right ascension in catalog (J2000)                                         \\
Catalog\_DEdeg                     & deg                  & declination in catalog (J2000)                                             \\
Xray\_Name           &         & X-ray source name in IAU format \\
ME        & keV & X-ray Median Energy in the total 0.5--8.0 keV band\\
LogLx   & [erg~s$^{-1}$] & absorption-corrected X-ray Luminosity in the 0.5--8.0 keV band \\
PhotonFlux   & [cm$^{-2}$~s$^{-1}$] & log incident photon flux in the 0.5--8.0 keV band \\
  \multicolumn{3}{l}{\raggedleft {\bf Counterparts in Other Catalogs}}\\
HD & & OB star name in the HD catalog \citep{1936AnHar.100....1C} \\
CPD-41d                      &                         &  OB star name in the CPD catalog \citep{1897AnCap...4....1G}        \\
Segg                      &                         &  OB star name in the \citet{1968ZA.....68..142S} catalog                           \\
SBL98                      &                         &  OB star name in the \citet{1998AJ....115..734S} catalog                         \\
BVF99                      &                         &  OB star name in the \citet{Baume99} catalog                     \\
SpType                     &                         &  spectral type tabulated by \citet{2006MNRAS.372..661S}  \\
r\_SpType                     &                         &  references for Spectral type given by \citet{2006MNRAS.372..661S}  \\
DSM2016\_Name                     &                         &  source name in the \citetalias{2016arXiv160708860D} catalog \\
DSM2016\_Group                     &                         &  membership classification in the \citetalias{2016arXiv160708860D} catalog\\
VVVv                      &                         &   source name in the catalog of $K_s$ variables in VVV                         \\
  \multicolumn{3}{l}{\raggedleft {\bf IR and Optical Photometry}}\\
VVV\_RAdeg                      &                         &   right ascension of VVV source (J2000)                          \\
VVV\_Dedeg                      &                         &   declination of VVV source (J2000)                          \\
Ks\_VVV                      &          mag               &  VVV $K_s$-band magnitude                            \\
e\_Ks\_VVV                      &          mag               &  1$\sigma$ uncertainty                           \\
flag\_Ks\_VVV                      &    mag                     &   CASU flag                          \\
H\_VVV                      &             mag            &  VVV $H$-band magnitude                           \\
e\_H\_VVV                      &         mag                &  1$\sigma$ uncertainty                           \\
flag\_H\_VVV                      &     mag                    & CASU flag                            \\
J\_VVV                      &       mag                  &  VVV $J$-band magnitude                            \\
e\_J\_VVV                      &          mag               &   1$\sigma$ uncertainty                          \\
flag\_J\_VVV                      &     mag                    &  CASU flag                           \\
Y\_VVV                      &          mag               &  VVV  $Y$-band magnitude                           \\
e\_Y\_VVV                      &         mag                &  1$\sigma$ uncertainty                           \\
flag\_Y\_VVV                      &    mag                     &  CASU flag                           \\
Z\_VVV                      &         mag                &    VVV $Z$-band magnitude                          \\
e\_Z\_VVV                      &        mag                 &     1$\sigma$ uncertainty                        \\
flag\_Z\_VVV                      &    mag                     &   CASU flag                          \\
2MASS\_ID                      &                         &      source name in the 2MASS catalog                       \\
J\_2M                      &         mag                &    2MASS $J$-band magnitude                         \\
e\_J\_2M                      &   mag                      &    1$\sigma$ uncertainty                         \\
H\_2M                      &          mag               &      2MASS $H$-band magnitude                       \\
e\_H\_2M                      &     mag                    &   1$\sigma$ uncertainty                          \\
Ks\_2M                      &          mag               &            2MASS $K_s$-band magnitude                 \\
e\_K\_2M                      &     mag                    &     1$\sigma$ uncertainty                        \\
Qflg                      &                         &       2MASS flag                      \\
Rflg                      &                         &        2MASS flag                     \\
Bflg                      &                         &        2MASS flag                     \\
Cflg                      &                         &      2MASS flag                       \\
Xflg                      &                         &       2MASS flag                      \\
Aflg                      &                         &       2MASS flag                      \\
J\_synth                      &     mag                    &   merged 2MASS and VVV photometry in the 2MASS system\\
H\_synth                     &        mag                 &   merged 2MASS and VVV photometry in the 2MASS system  \\
Ks\_synth                      &       mag                  &  merged 2MASS and VVV photometry in the 2MASS system  \\
VPHAS\_ID                      &                         &   source name in the VPHAS+ catalog  \citep{2014MNRAS.440.2036D}                        \\
umag                      &        mag                 &   VPHAS $u$-band magnitude                          \\
e\_umag                     &    mag                     &   1$\sigma$ uncertainty                          \\
gmag                     &         mag                &    VPHAS $g$-band magnitude                         \\
e\_gmag                     &     mag                    &  1$\sigma$ uncertainty                           \\
r2mag                     &          mag               &    VPHAS $r2$-band magnitude                         \\
e\_r2mag                     &      mag                   &   1$\sigma$ uncertainty                          \\
Hamag                     &         mag                &       VPHAS $H\alpha$-band magnitude                      \\
e\_Hamag                     &    mag                     &  1$\sigma$ uncertainty                           \\
rmag                     &             mag            &     VPHAS $r$-band magnitude                        \\
e\_rmag                     &     mag                    &  1$\sigma$ uncertainty                           \\
imag                     &           mag              &        VPHAS $i$-band magnitude                     \\
e\_imag                     &     mag                    &    1$\sigma$ uncertainty                         \\
SSB2013\_ID                     &                         & source name in the \citet{2013AJ....145...37S} catalog\\
Vmag                     &         mag                &    $V$-band magnitude in the CIT system                     \\
e\_Vmag                     &    mag                     &   1$\sigma$ uncertainty                          \\
V\_I                     &     mag                    &   $V-I$ color                          \\
eV\_I                     &     mag                    &   1$\sigma$ uncertainty                          \\
B\_V                     &       mag                  &      $B-V$ color                       \\
e\_B\_V                     &      mag                   &  1$\sigma$ uncertainty                           \\
U\_B                     &       mag                  &    $U-B$ color                         \\
e\_U\_B                     &    mag                     &    1$\sigma$ uncertainty                         \\
Ha\_SSB                     &          mag               &    $H\alpha$ magnitude                      \\
eHa\_SSB                     &    mag                     &    1$\sigma$ uncertainty                         \\
GLIMPSE\_ID                     &                         &   source name in the GLIMPSE catalog \citep{2003PASP..115..953B}   \\
3\_6mag                     &        mag                 &   magnitude in the 3.6~$\mu$m band                          \\
e\_3\_6mag                     &     mag                    &    1$\sigma$ uncertainty                         \\
q\_3\_6mag                     &                         &    GLIMPSE flag                         \\
4\_5mag                     &       mag                  &    magnitude in the 3.6~$\mu$m band                         \\
e\_4\_5mag                     &     mag                    &    1$\sigma$ uncertainty                         \\
q\_4\_5mag                     &                         &         GLIMPSE flag                    \\
5\_8mag                     &           mag              &    magnitude in the 3.6~$\mu$m band                         \\
e\_5\_8mag                     &        mag                 &    1$\sigma$ uncertainty                         \\
q\_5\_8mag                     &                         &          GLIMPSE flag                   \\
8\_0mag                     &           mag              &    magnitude in the 3.6~$\mu$m band                         \\
e\_8\_0mag                     &        mag                 &   1$\sigma$ uncertainty                          \\
q\_8\_0mag                     &                         &       GLIMPSE flag                      \\
  \multicolumn{3}{l}{\raggedleft {\bf Stellar Properties}}\\
Ak                     &           mag              &   estimated absorption in the $K_s$-band ($A_K$)                         \\
age\_vi                     &       Myr                  &  age estimated from the $V$ vs.\ $V-I$ diagram                           \\
logM\_3\_2Myr                     &        [solar mass]                 &  estimated mass for an age of 3.2~Myr   \\
logM\_6\_4Myr                     &      [solar mass]                   &  estimated mass for an age of 6.4~Myr                           \\
logL\_3\_2Myr                     &      [solar luminosity]                   &  estimated luminosity for an age of 3.2~Myr                           \\
logL\_6\_4Myr                     &     [solar luminosity]                     &  estimated luminosity for an age of 6.4~Myr                           \\
Ks\_excess                     &                         &   Indicator of $K_s$-band excess                          \\
classI                     &                         &   Indicator of classification as a Class~I YSO                          \\
classII                     &                         &  Indicator of classification as a Class~II YSO                           \\
IR\_excess                     &                         &  Indicator of infrared excess  \\
\enddata
\vspace{5mm}
\end{deluxetable*}

\begin{figure*}[t]
\centering
\includegraphics[width=1\textwidth]{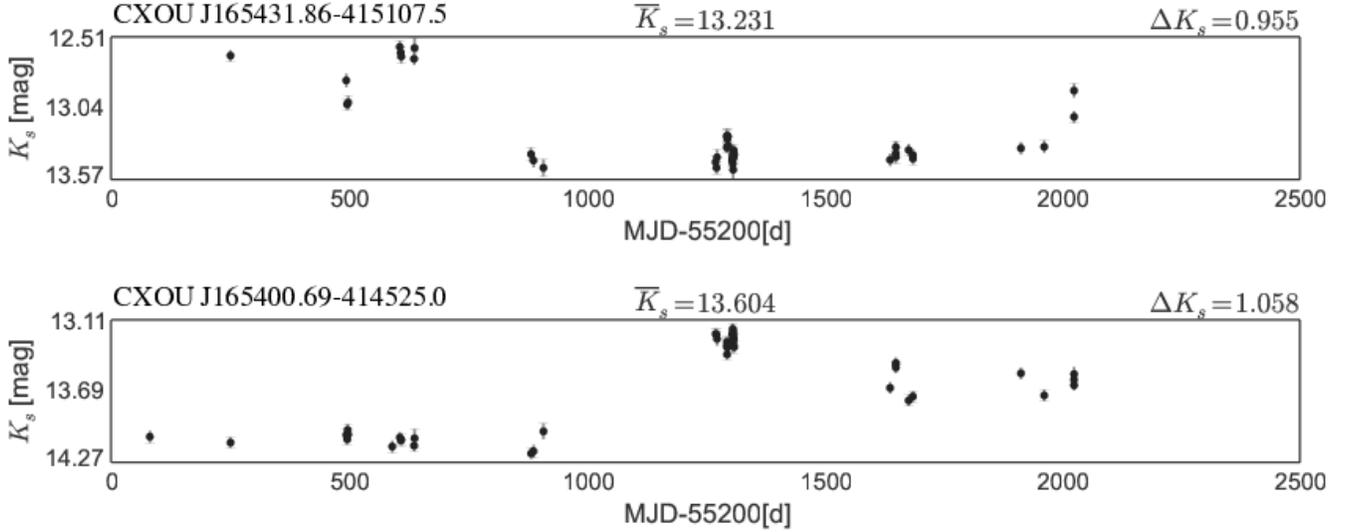} 
\caption{$K_s$ band light curve for the VVV variables CXOVVV~J165431.88-415107.6 (top) and CXOVVV~J165400.66-414524.9 (bottom) which lie within the {\it Chandra} field of view and have X-ray counterparts. 
 \label{vvv_variable.fig}}
\end{figure*}

\subsection{$K_s$-band Variable Stars \label{ksvar.sec}}

Pre--main-sequence stars can show optical and NIR variability up to several magnitudes due to rotational modulation by star spots, accretion from a circumstellar disk, and variable extinction from the circumstellar disk \citep{1945ApJ...102..168J, 1994AJ....108.1906H}. A study of variables in $\rho$~Ophiuchus by \citet{2014ApJS..211....3P} found $\sim$10\% to be variable in $JHK_s$ bands with amplitudes between 0.04 and 2.3~mag, and these stars showed both periodic and aperiodic behavior. Periodic variables were mostly associated with rotational modulation of a cool star spot, but in a few cases were associated with accretion hotspots or eclipses by a disk. Aperiodic variability was typically associated with variability in accretion rate or extinction. \citet{2015AJ....150..132R} studied NIR variables in the Orion Nebula Cluster, finding that protostars had the greatest amplitude of variability ($\sim$0.6~mag), followed by disk-bearing sources ($\sim$0.2~mag), and disk-free ($\sim$0.1~mag) stars. Most variability due to cool starspots will have amplitudes $<$0.2~mag in the $K$-band \citep[e.g.,][]{2001AJ....121.3160C,2013ApJ...773..145W}, below the level that can be reliably detected using the VVV data. Instead, the majority of the pre--main-sequence VVV variables are expected to have aperiodic variability due to variable absorption, variable accretion, and hot starspots. 

\begin{deluxetable*}{llrrrrc}[t!]
\tablecaption{VVV $K_s$-band Variables \label{variabes.tab}}
\tabletypesize{\small}\tablewidth{0pt}
\tablehead{
\colhead{VVVv} & \colhead{X-ray Source} & \colhead{RA} &\colhead{Dec} & \colhead{$\overline{K}_{s}$} & \colhead{$\Delta K_s$} & \colhead{$\log \chi^2$}\\
 \colhead{} & \colhead{(CXOU J)} & \colhead{J2000} & \colhead{J2000} & \colhead{mag} & \colhead{mag} & \colhead{}\\
\colhead{(1)} &  \colhead{(2)} &  \colhead{(3)} &  \colhead{(4)} & \colhead{(5)} & \colhead{(6)} & \colhead{(7)}
}
\startdata
165355.78-411441.5  &                  & 253.4824167 & -41.2448694 & 14.38& 0.48 &2.0\\
165357.41-414912.8 &165357.41-414912.8 &  253.4892083  &-41.8202194& 13.81& 0.44 &1.6\\
165359.78-421100.2 &                  &  253.4990833 & -42.1833889 &15.38 &0.81& 2.8\\
165400.65-414524.6 &165400.69-414525.0 & 253.5027083 & -41.7568306 &13.60 &1.06 &4.0\\
165401.82-411853.9 &                  &  253.5075833 & -41.3149611& 13.87 &0.68 &2.6\\
165402.98-404429.3 &                  &  253.5124167 & -40.7414806 &16.29 &1.23& 3.4\\
165403.00-423555.4 &                   & 253.5125000 & -42.5987111& 14.50 &0.36 &1.1\\
165405.57-411701.1 &                   & 253.5232083 & -41.2836306& 13.57& 1.33 &3.9\\
165410.34-413728.8 &                   & 253.5430833 & -41.6246611& 14.58 &0.79& 3.3\\
165410.70-415852.0 & 165410.72-415852.1 & 253.5445833 & -41.9811194 &13.57 &0.34& 0.9\\
\enddata
\tablecomments{ $K_s$-band variables in VVV tiles ``d148'' and ``d110''. Column~1: VVV variable (VVVv) designation. Column~2: Probable cluster member designation (CXOU).  Columns~3--4: Celestial coordinates. Column~5: Mean $K_s$-band magnitude. Column~6: Amplitude of the $K_s$-band variability. Column~7: $\log \chi^2$ statistic. (This table is available in its entirety in a machine-readable form in the online journal. A portion is shown here for guidance regarding its form and content.)}
\end{deluxetable*}

Variability in the $K$-band has been suggested by \citet{1999AJ....118..558K} as a method of identifying candidate young stellar objects. Like X-ray selection, this method does not require that young stars have circumstellar disks, which is necessary for an object to be classified as a young star based on infrared excess or H$\alpha$ emission. Identification of $K_s$-band variables is one of the primary objectives of the VVV survey, so the VVV variability is an excellent method of searching for NGC~6231 cluster members with large angular separations from the center of the cluster. Due to the large numbers of field-stars in the VVV survey, $K_s$ variability alone is not a definitive indicator of cluster membership, so objects identified this way should be regarded only as member candidates. In a study of high-amplitude $K_s$ band variables ($\Delta K_s>1$~mag) in the VVV, the fraction of pre--main-sequence stars was estimated to be 50\%, while many others are asymptotic giant branch (AGB) stars \citep{2016arXiv160206269C,2016arXiv160206267C}. The variables clustered at the location of NGC~6231 are more promising candidates than distributed variables, and may provide information about the structure of NGC~6231 on larger spatial scales than is possible with the X-ray observations. 

We search the $2.\!^\circ3\times1.\!^\circ5$ box around NGC~6231 covered by VVV tiles ``d148'' and ``d110'' for $K_s$-band variables. The procedures used to identify these sources are more fully presented by N.\ Medina (in preparation). Briefly, the subset from which variables are drawn includes point sources on the VVV tiles with more than 25 epochs.
For these stars, a variety of statistical measures are calculated: $\log \chi^2$ \citep[e.g.,][]{2014AJ....148...92R}, the \citet{von1941distribution} $\eta$ index, the \citet{1996PASP..108..851S} $J$ and $K$ indices for a single band, and the small kurtosis $\gamma$ and skewness $\kappa$ indices \citep{2011ApJ...733...10R}. The variable $\log \chi^2$ characterizes the statistical significance of variability without taking into account autocorrelation, while others like von Neumann's $\eta$ and Stetson's $J$ and $K$ identify light curves with autocorrelation, and the use of multiple statistics can be effective for characterization of variability \citep{2009MNRAS.400.1897S}.  A two-component normal mixture model is used to separate high-amplitude variables from other stars using these indices, with  $\eta$ and $\log \chi^2$ playing the most decisive roles. The variability analysis was performed twice for the small region of overlap between the tiles, once using the photometry from ``d148'' and again using the photometry from ``d110.'' A total of 22 variables were found in this region, 9 of which were identified for both of the tiles, and 13 of which were identified for only one tile.

\begin{figure*}[t]
\centering
\includegraphics[width=0.7\textwidth]{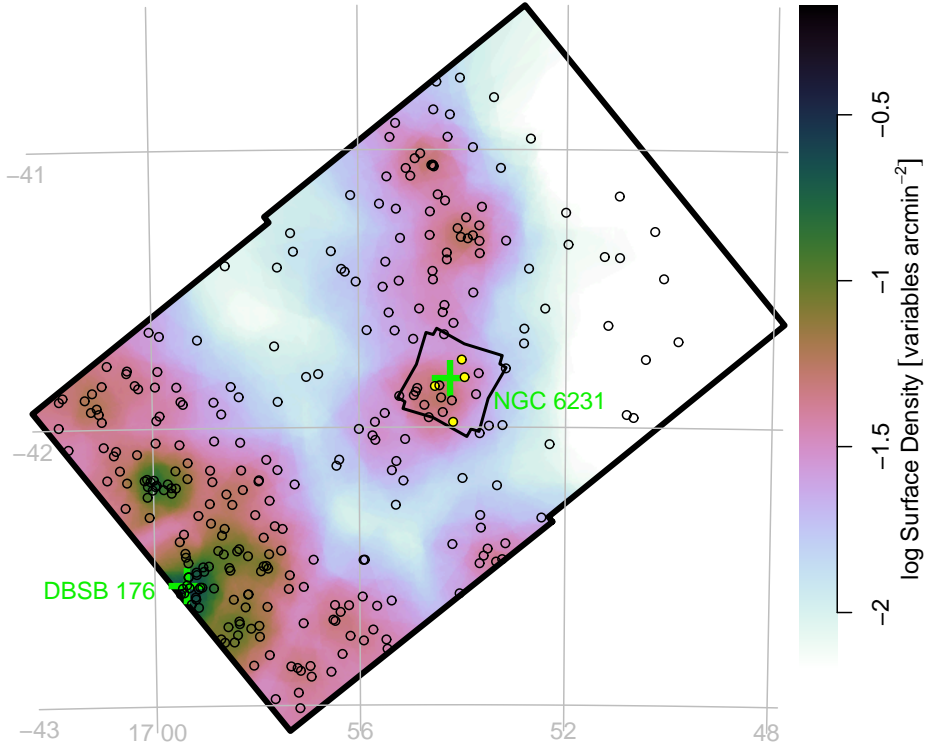} 
\caption{The spatial distribution of $K_s$ variable stars (black circles) in the VVV tiles ``d148'' and ``d110.'' A smoothed surface density map of these objects is shown using the color scale. The map shows that some $K_s$ variables are associated with clusters like NGC~6231, while others are distributed in the field. A gradient in surface density increases toward the Galactic Plane (lower left). The field of view of the tiles and the field of view of {\it Chandra} are outlined in black, and the four $K_s$ variables with X-ray counterparts are highlighted in yellow. Known clusters within the field of view are indicated.  
 \label{vvv_variable_density.fig}}
\end{figure*}

Table~\ref{variabes.tab} provides the catalog of 295 variables selected by the method described above, and provides coordinates, mean $K_s$-band magnitude, $\Delta K_s$, and $\log \chi^2$ for the sources. Out of the full catalog of 295 variables, $\sim$40 appear to be clustered at the approximate location of NGC~6231, while 16 lie within the {\it Chandra} field of view. Out of these 16, 4 were detected by the X-ray observations, including CXOU~J165357.41-414912.8, CXOU~J165400.69-414525.0, CXOU~J165410.72-415852.1, and CXOU J165431.86-415107.5. Figure~\ref{vvv_variable.fig} shows example light curves of two of these variables, both of which appear have aperiodic variations in the VVV data.

The spatial distribution of these sources is shown in Figure~\ref{vvv_variable_density.fig}. The data are adaptively smoothed using the {\it adaptive.density} algorithm from the R package Spatstat \citep{baddeley2005spatstat}. There is a local peak in density of stars that corresponds to the NGC~6231 cluster. In addition, there appears to be a population of unclustered variable objects, and a gradient in surface density, with increasing density near the Galactic Plane to the lower left of the image. Several other clusterings of variables lie to the north, south, and south east of NGC~6231---one of these was associated with the star cluster  DBSB~176, associated with \objectname[IRAS 16558-4228]{IRAS~16558-4228} \citep{2007ApJ...659.1360W,2016A&A...585A.101K}. Although DBSB~176 likely has many fewer total cluster members than NGC~6231, it has a larger number of members with $K_s$ variability, which may be an effect of it having a younger stellar population. Mid-infrared images of DBSB~176 show significant nebulosity in the region, suggesting active star formation. Other clusterings of $K_s$ variables seen in these VVV tiles may also be candidate star clusters, and these will be further investigated in Paper~II.

\newpage

\section{Properties of Cluster Members\label{prop.sec}}

\subsection{X-ray Properties \label{xray2.sec}}

\begin{figure*}
\centering
\includegraphics[width=0.7\textwidth]{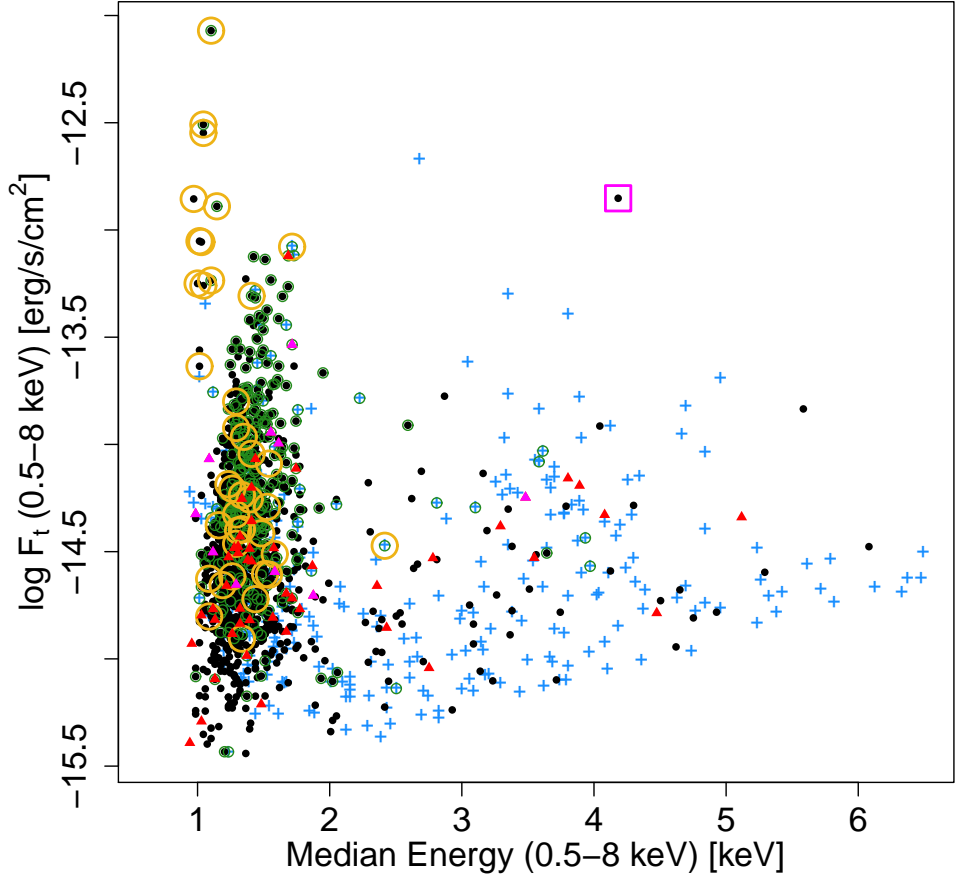} 
\caption{The X-ray ``color-magnitude diagram'' of NGC~6231 with total-band flux ($0.5-8.0$~keV) vs.\ total-band median energy. X-ray sources with $ZYJHK_s$ counterparts (black circles) and those without (blue crosses) are indicated. Green circles indicate that an X-ray source is variable, and orange circles indicate that it is matched to a object in a catalog of known O- and B-type cluster members. Likely field-star contaminants in the foreground (magenta triangles) and background (red triangles) are indicated. The magenta box marks the Wolf-Rayet star. \label{xcmd.fig}}
\end{figure*}

Figure~\ref{xcmd.fig} shows the X-ray ``color-magnitude diagram'' for all X-ray sources in the NGC~6231 {\it Chandra} field (both cluster members and contaminants) using observed flux in the total (0.5--8.0~keV) band ($F_t$) for the ``magnitude'' and median energy of observed X-ray photons in that band ($ME$) as the ``color.''  Black circles mark X-ray sources with NIR counterparts, while blue crosses mark sources without NIR counterparts. In addition, marks are shown for sources with X-ray variability (green circles), O or B spectral types (orange circles), and the Wolf-Rayet classification (magenta square). Likely field-star contaminants are also indicated (magenta and red triangles for foreground and background stars, respectively).
The bulk of the black, green, and orange points have $ME$ between 1 and 2~keV, while most of the blue points have $ME>2$~keV. 
 X-ray ``color-magnitude diagrams'' for 10 other star-forming regions are shown by \citet{2013ApJS..209...27K}.  They are similar to the NGC~6231 diagram with different ratios of lightly and heavily obscured X-ray sources. 

X-rays from low-mass pre--main-sequence stars are due to tens-of-millions of kelvins gas in the stellar coronae, which is heated by magnetic reconnection. Typical temperature components range from $kT\sim 0.8$~keV to 4~keV with flares up to 10~keV \citep{2005ApJS..160..319G,2008ApJ...688..418G}, which produce X-ray sources with median energies (unabsorbed) ranging from 1.0 to 1.5~keV \citep{2010ApJ...708.1760G}. The majority of the probable cluster members in NGC~6231 are this type of X-ray source, which mostly lie along a fairly narrow locus on the X-ray ``color-magnitude diagram.'' A slight increase in spectral hardness with increasing flux is related to an increase in plasma temperatures for more massive pre--main-sequence stars \citep{2010ApJ...708.1760G,2013ApJS..209...27K}.

X-ray emission from massive stars is typically greater than, or similar to, the X-ray emission of the most X-ray luminous pre--main-sequence stars \citep[e.g.,][]{2011ApJS..194....6P}. X-rays from both O and early B-type stars originate in stellar winds \citep{1980ApJ...241..300L,1988ApJ...335..914O,1999ApJ...520..833O,2008MNRAS.386.1855C}. Wind shocks due to the line-deshadowing instability will typically lead to soft ($ME\leq1$~keV), nearly constant X-ray emission, while the detection of a harder and variable X-ray component for some massive stars could be explained by colliding winds in a binary system \citep{2000ApJ...538..808Z} or magnetically channeled winds \citep{1997A&A...323..121B,2005ApJ...628..986G}.  
On the X-ray ``color-magnitude diagram'' all of the O stars and some early-B stars lie at the upper end of the flux distribution and have X-ray median energies $ME$ ($\sim$1~keV), consistent with X-ray emission with a wind origin. However, a number of X-ray sources matched to B stars lie along the locus for pre--main-sequence stars. Section~\ref{lindroos.sec} provides a more detailed discussion of pre--main-sequence binary companions to B stars.  

Absorption by interstellar material will harden the X-ray spectrum and decrease the X-ray flux, shifting sources to the lower-right on the X-ray ``color-magnitude diagrams.'' This absorption is primarily due to He and inner shell electrons in C, N, O, Ne, Si, S, Mg, Ar, and Fe atoms, which may be in any phase \citep{2000ApJ...542..914W}. In embedded star-forming regions, like those in MYStIX, a substantial population of the pre--main-sequence stars may have $ME>2$~keV due to absorption by the natal molecular cloud. In the case of NGC~6231, the lack of substantial absorption from clouds means that pre--main-sequence stars with high $ME$ likely have significant local absorption. 
A sample of 50 objects from the CXOVVV catalog with unusually hard X-ray spectra are discussed further in Section~\ref{hard.sec}.

The simulations of X-ray contaminants predict that foreground and background field stars will have $ME$ and $F_t$ values similar to those of cluster members. The distributions of stars classified as foreground or background objects in Figure~\ref{xcmd.fig} overlap the lightly absorbed cluster members, but some background sources have $ME$ up to 5~keV. The simulated extragalactic sources typically have $2<ME<4.5$~keV. On Figure~\ref{xcmd.fig}, most X-ray sources without NIR counterparts have $ME$ in this range, and are thus probable extragalactic sources. Overall, 199 X-ray sources without NIR counterparts have $ME>2$~keV compared to the 100$\pm$10 simulated extragalactic X-ray sources. However, other types of cosmic object can also produce hard X-ray sources without NIR counterparts.  One of the brightest sources in our X-ray catalog is \object[2XMM J165334.4-414423]{CXOU~J165334.41-414423.6} ($=$~2XMM~J165334.4$-$414423), with $\log F_t=-12.7$ [erg~s$^{-1}$~cm$^{-2}$], $ME=2.7$~keV, and no NIR match. \citet{2014ApJ...780...39L} classified this object as a periodically varying magnetic cataclysmic variable that is unrelated to NGC~6231.


\subsection{NIR Properties \label{nir2.sec}}

\begin{figure*}[t!]
\centering
\includegraphics[width=1.0\textwidth]{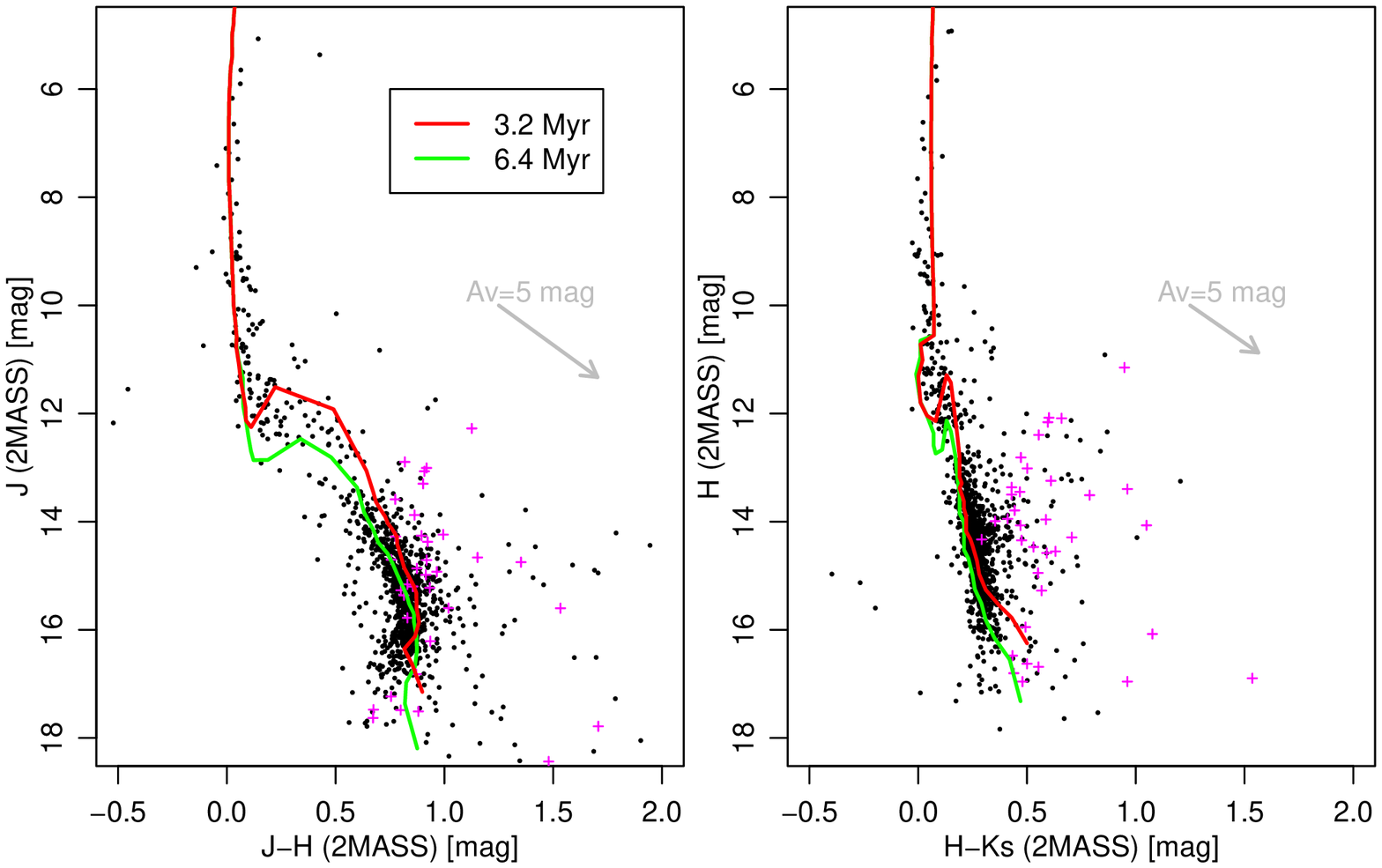} 
\caption{Color-magnitude diagrams of the probable cluster members are shown using the 2MASS $JHK_s$ system. Stellar photometry comes from 2MASS and VVV $JHK_s$ photometry converted using color equations. Lines indicate the 3.2 and 6.4~Myr pre--main-sequence isochrone models from \citet{2000A&A...358..593S} for an assumed distance modulus of 11.0 and an absorption of $A_V=1.6$~mag. Sources with infrared excess are marked with magenta crosses. The gray arrows are the $A_V=5$~mag reddening vectors of \citet{1985ApJ...288..618R}. 
 \label{cmd.fig}}
\end{figure*}

\begin{figure}[h]
\centering
\includegraphics[width=0.45\textwidth]{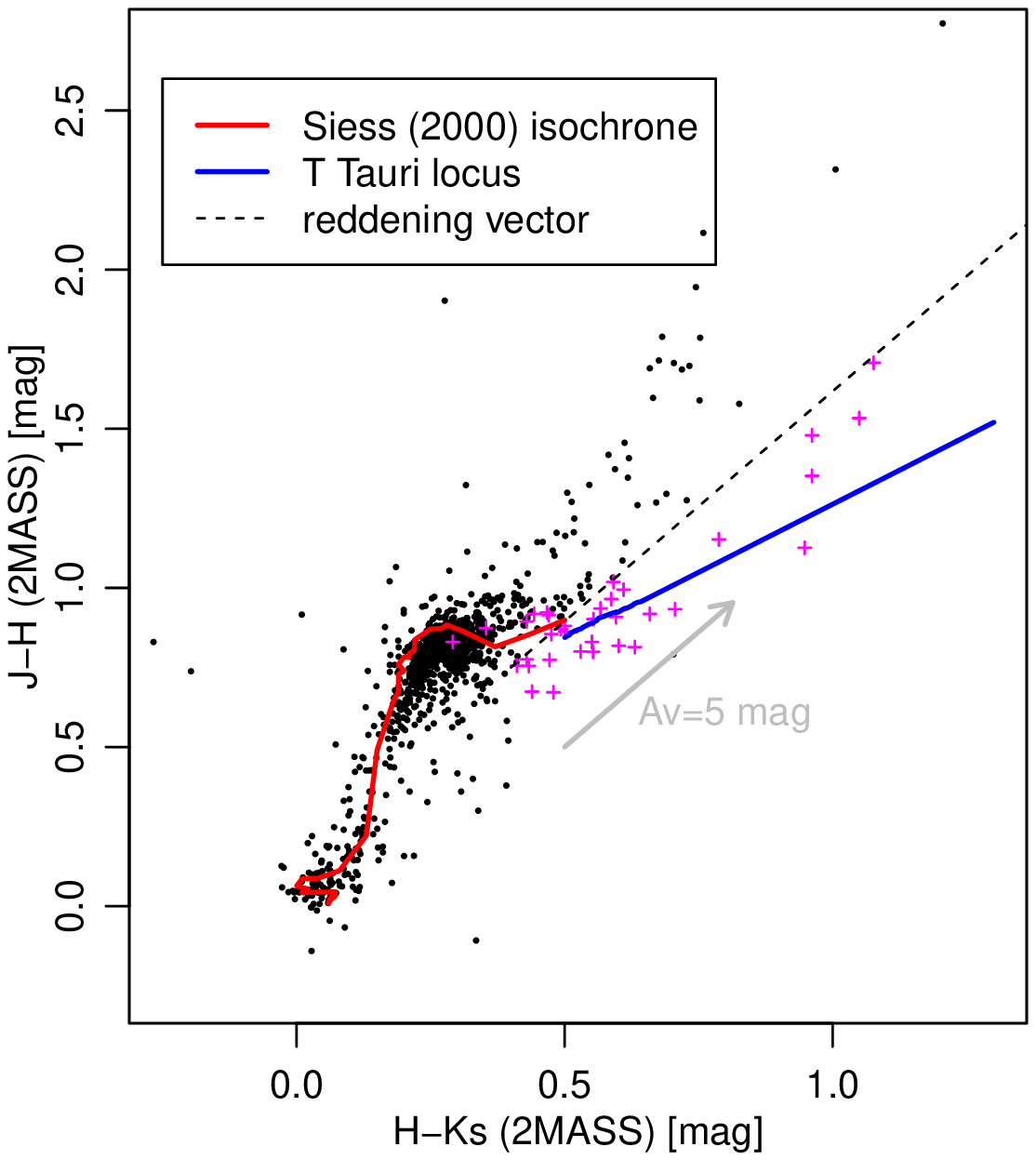} 
\caption{A color-color diagram of the probable cluster members is shown using the 2MASS $J-H$ vs.\ $H-K_s$ colors. Stellar photometry comes from 2MASS and VVV $JHK_s$ photometry converted using color equations. The reddened 3.2 and 6.4~Myr \citet{2000A&A...358..593S} isochrones lie on an identical locus in this color space, shown in red. The reddened T Tauri locus from  \citet{2014ApJ...787..108G} is shown in blue. A dashed line indicates a reddening vector starting from the colors of a 0.1~$M_\odot$ star, which can be used to separate stars with and without infrared excess. Sources with infrared excess are marked with magenta crosses, most of which are obtained from this diagram because the {\it Spitzer} coverage of the field of view is very incomplete. 
 \label{ccd.fig}}
\end{figure}

Figures~\ref{cmd.fig} and~\ref{ccd.fig} show NIR color-magnitude diagrams and a color-color diagram for probable cluster members in NGC~6231. The stars are plotted on both diagram using either 2MASS photometry or VVV photometry converted to the 2MASS system using the \citet{2013A&A...552A.101S} color transformations. Model isochrones for 3.2~Myr and 6.4~Myr from \citet{2000A&A...358..593S} are shown, also converted to the 2MASS system, and reddened by $A_V=1.6$~mag.

 The color-magnitude diagrams ($J$ vs.\ $J-H$ and $H$ vs.\ $H-K$) have relatively narrow distributions of points, revealing that there is not a significant range in extinction within the cluster, unlike other very young clusters that are still embedded in molecular clouds. The width of the dispersion of points in color may be due to slight differential reddening, variability, binarity, age spreads, and photometric uncertainties. 
Although, it is not possible to differentiate the 3.2 and 6.4-Myr models for low and high-mass stars on these color-magnitude diagrams, intermediate mass stars in the hooked region of this isochrone are sensitive to age \citep{2013AJ....145...46L}. These stars are enclosed by the two isochrones on the $J$ vs.\ $J-H$ diagram, suggesting that the ages of these stars are between 3.2 and 6.4~Myr.

Absorption in the NIR is estimated from the $J-H$ vs.\ $H-K_s$ diagram using either 2MASS photometry or VVV photometry converted into the 2MASS photometric system. The \citet{2000A&A...358..593S} isochrone is used to provide intrinsic stellar colors for stellar photospheres. (The complete overlap in the intrinsic NIR colors for 3.2-Myr and 6.4-Myr isochrones means absorption estimates do not depend on assumed stellar age.) For stars with infrared excess, we use the T-Tauri locus presented by \citet{2014ApJ...787..108G} based on observations of Taurus stars by \citet{2010ApJS..186..111L}. To estimate absorption, stars are shifted along a dereddening vector, based on NIR absorption relations from \citet{1985ApJ...288..618R}, until they reach the locus of intrinsic colors. In principal, there can be up to two possible absorption solutions for a star (one low-mass, low-absorption solution and one high-mass, high-absorption solution). However, in practice, most stars have sufficiently low absorptions that the nearest part of the intrinsic-color locus likely corresponds to the correct mass range.

The mean value of absorption of the stars is $A_K=0.17$~mag (corresponding to $A_V=1.6$~mag). Typical calculated absorptions range from $A_K=0.086$~mag ($A_V=0.76$~mag; first quartile) to $A_K=0.21$~mag ($A_V=1.9$~mag; third quartile). A lack of correlation between $A_K$ and $ME$ in this range suggests that much of this scatter is due to uncertainties in photometry. 

A small fraction of sources in the CXOVVV catalog ($\sim$2\%) have $0.5<A_k<1$~mag ($4.5<A_V<9$~mag). The foreground cloud is incapable of producing absorptions this high, so these objects may either be field stars in the background, or may be objects with high local absorption (e.g., objects with absorption from circumstellar material). A more thorough discussion of reddened NIR sources and hard $ME$ sources is provided in Section~\ref{hard.sec}.

\vspace{-2mm}
\subsection{Mass Estimates}

Masses are estimated from each of the dereddened $J$, $H$, and $K_s$-band magnitudes using the mass--magnitude relations from the \citeauthor{2000A&A...358..593S} models. The $Z$ and $Y$ bands are excluded from this analysis because predictions for these bands are not provided by the models.

Given that NIR magnitudes decrease during the pre--main-sequence evolution of a star, masses estimated from these magnitudes will be dependent on the ages that are assumed for the stars, with younger ages systematically yielding lower masses and older ages systematically yielding higher masses. We estimate masses using two possible ages, an age of 3.2~Myr and an age of 6.4~Myr, in order to see the magnitude of this effect on stellar mass estimates. Masses estimated using 6.4~Myr are $\sim$1.4~times greater ($0.14\pm0.10$~dex) than masses estimated using 3.2~Myr. The magnitude of this effect is greatest for lower mass stars ($M<0.3$~$M_\odot$ with 0.2--0.3~dex differences) and there is no difference for stars that have reached the main-sequence. 

The bolometric luminosities calculated using these two models will also differ, with the greatest deviation occurring for the mass range $2<M<5.6$~$M_\odot$ where the maximum deviation is a factor of $>$10 due to inconsistencies in mass estimate. However, outside this mass range, the deviation in estimated bolometric luminosity is minor.

\subsection{Infrared Excess}

The $J-H$ vs.\ $H-K_s$ color-color diagram may also be used to define sources with infrared excess in the $K_s$ band. To define a region on this color-color diagram to identify $K_s$-excess stars we use two lines: a reddening vector that starts from the intrinsic colors of a 0.1-$M_\odot$ star and the T-Tauri locus. (We accept stars lying up to 0.1~mag below the T-Tauri locus due to uncertainties in photometry; however, sources lying significantly below this line may have bad photometry or be reddened massive stars.) Overall, 30 $K_s$-excess sources are identified out of 1143 probable cluster members plotted on the diagram, corresponding to an observed fraction of 2.6\%. Nevertheless, $K_s$ excess is only sensitive to the hot inner disks of stars, and observed disk fractions for young clusters typically increases when longer wavelength photometry is also included \citep[e.g.][]{2001ApJ...553L.153H}.

NGC~6231 lies at the edge of the GLIMPSE survey of the {\it Spitzer Space Telescope}, so a low fraction of cluster members on the south side of the field of view have photometry in the IRAC 3.6, 4.5, 5.8, and 8.0~$\mu$m bands. We use the criteria from \citet{2009ApJS..184...18G} to identify Class~I and II young stellar objects (YSOs) using 1) the $JHK_s[3.6][4.5]$ bands and 2) the $K_s[3.6][4.5][5.8][8.0]$ bands. Given that the NGC~6231 field does not have significant nebulosity, this will not strongly affect our ability to detect disks. However, crowding of the field and the complex Galactic line of sight make the presence of mid-infrared excess sources unrelated to the cluster more likely. Using \citeauthor{2009ApJS..184...18G}'s first method, 0 Class~I sources are identified and 8 Class~II sources are identified. Using their second method, 0 Class~I sources are identified and 13 Class~II sources are identified. Overall, a total of 16 YSOs are identified out of 92 probable cluster members with {\it Spitzer}/IRAC photometry, corresponding to an observed fraction of 15\%.

Based on the exponential fit to disk fractions in various star-forming regions by \citet{2009AIPC.1158....3M}, a 3.2-Myr-old cluster would be expected to have a disk fraction of $\sim$30\% and a 6.4-Myr-old cluster would be expected to have a disk fraction of $\sim$8\%. However, selection effects in the crowded mid-infrared images makes the sensitivity of our {\it Spitzer}/IRAC infrared-excess sample difficult to estimate.

\subsection{Estimates of Median Stellar Age\label{age.sec}}

Ages are difficult to estimate for individual pre--main-sequence stars because a star's placement on the Hertzsprung-Russell diagram, or on various color-magnitude diagrams, is affected by episodic accretion \citep{2009ApJ...702L..27B}, uncertainty in models of pre--main-sequence stellar interiors (due to convection [Chabrier et al.\ 2007], magnetic fields [Mohanty et al.\ 2009], and rotation [Jeffries 2009]), \nocite{2007A&A...472L..17C,2009ApJ...697..713M,2009IAUS..258...95J}
binarity, and variability \citep{2011MNRAS.418.1948J}. More complete discussions of these problems are provided by \citet{2012RAA....12....1P} and \citet{2014ApJ...787..108G}. Nevertheless, estimates of age that have high statistical uncertainty may still be capable of revealing real differences in the median ages of different groups of stars. For example, statistical methods can be used to find spatial gradients in average stellar age, revealing progressive star formation \citep[][]{2009ApJ...699.1454G,2012MNRAS.426.2917G}. We use two independent methods to estimate stellar ages, the $J$ vs.\ $L_X$ relations following \citet{2014ApJ...787..108G} and the $V$ vs.\ $V-I$ diagram following \citetalias{2016arXiv160708860D}.

\begin{figure*}[t]
\centering
\includegraphics[width=0.9\textwidth]{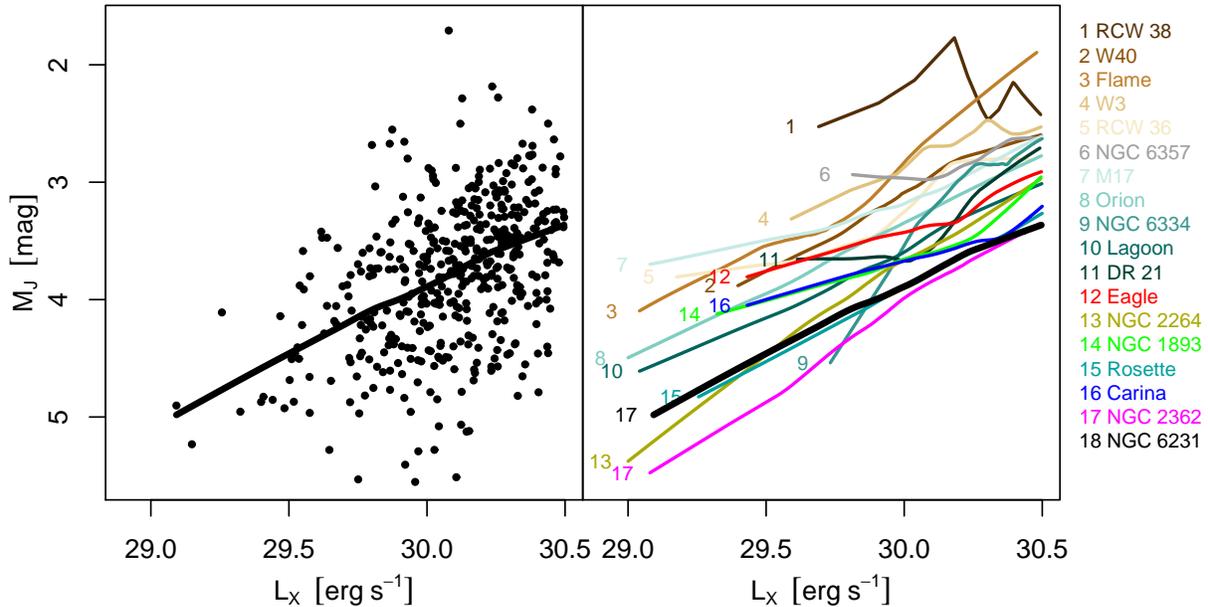} 
\caption{Left: X-ray luminosity is plotted against dereddened absolute $J$-band magnitude for NGC~6231 cluster members. The black line shows the local-regression (LOWESS) curve fit to the data. Right: Local-regression curves are shown for all MYStIX star-forming regions included in \citet{2014ApJ...787..108G} in comparison to NGC~6231. \citet{2014ApJ...787..108G} identified median age as being primarily responsible for the vertical stratification. (Stars with X-ray luminosities greater than $\log L_X>30.5$ are excluded by \citet{2014ApJ...787..108G} as many of these stars could be high-mass.)
 \label{lx_j.fig}}
\end{figure*}

For the star-forming regions in the MYStIX study (all younger than 5~Myr), \citet{2014ApJ...787..108G} found that relations between absolute absorption-corrected $J$-band magnitude ($M_J$) and X-ray luminosity ($L_X$) differ in different regions, and the differences can be attributed to median stellar age in each region (hereafter the age derived from the $M_J$ vs.\ $L_X$ diagram is denoted $Age_{JX}$). This can be shown in a plot of  $M_J$ (ordinate) vs.\ $L_X$ (absissa), with a non-parametric regression curve used to show the center of the distribution. Here, age produces a vertical shift, with older star-forming regions having higher values of $M_J$ for the same $L_X$ value compared to younger star-forming regions. This effect is shown in Figure~\ref{lx_j.fig}---the panel on the left shows absorption corrected absolute $J$-band magnitude vs.\ $L_X$ for low-mass\footnote{We also excluded highly absorbed sources ($A_V>3$~mag) that may include some background contaminants, but find their effect on the result is negligible.} NGC 6231 cluster members, which is fitted with a local-regression \citep[LOWESS;][]{cleveland1979robust,cleveland1981lowess} curve. The panel on the right shows LOWESS regression curves for 17 star-forming regions from the MYStIX project compared to the curve for NGC~6231.  This plot shows that NGC~6231 is most similar to the most evolved clusters included in the MYStIX study, including the Rosette Nebula Cluster, NGC~2264, and NGC~2362.

The stars used for the $Age_{JX}$ analysis described by \citet{2014ApJ...787..108G} must have a sufficient number of X-ray photons to obtain an estimate of $L_X$, they must have uncertainties $<$0.1~mag on $JHK_s$ colors and magnitudes, must have no $K_s$ excess, their X-ray luminosities must be $<$10$^{30.5}$~erg~s$^{-1}$, and they must have $J-H>0.5$. The last two requirements are used to remove stars with masses $\gtrsim$1.2~$M_\odot$ stars from the sample.  Overall, 456 stars contributed to the calculation. Individual $Age_{JX}$ values are imprecise estimates of stellar age due to the large known scatter in the $L_X$--$M$ relation, but $Age_{JX}$ can be more informative for determining the median age for groups of stars \citep{2014ApJ...787..109G,2014ApJ...787..108G}.

For NGC~6231 the median value of $Age_{JX}$ is $3.2\pm0.2$~Myr, where uncertainty of the median is calculated by bootstrap resampling. If the distance to NGC~6231 were 1.37~kpc, as estimated from the Gaia data, the estimated median age would increase to $3.7\pm0.2$~Myr. These values are consistent with the previously published age estimates that were inferred both from pre--main-sequence stellar evolution and from the main-sequence turn-off for O-type stars (\S\ref{previous.sec}). 

For probable cluster members with optical photometry, the $V$ vs.\ $V-I$ diagram may be used to estimate stellar ages based on the \citeauthor{2000A&A...358..593S} pre--main-sequence evolutionary models. This method may provide more precise age estimates for individual stars that have $V$ and $I$-band magnitudes because the $Age_{JX}$ estimates are subject to the large statistical scatter in the X-ray--stellar-mass relation. However, ages estimated in this way are still subject to the other systematic effects due to uncertainties in models and astrophysical effects that can affect stellar properties. $V-I$ colors are measured for approximately half of the probable cluster members. 

Ages from the $V$ vs.\ $V-I$ diagram are estimated for a variety of assumed absorptions, ranging from $A_V=0.5$ to 2.0~mag, in steps of 0.1~mag. For an $A_V=1.6$~mag, the median age is 3.3~Myr (1.9--4.7~Myr interquartile range). If the absorption is higher, the estimated median age would increase---for example, $A_V=2.0$~mag yields 4.0~Myr (2.2--6.4~Myr). On the other hand, if the absorption were lower the median age would decrease---for example, $A_V=1.0$~mag yields 2.6~Myr (1.7--3.7~Myr). A median age of 3.3~Myr is completely consistent with a median age found by the $Age_{JX}$ method of 3.2~Myr within the uncertainties on the estimates. Given that $A_V=1$~mag and $A_V=2$~mag are likely lower and higher, respectively, than the typical absorption of a star in the cluster, we conclude that most of the stars in the optical sample are less than 6.4~Myr old. Ages estimated with this method assuming an average extinction of $A_V=1.6$~mag are given in Table~\ref{nirprop.tab} (denoted $Age_{VI}$).

Previous work has suggested that the cluster has a significant age dispersion \citep{2013AJ....145...37S,2016arXiv160708860D}. Using the same photometric catalog as \citetalias{2016arXiv160708860D}, but with an increased sample of faint sources, we also find a distribution of $Age_{VI}$ estimates over several million years, regardless of our assumption about absorption. Therefore, the existence of a 6.4~Myr HMXB is not inconsistent with a younger median age. If the progenitor of the HMXB were among the first generation of O stars to go supernova 3~Myr ago, this could have led to the end of star formation around this time. Several numerical simulations of star-formation in collapsing molecular clouds have suggested that the star-formation rate may increase until the point that star formation is ended by destruction of molecular clouds \citep[e.g.,][]{2012MNRAS.420.1457H}. Such a scenario for NGC~6231 could lead to most stars having been formed not long before the end of star formation in the cluster, even if star formation started several million years earlier.

Given that there may be significant systematic uncertainties in the median age of the cluster and that there is evidence of an age spread of several million years, we estimate stellar properties (e.g.\ stellar mass) using both the lower median age estimate of 3.2~Myr and the higher age estimate of 6.4~Myr to investigate the effect of this assumption on the derived quantities.

\subsection{X-ray and Bolometric Luminosities and Stellar-Mass Relations \label{xrayir.sec}}

\begin{figure*}
\centering
\includegraphics[width=1.0\textwidth]{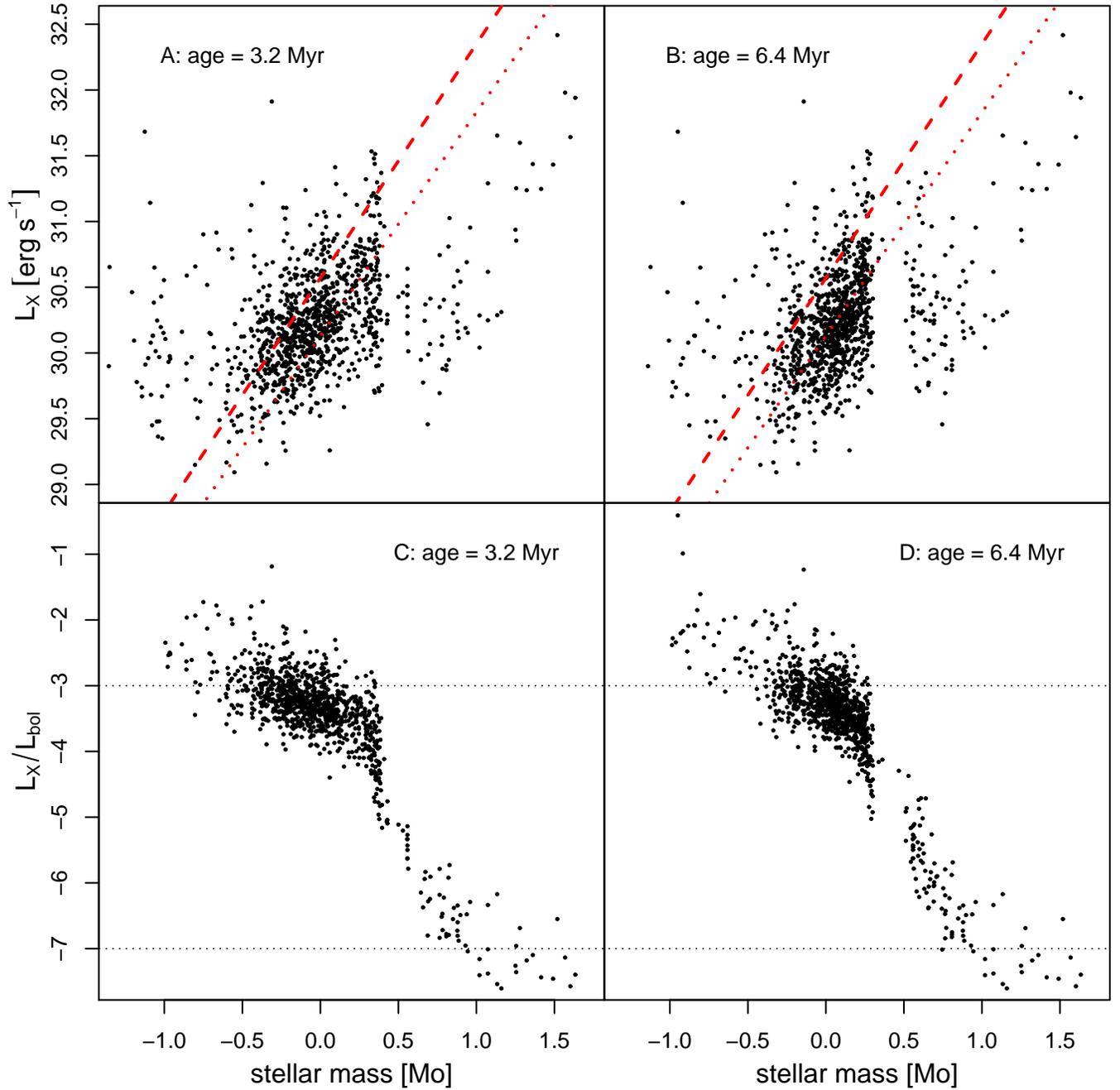} 
\caption{Top row: Absorption corrected X-ray luminosity (0.5--8.0\,keV band) versus stellar mass. Panel A shows masses derived assuming an age of 3.2~Myr and panel B shows masses derived assuming an age of 6.4~Myr. The $\log L_{X}$--$\log M$ regression lines found by \citet{2007A&A...468..425T} for young low-mass stars in the Taurus Molecular Cloud is indicated. The upper dashed line is the relation for weak-line T-Tauri stars and the lower dotted line is the relation for classical T-Tauri stars. Bottom row: $L_{X}/L_{bol}$ vs.\ stellar mass. Age assumptions for panels C and D are the same as for A and B, respectively. The levels $L_{X}/L_{bol}=10^{-3}$ and $10^{-7}$ are indicated.
 \label{lx_v_mass.fig}}
\end{figure*}

Figure~\ref{lx_v_mass.fig} shows the relationships between X-ray luminosity, bolometric luminosity ($L_{bol}$) and stellar mass. (Each graph is shown with each age estimate.)  For observed pre--main-sequence stars, the scatter in $L_X$ plotted as a function of $M$ is $\sim$0.5~dex, which can be partially attributed to uncertainty in both extinction-corrected X-ray luminosities and stellar masses. Two lines show the relations found by  \citet{2007A&A...468..425T} for weak-line T-Tauri stars (dashed line) and classical T-Tauri stars (dotted line) in the XMM-Newton Extended Survey of the Taurus Molecular Cloud in a sample of stars having a logarithmically averaged mean age of 2.4~Myr \citep[XEST;][]{2007A&A...468..353G}. These relations predict that weak-line T-Tauri stars will have higher X-ray luminosities than classical T-Tauri stars for the same mass because the presence of a disk has been observed to suppress X-ray emission. However, the locus of NGC~6231 stars on the plot is shifted down in X-ray luminosity relative to the XEST relation for weak-line T-Tauri stars. This shift has a magnitude of 0.2~dex (a factor of $\sim$1.5) for an age of 3.2~Myr or 0.4~dex (a factor of $\sim$2.5) for an age of 6.4~Myr. \citet{2016MNRAS.457.3836G} have shown that X-ray luminosity decreases for pre--main-sequence stars with radiative cores, compared to fully convective ones. The mass range for most of the observed low-mass stars in NGC~6231 is $0.5\lesssim M_\star \lesssim 2.5$, and for an age of 3.2~Myr, stars with $M>0.9$~$M_\odot$ will have developed radiative cores \citep{2000A&A...358..593S}, which may partially explain the lower X-ray flux.

The plots of $L_X/L_\mathrm{bol}$ vs.\ $M$ (similar to Figure~30 in \citetalias{2016arXiv160708860D}) show that most low-mass stars have $L_X/L_\mathrm{bol}$ slightly less than $10^{-3}$ (regardless of the age that is assumed). This luminosity ratio decreases to $\sim$10$^{-7}$ for high mass stars, which is the expected ratio for X-ray emission from a wind-shock mechanism. In between low- and high-mass stars, the precise mass at which the ratio begins to rapidly decline varies from $\sim$1.8 to 2.5~$M_\odot$ depending on assumptions about stellar age. For many objects with spectral types of A or B, X-ray emission and bolometric emission may be dominated by different components of a multiple star system.

\subsection{Estimates of Total Population \label{imf.sec}}

Despite the enlarged probable cluster member catalog, the majority of stellar-mass objects in NGC~6231 are still missing from the catalog. Completeness may be estimated through investigation of the observed mass distribution and X-ray luminosity distribution of probable cluster members, with comparison to results from other young clusters.

\subsubsection{Initial Mass Function}

\begin{figure*}
\centering
\includegraphics[width=0.9\textwidth]{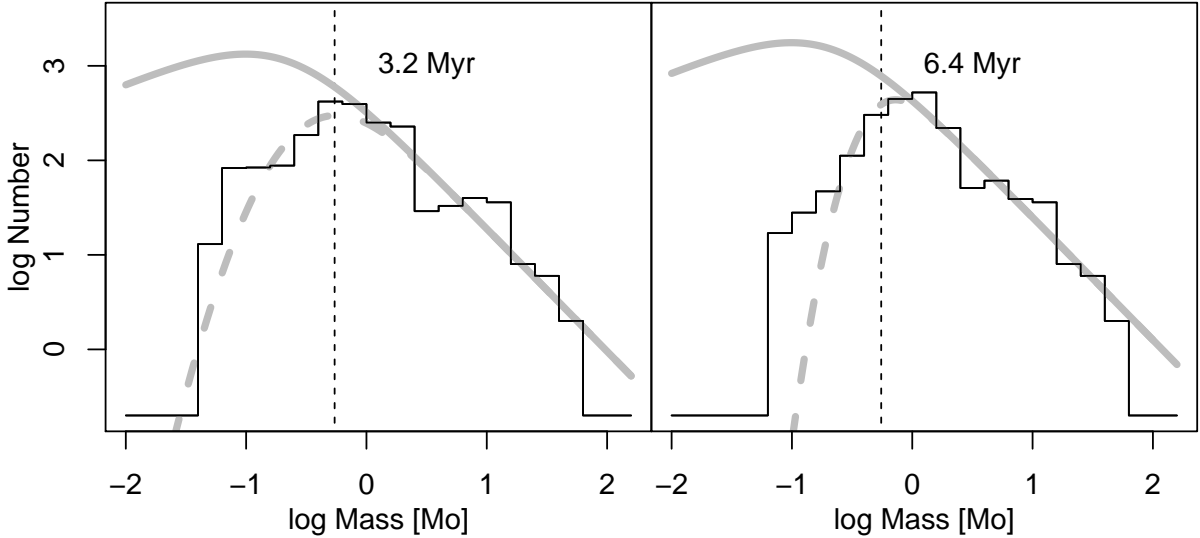} 
\caption{Histograms of stellar mass (the observed ``mass function'') for probable cluster members of NGC~6231 assuming an age of 3.6~Myr (left) or 6.4~Myr (right). The histogram counts are shown by the black lines. The observed mass function is fit with an IMF model \citep[gray line;][]{2013MNRAS.429.1725M}, which has been modified to account for incompleteness (dashed gray line). Completeness limits derived from these fits are shown by the dashed vertical lines. 
 \label{imf.fig}}
\end{figure*}

Figure~\ref{imf.fig} shows the observed mass functions for NGC~6231 assuming either an age of 3.2~Myr (left panel) or 6.4~Myr (right panel). These distributions are compared to theoretical initial mass functions (IMF), for which we use the  parameterization from \citet{2013MNRAS.429.1725M}. We expect the sample of pre--main-sequence stars to be complete above a certain mass limit. In addition we assume that the catalog of OB stars taken from the literature is complete. Stars with A and late-B spectral types are expected to be missing from X-ray surveys; however, some stars in this range may have X-ray emitting pre--main-sequence companions.

The completeness limit is estimated by fitting the observed mass function with a model in which the 50\%-completeness limit is a parameter. The model is generated by multiplying the \citeauthor{2013MNRAS.429.1725M} IMF by a completeness function that goes from 100\% detection probability for high-mass stars to 0\% detection probability for very low mass objects. We approximate the completeness function using the error function,
\begin{equation}
f(M)=0.5+0.5\,\mathrm{erf}[(\log M- \log M_{50\%})/w],
\end{equation}
where $M_{50\%}$ is the mass at which a star has 50\% chance of being included and $w$ characterizes how quickly the probability falls to 0. So, the model for the observed mass function is
\begin{equation}
\frac{\mathrm{d}N(M)}{\mathrm{d}\log M}=f(M)\,\Phi_\mathrm{IMF}(M),
\end{equation}
where $\Phi_\mathrm{IMF}$ is the \citeauthor{2013MNRAS.429.1725M} IMF. We note that the vast majority of stars with $M\ge M_{50\%}$ will be detected.

For both types of mass estimate, the 50\%-completeness limit was found to be 0.5~$M_\odot$, although the completeness rolls off more gradually for 3.2~Myr. For an age estimate of 3.2~Myr, the total number of stellar-mass objects (down to the hydrogen burning limit of 0.08~$M_\odot$) is estimated to be 5700 cluster members within the field of view. For an age of 6.4~Myr, a total of 7500 cluster members are estimated to lie within the field of view. The assumption of an older age yields a larger total population estimate because the older ages yield higher stellar masses. 

In both cases, the mass function above the completeness limits is consistent with a universal IMF. A small dip in the histogram at 2--4~$M_\odot$ stars is not likely astrophysical, because our mass estimation method (or the method used by \citetalias{2016arXiv160708860D}) is degenerate in this range, so some stars that would be assigned these stellar masses are incorrectly assigned masses in the previous bins. There is also an apparent deficit of $\sim$100~$M_\odot$ stars compared to the IMF prediction of $\sim$1. Although, this is entirely consistent with the $\sqrt{N}$ Poisson uncertainty, we also note that at least one supernova has occurred in this cluster.

The \citet{1998A&A...337..403B} pre--main-sequence evolutionary tracks are used to estimate masses below 0.1~$M_\odot$. While the overall shape of the IMF changes relatively little depending on the specific age estimate used, the number of objects assigned substellar masses (i.e.\ $M<0.08$~$M_\odot$) depends greatly on the age estimate because the hydrogen-burning limit is near the sensitivity limit of the observation. An assumed age of 3.2~Myr yields 26 substellar objects, with estimated masses ranging from 0.04--0.08~$M_\odot$. The median X-ray net counts is 4.4 (3.7--12~net counts interquartile range), and the median X-ray luminosity is $\log L_X=29.9$~[erg~s$^{-1}$] (29.8--30.4 interquartile range). These stars have moderate extinctions (median $A_K=0.03$~mag), typical X-ray hardnesses (median $ME=1.4$~keV), and none of them has indication of infrared excess. However, when an age of 6.4~Myr is assumed, only 4 of these sources are assigned masses less than 0.08~$M_\odot$.

\subsubsection{X-ray Luminosity Function}

The X-ray luminosity function (XLF) of pre--main-sequence stars in a cluster is related to the cluster's IMF due to the statistical link between X-ray luminosity and stellar mass, and it has been hypothesized that for pre--main-sequence populations less than $\sim$5~Myr old, the XLF may be universal \citep{2005ApJS..160..379F}. \citet{2015ApJ...802...60K} examine distributions of X-ray luminosity for pre--main-sequence stars in the MYStIX star-forming regions and find that there is no evidence of a change in the shape of the XLF and that the different regions had approximately the same XLF shape (for $L_X$ above the completeness limit).

\begin{figure*}
\centering
\includegraphics[width=0.8\textwidth]{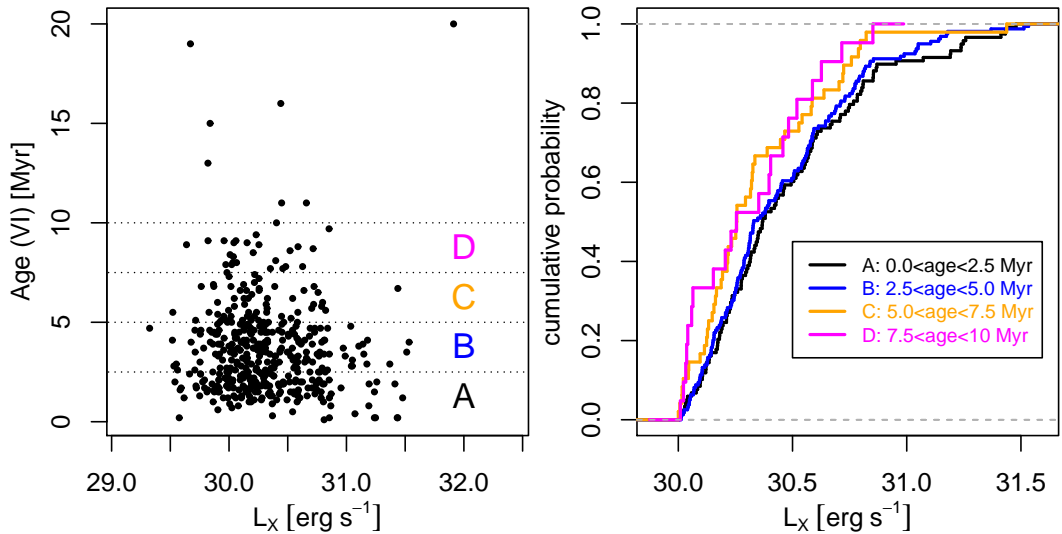} 
\caption{Left: X-ray luminosity is plotted against stellar ages derived from the $V$ vs.\ $V-I$ diagram. Four age strata are indicated by dashed lines, labeled ``A'' through ``D.'' Right: Cumulative distributions of X-ray luminosities of stars in each of these age strata, down to the X-ray completeness limit at $L_X=10^{30.0}$~erg~s$^{-1}$. There is little change in the shape of the XLF during the first 5~Myr years, but for older ages there is a deficit of high-X-ray luminosity stars. 
 \label{optage_lx.fig}}
\end{figure*}

The age spread in NGC~6231 provides an opportunity to examine how the distribution of X-ray luminosities are affected by age. Figure~\ref{optage_lx.fig} (left panel) is a scatter plot showing $L_X$ vs.\ $Age_{VI}$. We subdivide the points into four age groups: (a) 0--2.5~Myr, (b) 2.5--5~Myr, (c) 5-7.5~Myr, and (d) 7.5--10~Myr. The right panel shows cumulative distribution plots of $L_X$ in each of these strata for sources brighter than the completeness limit $L_X>10^{30.0}$~erg~s$^{-1}$. For the two youngest strata (a) and (b) the distributions are very similar, and the Kolmogorov--Smirnov (K--S) test shows no statistically significant distinction. The two older strata (c) and (d) are also relatively similar without statistically significant difference. However, when comparing stars 0--5~Myr old to stars 5--10~Myr old, the difference is moderately statistically significant with a $p$-value of 0.02. The older stars have a slightly fainter distribution of X-ray luminosities. We also note that only one star in the 5--10-Myr range has an X-ray luminosity greater than $10^{31.0}$~erg~s$^{-1}$, while many in the 0--5-Myr range do. 

No (or minor) variation in the shape of the XLF during the first 5~Myr is consistent with the results from \citet{2015ApJ...802...60K}, who find little variation in the XLF shapes of MYStIX star-forming regions, most of which are $<$5~Myr old. In contrast, \citet{2016MNRAS.457.3836G} indicate that some decline in X-ray luminosities would be expected during this phase.

\begin{figure}
\centering
\includegraphics[width=0.40\textwidth]{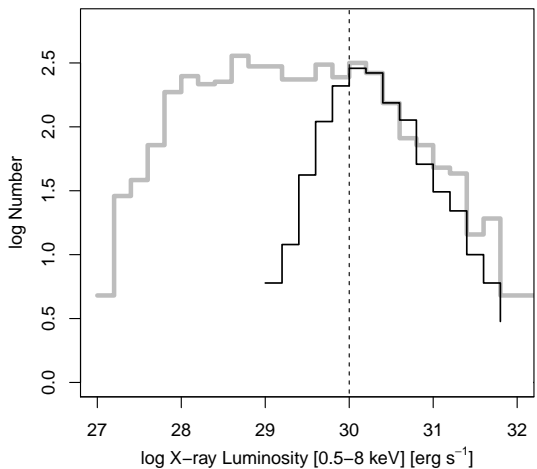} 
\caption{Histogram of X-ray luminosity (the observed ``X-ray Luminosity Function'') for probable cluster members of NGC~6231. The histogram counts are shown by the black lines. The observed luminosity function is fit by a scaled ``template'' XLF for low-mass stars from the Orion Nebula Cluster \citep{2005ApJS..160..379F}. The completeness limit is shown by the dashed vertical line. 
 \label{xlf.fig}}
\end{figure}

Figure~\ref{xlf.fig} shows the XLF for pre--main-sequence stars in NGC~6231 (black histogram). This XLF is compared to a scaled ``template'' XLF (gray histogram) based on the sample of 839 lightly-obscured, low-mass stars from the COUP study of the Orion Nebula Cluster \citep{2005ApJS..160..379F}.\footnote{We assume a distance to the Orion Nebula of 415~pc from \citet{2007A&A...474..515M}; however, an alternate distance has been recently suggested of 388~pc \citep{2016arXiv160904041K}.} The COUP sample is approximately complete down to stellar masses of 0.1--0.2~$M_\odot$.  On the X-ray luminous side of the distribution, the XLF of NGC~6231 declines more steeply than the COUP XLF. Above an X-ray luminosity of $L_X =10^{30.2}$~erg~s$^{-2}$, the distribution of X-ray luminosities for NGC~6231 has a power-law form with exponent $\alpha=-1.21\pm0.05$ while COUP has a power-law form with exponent $\alpha=-0.91\pm0.10$ \citep[cf.][]{2015ApJ...802...60K}. The Anderson-Darling test suggests these two distributions differ with a $p$-value of 0.007. Given that the Orion Nebula Cluster is a younger population \citep[mean age $\sim$2.5~Myr;][]{2011MNRAS.418.1948J}, the smaller number of very luminous sources in NGC~6231 relative to the Orion Nebula sample may result from a greater relative decline of X-ray luminosities for more massive stars than for less massive stars.

X-ray luminosity functions (XLFs) have been used to estimate stellar total populations of cluster members in young stellar clusters in cases where there is little difference in XLF shape \citep[e.g.,][and references therein]{2015ApJ...802...60K}. Although there does appear to be some variation in XLF shape for NGC~6231 (at least for stars older than 5~Myr) the differences are not huge and they may affect the brightest X-ray sources the most, which are a minority of the catalog. Our template ``universal XLF'' (gray histogram) is scaled vertically until it matches the NGC~6231 XLF down to the completeness limit, which is located at $\log L_X=30.0$~[erg~s$^{-1}$], requiring a scale factor of 4.9. Thus, the XLF analysis suggests that the number of pre--main-sequence stars in NGC~6231 is $839\times4.9\approx4100$ stars down 0.1--0.2~$M_\odot$. The difference expected from extrapolating down to 0.08~$M_\odot$ rather than 0.15~$M_\odot$ is a factor of $\sim$1.5, so, applying this as a correction factor, we obtain a total population of 6000 cluster members.  The estimates from the XLF and IMF methods are in approximate agreement considering uncertainties of $\sim$0.25~dex on estimates of total populations suggested by \citet{2015ApJ...802...60K}.

~~~
~~~
~~~
\subsection{Highly Absorbed Sources in the CXOVVV Catalog \label{hard.sec}}

The X-ray and NIR color-magnitude and color-color diagrams (Figures~\ref{xcmd.fig}, \ref{cmd.fig}, and \ref{ccd.fig}) all show small fractions of CXOVVV objects with high absorption. Highly absorbed objects are not atypical compared to other X-ray/infrared observations of very young stellar clusters or star forming regions \citep[see the X-ray ``color-magnitude diagrams'' in Figure~6 of][]{2013ApJS..209...27K}. However, in the case of NGC~6231, which neither  has sufficient cloud material to produce high extinctions nor has ongoing star formation, the most likely remaining explanations for these sources is that they are either cluster members with disks, background field-star contaminants, or background extragalactic contaminants spuriously matched to NIR sources.

We define a high-$ME$ source in the CXOVVV catalog to have $ME>2.0$ with at least 5~net counts that is classified as a probable cluster member due to X-ray variability or $J<17.0$~mag, $H<16.5$~mag, or $K<16.0$~mag.\footnote{The Wolf-Rayet star is also omitted.} Although most extragalactic sources are expected to have $J\gtrsim20$~mag, the color-magnitude diagram in Figure~\ref{cmd.fig} shows a few sources with $J\sim18$~mag with red colors that may be extragalactic contaminants, so our magnitude cuts are designed to avoid these sources. 

Using this definition, there are 50 high-$ME$ sources in NGC~6231, 27 with $2.0<ME<3.0$~keV, 20 with $3.0<ME<4.5$~keV, and 3 with $ME>4.5$~keV.
This represents 2.5\% of the 2148 X-ray selected probable cluster members. The frequency of these sources decreases rapidly with increasing $ME$, having a power-law-like distribution with an index of $-2.4\pm0.4$. About half of these X-ray sources have 5--10 net counts, while half have 10--100 net counts. Their spatial distribution resembles the underlying spatial distribution of X-ray sources, but there are too few objects to determine whether they are more strongly clustered than expected for contaminants.

If these high-$ME$ sources are mostly extragalactic contaminants rather than cluster members, then we would expect the distribution of their median energies to be similar to the high-$ME$ sources that are not classified as probable cluster members. We use the Anderson-Darling test to examine these two cases. Overall, for sources with $ME>2.0$~keV the Anderson-Darling test rejects the hypothesis that these samples have the same distribution ($p<0.0001$), but for sources with $ME>3.0$~keV the null hypothesis is marginally rejected ($p\approx0.04$). 

\begin{figure*}
\centering
\includegraphics[width=1\textwidth]{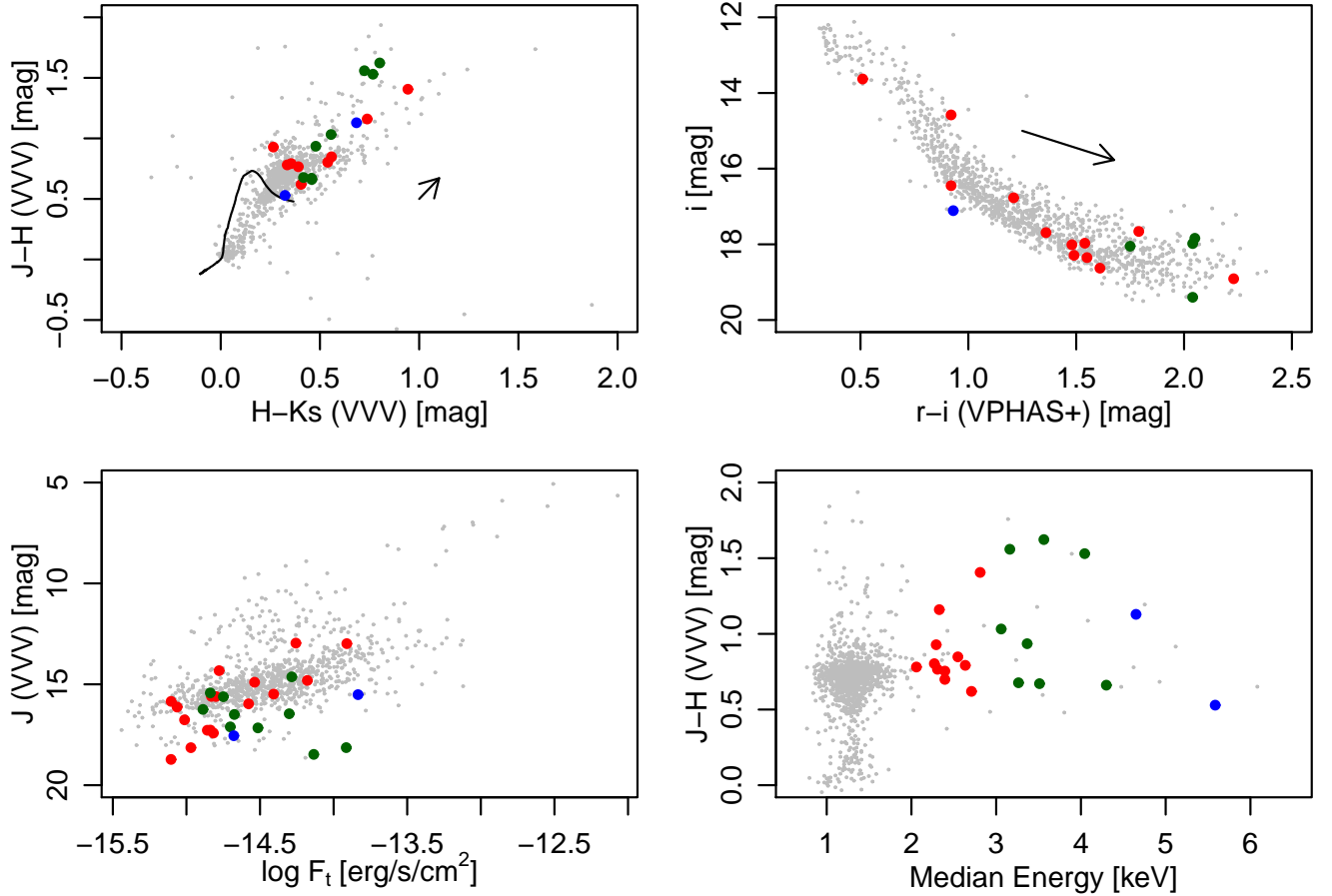} 
\caption{Distributions of all probable cluster members (gray points) and probable cluster members with $2.0<ME<3.0$~keV (red circles), $3.0<ME<4.5$~keV (green circles), and $ME>4.5$~keV (blue circles). Upper left: $JHKs$ color-color diagram. Upper right: VPHAS+ $ri$ color-magnitude diagram. Arrows in both diagrams indicate absorption of $A_V=1.6$~mag. Lower left: $J$ mag vs.\ X-ray flux. Lower right: $J-H$ color vs.\ X-ray median energy.
 \label{hard.fig}}
\end{figure*}

Another potential explanation is that these objects are background sources reddened by the interstellar medium behind the cluster. \citetalias{2016arXiv160708860D} note that the distribution of $J-H$ colors for field stars is bimodal, possibly indicating the presence of a cloud. However, the far-infrared {\it AKARI}-FIS 160-$\mu$m maps \citep{2007PASJ...59S.389K} show no indication of a cloud behind the cluster. In addition, the bimodality that \citetalias{2016arXiv160708860D} note exists over a wide range of Galactic longitudes, and, thus, is not likely to be a discrete cloud associated with the cluster. Whether the high-$ME$ objects lie behind the cluster can be examined with the VPHAS+ $r-i$ vs.\ $i$ color-magnitude diagram (Figure~\ref{hard.fig}). On this diagram, reddening is approximately parallel to the isochrone for low-mass stars, while stars that are more distant would be shifted vertically.  The high-$ME$ objects with $r$ and $i$ photometric measurements do not show any sign of a vertical shift relative to the cluster members, so they are most likely located at approximately the same distance and are possible candidate members. Nevertheless, not all high-$ME$ objects have photometry in these bands.

Figure~\ref{hard.fig} shows the high-$ME$ sources on scatter plots of X-ray, NIR, and optical source properties. Points are marked by sources' median energies, with $2.0<ME<3.0$~keV sources shown in red, $3.0<ME<4.5$~keV sources shown in green, and the $ME>4.5$~keV sources shown in blue. On the $J-H$ vs.\ $H-K_s$ color-color diagram, the points are shifted to the upper right (parallel to the direction of the reddening vector). The shift of the green sources is greater than the shift of the red sources, which would be expected if the high-$ME$ values are caused by absorption. In contrast, one of the sources with $ME>4.5$~keV does not have unusually red colors. The location of the red and green sources are consistent with $A_v=5$--20~mag of absorption.
These values of NIR extinction are similar to the extinctions required to produce the median energies of the sources---$ME=2.0$~keV corresponds to $A_V=6$~mag, $ME=3.0$~keV corresponds to $A_V=15$~mag, and $ME=4.5$~keV corresponds to $A_V=100$~mag. 
On the $r-i$ vs.\ $i$ diagram, most points lie within the locus of other cluster members, with green sources typically having lower masses than red sources.  

On the X-ray flux vs.\ $J$ diagram the full set of probable cluster members shows the positive correlation that is expected for cluster members. The locations of the high-$ME$ sources on this diagram are also consistent with this trend, but due to the large amount of scatter and the low number of sources the trend is not very clear. On the plot of X-ray median energy vs.\ $J-H$, most of the red sources are tightly grouped with $ME$ and $J-H$ larger than the majority of the cluster members. In contrast, the green sources have much more scatter in $J-H$ values, suggesting that some may be coincidental matches between background galaxies and unrelated foreground stars. The $J-H$ color of one of the blue sources is inconsistent with it being a highly absorbed cluster member.

On the NIR color-color diagram presented in Figure~\ref{ccd.fig}, objects with high extinction can be noted on the upper-right side of the plot. Overall, 40 objects have $A_V>5$~mag estimated from NIR photometry. These stars have $K_s$-band magnitudes ranging from 12 to 17~mag, but most are not detected in optical VPHAS+ bands. They have X-ray luminosities (assuming a distance of 1.59~kpc) ranging from $\log L_X=29.5$--31.4~[erg~s$^{-1}$]. They also have higher than average X-ray median energies, although only one third of them meet the definition for high-$ME$ object above. Although some of these objects have infrared excess in the $K_s$ band, most do not. These stars are not uniformly distributed in the {\it Chandra} field of view, but are loosely concentrated toward the center of the field.

While some of the highly absorbed sources may be cluster members, the sample likely does include contaminants. For any cluster member, the absorption producing the high $ME$ or high $A_V$ must be local. This absorption could come from either disks with fortuitously low inclination angles or from thick circumstellar disks.  Two examples of stars in the Orion Nebula Cluster with this effect have edge-on silhouette disks imaged with the {\it Hubble Space Telescope} \citep{2005ApJS..160..511K}.  X-ray source COUP~241 (Orion silhouette disk d053-717) with disk aspect ratio $R = 5.5$ has an X-ray $ME = 3.2$~keV corresponding to $\log N_H = 22.7$~cm$^{-2}$.  And, X-ray source COUP~419 (d114-426) with $R = 3.9$ has $ME = 5.3$~keV with $\log N_H = 23.7$~cm$^{-2}$. (Interestingly, the NIR colors of COUP~241 neither reveal $K$-band excess nor high absorption, so may be affected by reflected light.) 

The observed disk fraction among the highly absorbed sources in NGC~6231 is not high, only one high-$ME$ source has $K_s$ excess and only a small fraction of the high-$A_V$ sources do. However, not all stars with disks have $K_s$-band excess, which would be produced by the hot inner region of a circumstellar disk, and GLIMPSE photometry is not available for any of these sources to determine if any have mid-infrared excess.

\section{O and B-Star Systems \label{ob.sec}}

NGC~6231 is unique in being a very young stellar cluster with a large, nearly complete spectroscopic census of OB stars (Table~\ref{hm.tab}) that also has been observed with a deep X-ray observation. The catalog of O and B spectral types is obtained from \citet{2006MNRAS.372..661S}. (Out of the 108 WR+OB NGC~6231 stars in the Table, 95 are within the ACIS-I field of view.) Thus, the cluster makes an ideal testbed for statistical studies of X-ray emission from O and B stars/systems. In the {\it Chandra} X-ray catalog, 100\% (13 out of 13) of the known O stars/systems and 50\% (41 out of 82) of the known B stars/systems have X-ray counterparts within 2$^{\prime\prime}$. \citet{2006MNRAS.372..661S} study 30 of these systems using {\it XMM-Newton} data, and they conclude that X-rays come from stellar winds of individual stars, colliding winds systems, and pre--main-sequence binary companions. The enlarged sample provided by the {\it Chandra} data can give a more complete view of the distributions of OB-star X-ray properties.

The XPHOT X-ray fluxes from Section~\ref{xraydata.sec} are based on relations derived from pre--main-sequence stars, so the XPHOT fluxes for objects where most of the X-rays come from a stellar wind may be biased due to an incorrect X-ray spectral model. Here, we perform spectral fitting in XSPEC using a thermal plasma model, {\it apec} \citep{2001ApJ...556L..91S}, attenuated by an absorption model, {\it wabs} \citep{1983ApJ...270..119M}, with a metallicity of $Z=0.3$~$Z_\odot$ and \citet{1989GeCoA..53..197A} abundances.\footnote{We use typical metallicity and abundance assumptions for pre--main-sequence stars \citep[e.g.,][]{2005ApJS..160..319G} to aid comparison with the literature.} The results of this fitting are given in Table~\ref{hm.tab}. 

\subsection{O and B Populations in the X-ray \label{obxray.sec}}

\begin{figure*}[t]
\centering
\includegraphics[width=0.8\textwidth]{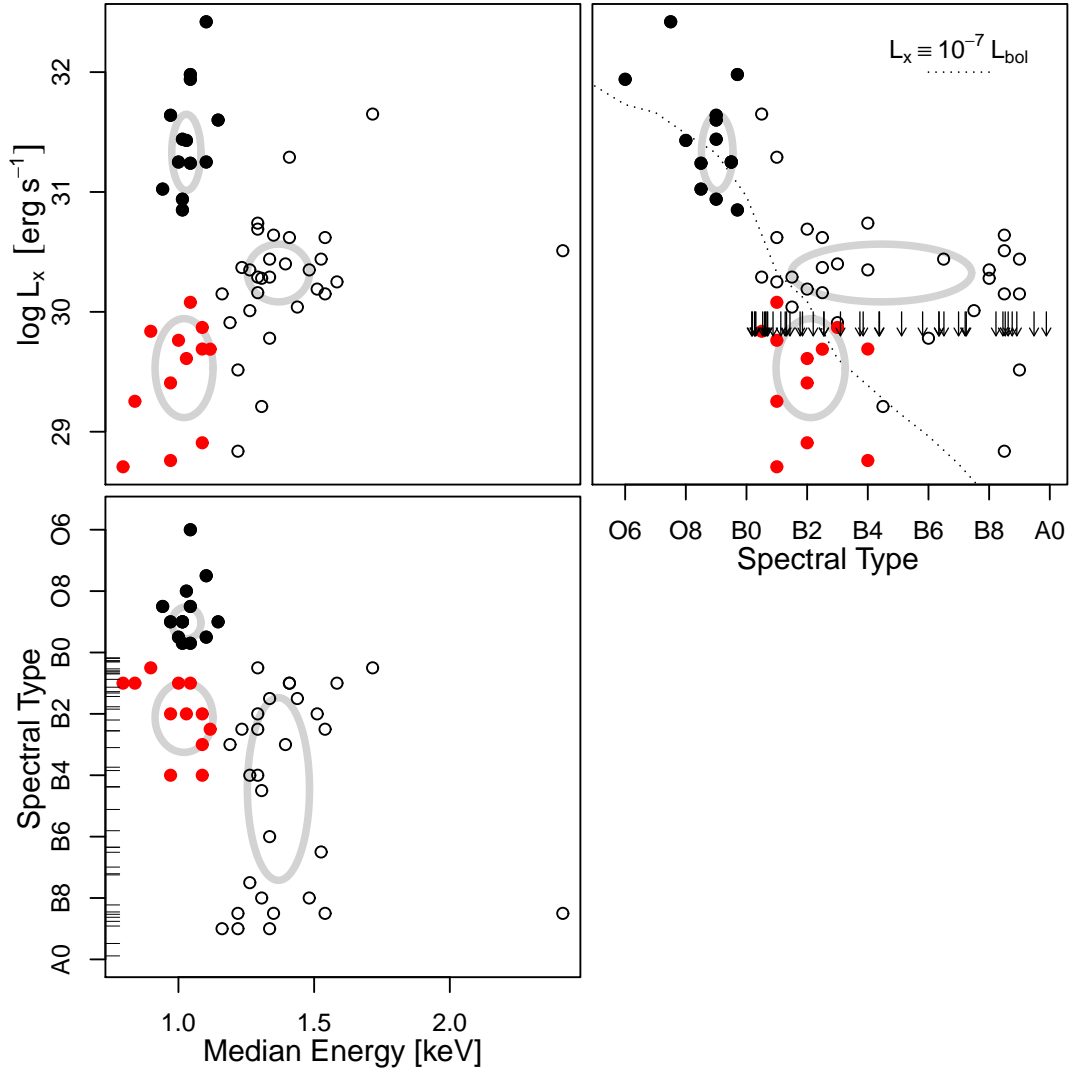} 
\caption{Scatter plots showing spectral type, X-ray luminosity, and X-ray median energy for O and B stars in NGC~6231. The three combinations of variables are $L_X$ vs.\ $ME$ (upper-left panel), $L_X$ vs.\ spectral type (upper-right panel), and spectral type vs.\ $ME$ (lower-left panel). X-ray detected O and B stars are shown as circles, while undetected B stars are shown as arrows (upper limits) or tic marks. The three clusters identified with the normal mixture model are shown by the gray ellipses. The most likely classification of each source is indicate by its color on the plot: black circles for ``O-star wind'' sources, open circles for ``Lindroos binary'' sources, and red circles for ``B-star wind'' sources. 
On the $L_X$ vs.\ spectral type plot, the dotted line shows a constant $L_X$--$L_\mathrm{bol}$ ratio for main-sequence OB stars.
\label{ob_properties.fig}}
\end{figure*}

Figure~\ref{ob_properties.fig} shows the O and B stars in the multivariate ($\log L_X$, $ME$, spectral type)-space. Given that these stars are only lightly absorbed (e.g., $N_H\sim3.5\times10^{21}$~cm$^{-2}$), $ME$ will be primarily an indicator of plasma temperature. For this absorption, a source with a temperature $T=1$~MK corresponds to $ME\sim0.7$~keV, 2.5~MK would correspond to 0.8~keV, 5~MK to 1~keV, 10~MK to 1.2~keV, 20~MK to 1.5~keV, 50~MK to 1.7~keV, and 100~MK to 1.9~keV. Each of the 54 X-ray-detected O and B stars are indicated by circles, while the undetected B stars are indicated by upper limits or tick marks at their spectral type. Upper limits on X-ray luminosity in the catalog are $L_X=10^{30.0}$~erg~s$^{-1}$. It is clear from a visual inspection that more than one distinct group of points is present. 

We use a mixture-model cluster analysis of the points in ($\log L_X$, $ME$, spectral type)-space to determine if multiple classes of object are suggested by the data. We use the {\it mclust} software  \citep{mclust02,mclust12} which implements a mixture model analysis using multivariate normal distributions. This analysis will provide information about 1) the number of clusters present in the data, 2) the properties of these clusters, and 3) the classifications of individual objects. 
The Bayesian information criterion \citep[BIC;][]{schwarz1978estimating} is a penalized likelihood used for model selection where different models have different numbers of parameters (in this case different numbers of clusters). The best model will be the one with the lowest value of the BIC, improvements of $\Delta\mathrm{BIC}>6$ are considered strong evidence, and improvements of $\Delta\mathrm{BIC}>10$ are considered very strong evidence \citep{jeffreys1961theory,KassRaftery95}. 

For the O and B stars on the plot, the best model (BIC=143) includes 3 clusters, which are indicated by the gray ellipses on Figure~\ref{ob_properties.fig}. The BIC value for 1 cluster is 187, for 2 clusters is 149, for 4 clusters is 150, and for 5 clusters is 175, so there is strong evidence ($\Delta BIC=6$) of 3 clusters and very strong evidence ($\Delta BIC=44$) of more than one cluster.

One group of points found by the {\it mclust} analysis (colored black in Figure~\ref{ob_properties.fig}) includes every O-star system, for which X-rays are generated by either the winds of individual stars or colliding stellar winds. The defining features of this subset are soft ($\sim$1~keV) median energies and high luminosities ($\log L_X>30.6$ [erg~s$^{-1}$]. On the $L_X$ vs.\ spectral type diagram, a gray line indicating a constant $L_X=10^{-7}$~$L_\mathrm{bol}$ ratio is shown, and these points are mostly distributed near this line. This is similar to the results of other studies of X-ray emission from stars with strong winds \citep[e.g.,][their Figure~9]{2005ApJS..160..557S}. 

B-type stars have been less well studied in the X-ray than O-type stars, so the origin of X-ray emission from these stars is less certain. Spectroscopic X-ray studies of several early-B stars, including \objectname[tau Sco]{$\tau$~Sco} \citep[B0.2V;][]{2003ApJ...586..495C} and \objectname[bet Cru]{$\beta$~Cru~A} \citep[B0.5III+B2V;][]{2008MNRAS.386.1855C}, have revealed that X-ray emission comes from a stellar wind, which is indicated by forbidden-to-intercombination line ratios showing that the X-ray emitting gas is located several stellar radii above the stellar photospheres and, in the latter case, Doppler broadened X-ray lines. However, it is difficult to explain some effects (e.g., lower velocities and higher X-ray emitting wind fraction) using the standard wind-shock paradigm alone. The 41 X-ray detected B stars in NGC~6231 are subdivided into two groups by the mixture-model cluster analysis: 12 are in a group with soft X-ray spectra, while 29 have harder spectra. These two groups are approximately separated by a cut at $ME<1.15$~keV.  

The median energies of the former group (red circles on Figure~\ref{ob_properties.fig}) are identical to those of O stars and are significantly softer than the typical pre--main-sequence star's median energy; some of these objects have median energies as low as $ME\sim0.7$~keV. For comparison, the star $\beta$~Cru~A has a median energy of $0.6$~keV (similar to the softest of the NGC~6231 B stars) and $\tau$~Sco has a median energy of 1.0~keV. The B4 limit on spectral type for this group also coincides with the theoretical transition between a fast wind and a weak wind \citep[][and references therein]{2005ApJS..160..557S}. We thus suggest that these early-B stars have X-ray emitting winds. 

The X-ray luminosity of this group of probable shocked-wind B stars ($L_X\lesssim30$ [erg~s$^{-1}$]) is more than a factor of 10 lower than the X-ray luminosity of the O stars. If the X-rays from these stars are wind origin, then there is a gap in the X-ray luminosity distribution for stellar-wind X-ray sources. This may indicate that a different model for X-ray emission from the stellar winds of B stars, as has also been suggested by the spectroscopic studies mentioned above. The X-ray luminosities of $\tau$~Sco and $\beta$~Cru~A ($\log L_X=31.5$ and 30.4 [erg~s$^{-1}$], respectively) are similar to or lower than the X-ray luminosities of the O stars in NGC~6231, but higher than its B stars. However, the gray line in Figure~\ref{ob_properties.fig}, indicating a constant $L_X=10^{-7} L_\mathrm{bol}$ ratio, passes through the cluster of O stars on this graph (marked black), but passes above most of the B-star wind-emission candidates (marked red). 
Thus, a ratio of $10^{-7}$ may be to large by 0.5~dex or more.

A third group of B-type stars (open circles on Figure~\ref{ob_properties.fig}), have X-ray median energies higher than 1.15~keV. The spectral type for this group range from B0.5 to B9.5, with no apparent preference for early or late B spectral types. Their X-ray luminosities are mostly in a well-defined range $29.7<\log L_X<30.7$~[erg~s$^{-1}$], with X-ray median energies $1.15<ME<1.6$~keV. 
X-ray median energies in this range are harder than the typical X-ray median energies expected for X-rays generated by shock-heated plasma from the line-deshadowing instability in OB star winds. It has been suggested that hard components in X-ray spectra could be produced by colliding winds from OB+OB binaries or magnetically confined wind shocks \citep{2015arXiv150906482U}. However, many of the O-star systems in NGC~6231 are known to have colliding winds \citep{2004MNRAS.350..809S,2006MNRAS.371...67S,2005A&A...441..213S,2008NewA...13..202S}, but they still have $ME<1.15~$keV. The  \objectname[tet01 Ori C]{O-star $\theta^1$~Ori~C} has been suggested as prototypical example of a magnetically confined wind shock system, but it's median energy, if it were located in NGC~6231 rather than the Orion Nebula Cluster (i.e.\ adjusting for absorption and observational effects), would be 1.12~keV.  Furthermore, two of these sources, CXOU~J165354.52-415214.9 ( \objectname[V* V945 Sco]{CPD-41~7706}; B1V+B1Ve) and CXOU~J165436.10-415338.6 (\objectname[V* V947 Sco]{CPD-41~7753}; B0.5V), have large X-ray flares, which almost certainly come from low-mass companions. Thus, we conclude that the NGC~6231 B stars in this category are likely to have pre--main-sequence binary companions that are responsible for the X-ray emission.
 
We test whether the distributions of X-ray luminosity and median energy of the B-stars in the third group are different from the distributions of low-mass X-ray selected cluster members with $ME>1.15$~keV. For X-ray luminosity the K-S $p$-values is 0.90 (no statistically significant difference), while the median energy the K-S $p$-value is 0.07 (very weak evidence of a difference). This result indicates that X-ray properties of the third group of B stars are consistent with X-rays from pre--main-sequence companions that were randomly drawn from the general population of pre--main-sequence stars in the cluster.

\subsection{Lindroos Binary Fraction \label{lindroos.sec}}

OB-star systems which include a pre--main-sequence companion are sometimes known as Lindroos binaries after the catalogs of \citet{1985A&AS...60..183L,1986A&A...156..223L}. Low-mass binary companions are difficult to identify in the vicinity of B-type stars in optical or NIR wavelengths due to high contrast ratios. But in the X-ray band, pre--main-sequence stars may significantly contributed to or dominate the X-ray luminosity of the system. For example the $\beta$~Cru system, mentioned above, consists of both a B-star binary ($\beta$~Cru~A; 5 year period), and a pre--main sequence star with a $\sim$400-au projected separation \citep[$\beta$~Cru~D;][]{2008MNRAS.386.1855C}. The pre--main-sequence companion is harder (1.0~keV) than the B-star system. At a distance of $\sim$108~pc, components A and D are individually resolved by {\it Chandra}; however, if the system were at the distance of NGC~6231, the system would appear as a single source with $ME=0.73$~keV. 

Several studies have used X-ray counterparts to B stars to characterize Lindroos binary populations, including work by \citet{1993ApJ...402L..13S}, \citet{1994A&A...292L...5B}, \citet{2000A&A...359..227H,2001A&A...373..657H}, \citet{2003A&A...407.1067S}, \citet{2001A&A...372..152H}, and \citet{2010ApJ...725.2485K}. This interpretation was also used in the deep COUP X-ray study of the Orion Nebula Cluster \citep{2005ApJS..160..557S} and the Carina Nebula \citep{2011ApJS..194....5G,2011ApJS..194....7N,2011ApJS..194...13E}. 

If we consider the B stars with $ME>1.15$ (the open circles in Figure~\ref{ob_properties.fig}) to be our candidate Lindroos binaries, then the cluster has 29 such candidates, compared to 42 B stars with no X-ray counterpart, or 83 total known B stars (within the {\it Chandra} field of view). The mass completeness limit\footnote{This is the 50\%-completeness limit, defined as the mass of a pre--main-sequence stars at which it has a 50\% probability of being included in our {\it Chandra} catalog. Almost all pre--main-sequence stars with mass greater than this limit will be included in the catalog.} calculated for X-ray selection of pre--main-sequence stars in Section~\ref{imf.sec} is 0.5~$M_\odot$, so, assuming that pre--main-sequence stars in Lindroos binaries have the same X-ray--mass relation as isolated pre--main-sequence stars, we can apply this completeness limit here as well. Thus, between 35\% (Lindroos-binary candidates divided by the total number of B stars) and 41\% (omitting wind-emission candidates) of B stars have possible pre--main-sequence binary companions with masses greater than $\sim$0.5~$M_\odot$. 

Spectral type has no apparent effect on whether a B-star is a candidate Lindroos binary---we compare the distribution of spectral types for B stars without X-ray counterparts and Lindroos binary candidates using the two-sample Kolmogorov-Smirnov (K-S) test and find a $p$-value of 0.30 (indicating no statistically significant difference). This effect suggests that the pre--main-sequence companion masses are randomly drawn from the same mass distribution as the isolated pre--main-squence stars, at least for the $M>0.4$~$M_\odot$ range. If the random drawing extended down to the hydrogen burning mass limit, but we are only seeing the most X-ray luminous pre--main-sequence companions, the fraction of B stars that are Lindroos binaries could be much larger. A recent study by \citet{2016arXiv160605347M} also suggests a high fraction of B stars have low-mass binary companions randomly drawn from the IMF. 

\startlongtable
\onecolumngrid
\begin{deluxetable}{llrrrrrrrrrr}
\tablecaption{Massive Stars \label{hm.tab}}
\tabletypesize{\tiny}\tablewidth{0pt}
\tablehead{
\colhead{Name} & \colhead{SpTy} &\colhead{CXOU J} & \colhead{$NC$} & \colhead{$ME$} & \colhead{$P_\mathrm{KS}$} & \colhead{$\log N_\mathrm{H}$} & \colhead{$kT$} &\colhead{$\log F_t$} & \colhead{$\log F_{tc}$} & \colhead{$X^2$} & \colhead{d.o.f.}\\
 \colhead{} &\colhead{} & \colhead{} & \colhead{counts} & \colhead{keV} & \colhead{\%} & \colhead{cm$^{-2}$} & \colhead{keV}& \colhead{erg~s$^{-1}$~cm$^{-2}$}& \colhead{erg~s$^{-1}$~cm$^{-2}$}& \colhead{}& \colhead{}\\
\colhead{(1)} &  \colhead{(2)} &  \colhead{(3)} &  \colhead{(4)} & \colhead{(5)} &\colhead{(6)} &  \colhead{(7)} & \colhead{(8)} &  \colhead{(9)} &  \colhead{(10)} &  \colhead{(11)} & \colhead{(12)}
}
\startdata
     NGC 6231 724 &            B0V/IV &                           &         &        &                                               &  \\
     NGC 6231 723 &               B3V &                           &         &        &                                               &  \\
     NGC 6231 726 &               B1V &                           &         &        &                                               &  \\
        HD 326326 &               B2V &        165320.24-414825.5 &    13.2 &    1.0 & $>$0.05                             & 22.43$\pm$0.17 & 0.09$\pm$0.00 & -14.97$\pm$0.21 & -9.94 & 13.3 & 9 \\
        HD 326327 &     B1.5IVe+shell &                           &         &        &                                               &  \\
      NGC 6231 30 &               B7V &                           &         &        &                                               &  \\
      NGC 6231 41 &               B4V &        165344.03-415036.9 &   202.3 &    1.3 & 0.04                           & 20.47$\pm$1.18 & 1.92$\pm$0.28 & -13.86$\pm$0.04 & -13.83 & 19.3 & 31 \\
        HD 326328 &               B1V &                           &         &        &                                               &  \\
      NGC 6231 33 &               B7V &                           &         &        &                                               &  \\
      NGC 6231 54 &          B6III/IV &        165347.98-415505.0 &    14.4 &    1.3 & $>$0.05                        & \nodata & 1.33$\pm$0.46 & -15.09$\pm$0.16 & -15.09 & 4.9 & 5 \\
      NGC 6231 14 &              B4IV &                           &         &        &                                               &  \\
     NGC 6231 300 &            B8.5V? &        165349.71-415013.8 &    22.2 &    1.5 & $>$0.05                        & \nodata & 3.73$\pm$8.37 & -14.58$\pm$0.16 & -14.58 & 2.3 & 4 \\
     NGC 6231 265 &             B8.5V &        165351.19-415053.9 &   136.8 &    1.4 & $<$0.0001                      & \nodata & 5.37$\pm$2.53 & -13.91$\pm$0.06 & -13.91 & 19.6 & 20 \\
        HD 152200 &             O9.7V &        165351.65-415032.7 &   367.8 &    1.0 & $>$0.05                           & 21.83$\pm$0.06 & 0.27$\pm$0.04 & -13.70$\pm$0.11 & -12.51 & 19.6 & 24 \\
      CPD-41 7706 &          B1V+B1Ve &        165354.52-415214.9 &   388.1 &    1.4 & 0.0003                           & 20.82$\pm$0.56 & 2.39$\pm$0.47 & -13.34$\pm$0.04 & -13.30 & 29.6 & 29 \\
        HD 152219 & O9.5III+B1-2III/V &        165355.61-415251.5 &   901.7 &    1.0 & $>$0.05                           & 21.71$\pm$0.05 & 0.35$\pm$0.03 & -13.31$\pm$0.03 & -12.51 & 69.0 & 53 \\
      NGC 6231 24 &               B4V &                           &         &        &                                               &  \\
     NGC 6231 283 &               B3V &        165356.21-414815.8 &    32.5 &    1.2 & $>$0.05                          & \nodata & 1.54$\pm$1.20 & -14.74$\pm$0.19 & -14.74 & 5.9 & 7 \\
     NGC 6231 255 &               B6V &                           &         &        &                                               &  \\
      CPD-41 7711 &           B2V+B2V &                           &         &        &                                               &  \\
        HD 152235 &              B1Ia &        165358.90-415939.3 &    22.5 &    1.0 & $>$0.05                             & 22.13$\pm$0.23 & 0.14$\pm$0.09 & -14.51$\pm$0.06 & -11.58 & 3.7 & 6 \\
     NGC 6231 284 &             B4.5V &                           &         &        &                                               &  \\
     NGC 6231 249 &             B4.5V &        165359.33-415303.6 &     7.0 &    1.3 & $>$0.05                                       & \nodata & \nodata & \nodata & \nodata & \nodata & \nodata \\
        HD 152218 &        O9IV+O9.7V &        165400.02-414252.6 &   892.5 &    1.0 & $>$0.05                           & 21.48$\pm$0.05 & 0.64$\pm$0.03 & -13.11$\pm$0.02 & -12.71 & 81.2 & 55 \\
     CPD-41 7712 &             B0.5V &        165400.40-415243.3 &    35.7 &    1.5 & $>$0.05                            & 20.55$\pm$3.11 & 6.24$\pm$9.98 & -14.40$\pm$0.26 & -14.38 & 8.9 & 10 \\  
     NGC 6231 276 &              B8Vp &        165401.52-414923.7 &    49.4 &    1.5 & 0.001                            & 21.94$\pm$0.27 & 1.29$\pm$0.44 & -14.59$\pm$0.20 & -14.10 & 9.2 & 12 \\
      CPD-41 7715 &              B2IV &        165401.72-415112.1 &    29.0 &    1.3 & $>$0.05                             & 21.92$\pm$0.15 & 0.79$\pm$0.60 & -14.83$\pm$0.20 & -14.07 & 0.4 & 4 \\ 
        HD 152234 &        O9.7Ia+O8V &        165401.84-414823.0 &  3394.6 &    1.0 & $<$0.001                         & 21.45$\pm$0.03 & 0.74$\pm$0.02 & -12.62$\pm$0.01 & -12.28 & 201.1 & 103 \\
     NGC 6231 259 &               B2V &        165403.15-415149.1 &   159.1 &    1.3 & $<$0.002                           & 21.60$\pm$0.25 & 1.29$\pm$0.17 & -14.04$\pm$0.06 & -13.74 & 15.4 & 24 \\
       NGC 6231 1 &               B4V &        165403.58-414253.2 &    21.0 &    1.1 & $>$0.05                             & 20.94$\pm$2.24 & 0.45$\pm$0.53 & -14.89$\pm$0.11 & -14.74 & 5.7 & 6 \\
        HD 152233 &      O6III(f)+O9? &        165403.60-414730.0 &   1326 &    1.1 & $>$0.05                           & 21.66$\pm$0.17 & 0.52$\pm$0.22 & -13.57$\pm$0.20 & -12.94 & 32.8 & 42 \\ 
     NGC 6231 194 &             B3.5V &                           &         &        &                                               &  \\
      CPD-41 7719 &               B1V &        165405.09-415006.9 &    34.5 &    1.0 & $>$0.05                             & 21.08$\pm$0.97 & 0.79$\pm$0.19 & -14.79$\pm$0.19 & -14.63 & 4.4 & 7 \\
     NGC 6231 274 &               B3V &                           &         &        &                                               &  \\
      CPD-41 7722 &               B1V &                           &         &        &                                               &  \\
      CPD-41 7723 &               B1V &                           &         &        &                                               &  \\
      CPD-41 7721 &               O9V &        165406.71-415107.0 &     9.7 &    1.2 & $>$0.05                                       &  \\ 
      CPD-41 7721 &               B1V &                           &         &        &                                               &  \\
      CPD-41 7724 &             B0.5V &        165406.95-414923.4 &    44.7 &    1.3 & $<$0.0001                           & 21.04$\pm$2.24 & 1.27$\pm$0.43 & -14.58$\pm$0.18 & -14.48 & 12.5 & 10 \\
      CPD-41 7725 &               B0V &                           &         &        &                                               &  \\
        HD 326340 &             B0.5V &                           &         &        &                                               &  \\
     NGC 6231 273 &             B9IVp &        165409.14-415012.8 &    12.6 &    1.2 & 0.006                             & 22.48$\pm$0.42 & 0.11$\pm$0.03 & -15.30$\pm$0.09 & -10.68 & 0.0 & 0 \\
      CPD-41 7727 &             B0.5V &                           &         &        &                                               &  \\
        HD 152248 & O7.5III(f)+O7III(f) &        165410.06-414930.1 &  5447.0 &    1.1 & $<$0.0001                         & 21.63$\pm$0.02 & 0.72$\pm$0.02 & -12.22$\pm$0.01 & -11.72 & 528.8 & 149 \\
     NGC 6231 209 &             B2.5V &        165410.72-414747.4 &    93.4 &    1.2 & $>$0.05                           & 20.62$\pm$2.43 & 1.43$\pm$0.40 & -14.32$\pm$0.09 & -14.28 & 12.9 & 13 \\
     NGC 6231 374 &               B2V &        165410.97-414939.0 &     9.2 &    1.0 & $>$0.05                                       & \nodata & \nodata & \nodata & \nodata & \nodata & \nodata \\
      CPD-41 7730 &               B1V &                           &         &        &                                               &  \\
        HD 152249 &         O9Ib((f)) &        165411.63-415057.3 &  1962.3 &    1.0 & $>$0.05                           & 21.77$\pm$0.03 & 0.26$\pm$0.02 & -12.88$\pm$0.02 & -11.76 & 92.2 & 74 \\
     NGC 6231 243 &             B8.5V &        165412.28-415237.6 &     3.3 &    1.2 & $>$0.05                                       & \nodata & \nodata & \nodata & \nodata & \nodata & \nodata \\
      NGC 6231 75 &             B8.5V &                           &         &        &                                               &  \\
      CPD-41 7733 &         O8.5V+B3? &        165413.24-415032.6 &    274 &    0.9 & $>$0.05                           & 21.47$\pm$0.62 & 0.17$\pm$0.12 & -14.10$\pm$0.14 & -13.25 & 10.9 & 13 \\ 
        HD 326329 &             O9.5V &        165414.10-415008.5 &   327.7 &    1.1 & 0.006                          & 21.60$\pm$0.08 & 0.82$\pm$0.07 & -13.38$\pm$0.03 & -12.95 & 27.9 & 23 \\
      CPD-41 7734 &              B0Vn &                           &         &        &                                               &  \\
     NGC 6231 236 &             B2.5V &        165414.72-415111.0 &    59.7 &    1.5 & 0.02                    & \nodata & 57.55$\pm$898.48 & -13.96$\pm$0.96 & -13.96 & 10.6 & 7 \\
      NGC 6231 78 &           B8.5III &        165415.14-415527.7 &    22.6 &    2.4 & 0.005                    & \nodata & 61.04$\pm$1317.10 & -14.35$\pm$1.02 & -14.35 & 4.3 & 5 \\
     NGC 6231 235 &             B8.5V &                           &         &        &                                               &  \\
      CPD-41 7736 &              B1Vn &        165415.73-414932.3 &    35.5 &    1.0 & $>$0.05                          & \nodata & 1.18$\pm$0.82 & -14.94$\pm$0.19 & -14.94 & 0.3 & 1 \\ 
      CPD-41 7737 &               B2V &        165416.29-415026.5 &     4.4 &    1.1 & $>$0.05                                       & \nodata & \nodata & \nodata & \nodata & \nodata & \nodata \\
     NGC 6231 189 &             B8.5V &                           &         &        &                                               &  \\
     NGC 6231 234 &              B8Vn &                           &         &        &                                               &  \\
     NGC 6231 213 &             B2IVn &                           &         &        &                                               &  \\
     NGC 6231 225 &               B9V &        165418.13-415016.4 &    79.6 &    1.3 & 0.03                            & 21.86$\pm$0.14 & 1.02$\pm$0.29 & -14.36$\pm$0.10 & -13.81 & 8.7 & 11 \\
        HD 326330 &            B1V(n) &        165418.33-415135.2 &     3.5 &    0.8 & $>$0.05                                       & \nodata & \nodata & \nodata & \nodata & \nodata & \nodata \\
        HD 326339 &           B0.5III &                           &         &        &                                               &  \\
    HD 152270 & WC7+O5-8   &165419.69-414911.5  & 493 & 1.2  & $>$0.05 & \nodata& \nodata& \nodata& \nodata& \nodata& \nodata\\
      CPD-41 7742 &         O9V+B1.5V &        165419.83-415009.3 &  1840.5 &    1.1 & $<$0.0001                          & 21.62$\pm$0.04 & 0.90$\pm$0.03 & -13.03$\pm$0.01 & -12.61 & 145.2 & 92 \\
     NGC 6231 227 &              B7Vn &                           &         &        &                                               & \\
     NGC 6231 334 &              B1Vn &                           &         &        &                                               &  \\
      CPD-41 7743 &             B0.5V &                           &         &        &                                               &  \\
      NGC 6231 89 &           B2.5III &        165421.35-415536.4 &    23.0 &    1.1 & $>$0.05                             & 22.04$\pm$0.25 & 0.22$\pm$0.21 & -14.99$\pm$0.13 & -13.07 & 3.6 & 4 \\
     NGC 6231 223 &              B6Vn &                           &         &        &                                               &  \\
     NGC 6231 217 &              B6Vn &                           &         &        &                                               &  \\
     NGC 6231 186 &             B2.5V &        165422.84-414523.4 &    43.9 &    1.3 & $>$0.05                            & 21.88$\pm$0.19 & 0.93$\pm$0.34 & -14.64$\pm$0.12 & -14.01 & 7.8 & 10 \\
     NGC 6231 108 &               B2V &                           &         &        &                                               &  \\
     NGC 6231 165 &             B7.5V &        165425.99-414707.6 &    27.1 &    1.3 & $>$0.05                             & 21.67$\pm$0.42 & 0.99$\pm$0.51 & -14.81$\pm$0.21 & -14.38 & 5.2 & 5 \\
        HD 326331 &        O8III((f)) &        165425.97-414955.7 &   749.6 &    1.0 & $>$0.05                           & 21.64$\pm$0.05 & 0.47$\pm$0.06 & -13.18$\pm$0.03 & -12.56 & 62.7 & 49 \\
     NGC 6231 184 &             B8.5V &                           &         &        &                                               &  \\
      CPD-41 7744 &             B1.5V &        165426.54-414951.0 &    24.3 &    1.4 & $>$0.05                             & 21.50$\pm$1.42 & 1.33$\pm$0.87 & -14.98$\pm$0.24 & -14.74 & 3.5 & 3 \\
     NGC 6231 222 &              B4Vn &        165427.88-415013.3 &     3.3 &    1.0 & $>$0.05                                       & \nodata & \nodata & \nodata & \nodata & \nodata & \nodata \\
     NGC 6231 160 &             B3III &        165428.95-414826.0 &    52.2 &    1.4 & 0.02                             & 21.83$\pm$0.43 & 1.16$\pm$0.63 & -14.45$\pm$0.20 & -13.98 & 9.3 & 6 \\
      CPD-41 7746 &             B0.5V &                           &         &        &                                               &  \\
        HD 152314 &       O8.5V+B+... &        165432.00-414818.8 &   528.0 &    1.0 & $>$0.05                           & 21.63$\pm$0.07 & 0.48$\pm$0.07 & -13.34$\pm$0.04 & -12.73 & 32.6 & 36 \\
      NGC 6231 96 &             B1.5V &        165432.21-415652.4 &    29.2 &    1.3 & $>$0.05                          & \nodata & 3.43$\pm$3.16 & -14.29$\pm$0.22 & -14.29 & 6.0 & 9 \\
     NGC 6231 115 &               B3V &                           &         &        &                                               &  \\
     NGC 6231 152 &               B5V &                           &         &        &                                               &  \\
        HD 326332 &            B1III* &        165435.79-415011.6 &    11.3 &    0.8 & $>$0.05                                       & \nodata & \nodata & \nodata & \nodata & \nodata & \nodata \\
      CPD-41 7753 &             B0.5V &        165436.10-415338.6 &   636.8 &    1.7 & $<$0.0001                           & 21.28$\pm$0.15 & 7.55$\pm$2.44 & -13.09$\pm$0.03 & -13.02 & 51.3 & 54 \\
      CPD-41 7755 &               B1V &        165439.87-415338.7 &   104.7 &    1.4 & 0.002                       & \nodata & 2.74$\pm$1.16 & -14.08$\pm$0.07 & -14.08 & 17.4 & 17 \\
     NGC 6231 173 &               B9V &                           &         &        &                                               &  \\
        HD 326333 &            B1V(n) &                           &         &        &                                               &  \\
     NGC 6231 123 &             B6.5V &                           &         &        &                                               &  \\
     NGC 6231 121 &             B6.5V &        165444.22-415627.3 &    49.2 &    1.5 & $<$0.0001                           & 21.79$\pm$0.42 & 1.22$\pm$0.53 & -14.56$\pm$0.21 & -14.14 & 15.2 & 19 \\
     NGC 6231 172 &               B9V &        165444.38-414642.7 &    55.3 &    1.2 & $>$0.05                        & \nodata & 1.36$\pm$0.39 & -14.45$\pm$0.14 & -14.45 & 6.7 & 8 \\
     NGC 6231 175 &               B9V &                           &         &        &                                               &  \\
     NGC 6231 127 &          B8III-IV &        165447.39-415250.6 &    37.3 &    1.3 & $>$0.05                          & 18.74$\pm$258.34 & 1.58$\pm$0.35 & -14.52$\pm$0.10 & -14.52 & 6.9 & 11 \\
     NGC 6231 142 &               B4V &        165447.73-415047.7 &    71.9 &    1.3 & $>$0.05                           & 21.77$\pm$0.16 & 0.86$\pm$0.37 & -14.39$\pm$0.15 & -13.82 & 13.5 & 10 \\
        HD 326334 &               B3V &                           &         &        &                                               &  \\
     NGC 6231 147 &             B9.5V &                           &         &        &                                               &  \\
     NGC 6231 146 &               B3V &        165458.28-414917.5 &    19.3 &    1.1 & $>$0.05                             & 21.90$\pm$1.10 & 0.09$\pm$0.02 & -14.10$\pm$0.02 & -11.58 & 5.2 & 6 \\
\enddata
\tablecomments{{\it Chandra} X-ray properties of OB stars. Column~1: Star name. Column~2: Spectral type from \citet{2006MNRAS.372..661S} and references therein. Column~3: {\it CXO} designation. Column~4: Net X-ray counts in the total (0.5--8.0~keV) band. Column~5: Median energy of X-ray photons in the (0.5--8.0~keV) band. Column~6: The K-S test for X-ray variability, where $P_\mathrm{KS}$ is the null-hypothesis probability that the flux is constant.  Column~7-8: Hydrogen column density and plasma temperature parameters from the $wabs\times apec$ model fit. Columns~9--10: Observed X-ray flux and absorption-corrected X-ray flux from the model fit. Columns~11--12: $X^2$ and the number of degrees of freedom for the model fit. X-ray properties for stars that were not detected by {\it Chandra} are left blank, while missing X-ray properties for {\it Chandra} sources are indicated by ``...''.  }
\end{deluxetable}

\section{Summary and Conclusion \label{conclusion.sec}}

NGC~6231 is a well-known young stellar cluster that makes an excellent testbed for studies of young stars and early cluster dynamics. It contains rich populations of lower mass pre--main-sequence stars, main-sequence OB stars, post-main sequence supergiants, and a Wolf-Rayet system.  We present the a catalog of 2148 probable cluster members, the largest available catalog for NGC~6231, based on an analysis of archival {\it Chandra} X-ray data, VVV photometry, and public catalogs of the region. X-ray selection using {\it XMM Newton}  and the {\it Chandra X-ray Observatory} has been the most effective methods of identifying the low-mass cluster population, and our study builds on X-ray studies by \citet{2004MNRAS.350..809S,2005A&A...441..213S,2006MNRAS.371...67S,2006A&A...454.1047S,2006MNRAS.372..661S,2007ApJ...659.1582S,2007MNRAS.377..945S,2008MNRAS.386..447S,2008NewA...13..202S} and \citetalias{2016arXiv160708860D}. We will use the large catalog of low-mass probable cluster members presented here for analysis of the structure of NGC~6231 in Paper~II. A summary of our methods, catalogs, and results are listed below. 
\begin{enumerate}

\item The study is based on reanalysis of an archival 122-ks {\it Chandra} ACIS-I observation of NGC~6231. We use the data-reduction methodology from the MYStIX project, which has been optimized for the detection of faint X-ray sources, to create a list of 2411 point sources  (\S\ref{xraydata.sec}). This source list contains $\sim$1.5 times more point sources than detected by \citetalias{2016arXiv160708860D}, which is similar to the improvement seen in other regions \citep{2010ApJ...714.1582B,2013ApJS..209...27K}. The X-ray sources are matched to proprietary NIR data from the VVV survey, which provides deep $ZYJHK_s$ photometry and $\sim$30~epochs of $K_s$ photometry over a five-year period (\S\ref{nir.sec}). The data are also matched to the 2MASS catalog, the VPHAS+ catalog, the {\it Spitzer} GLIMPSE catalogs, and an optical catalog by \citet{2013AJ....145...37S}. NIR color-magnitude diagrams ($Y$ vs.\ $Z-Y$ and $K_s$ vs. $H-K_s$; Figure~\ref{new_sources.fig}) compare the distribution of X-ray selected ``probable cluster members'' to non members, showing that the new sample contains many new true positive not included in previous studies, while adding few false positives.

\item The classification of X-ray-selected probable cluster members is based on NIR matches, X-ray variability, X-ray median energy, and filtering of likely foreground and background field stars on optical color-magnitude diagrams (\S\ref{members.sec}). Overall, in the {\it Chandra} field, 2093 X-ray sources are classified as probable cluster members, 106 OB stars are classified as probable cluster members, and 123 X-ray source are classified as likely field stars. The catalog of 2148 probable cluster members (given the IAU prefix ``CXOVVV'') are provided in Table~\ref{nirprop.tab} along with inferred values of stellar age, mass, bolometric luminosity, and infrared excess. 

\item Using the variability criteria developed by N.\ Medina (in preparation), 295 VVV sources showing significant $K_s$ variability are identified (\S\ref{ksvar.sec}) in a 3.5-square-degree region around NGC~6231. Sixteen $K_s$ variables are located within the {\it Chandra} field of view, of which 4 have X-ray counterparts. Previous analysis suggests approximately half of these objects are pre--main-sequence stars, while many others are asymptotic giant branch (AGB) stars \citep{2016arXiv160206269C,2016arXiv160206267C}. The NGC~6231 cluster is associated with a clustering of $K_s$ variables, and the presence of several young stellar clusters outside the {\it Chandra} field are indicated by overdensities in the data. The catalog of $K_s$ variables is provided in Table~4.

\item Age estimates are obtained using both stellar evolution on the $V$ vs.\ $V-I$ color-magnitude diagram ($Age_{VI}$) and evolution of X-ray luminosity vs.\ dereddened $J$-band luminosity ($Age_{JX}$), with both methods suggest that the median stellar age is $\sim$3.2--3.3~Myr (\S\ref{age.sec}). These estimates of median age are consistent with previous age estimates, but systematic effects on age estimates may contributed to considerable uncertainty. The distribution of  $Age_{VI}$ values suggests a significant age spread which has also been noted by \citetalias{2016arXiv160708860D} and \citet{2013AJ....145...37S}. A large age spread would also be consistent with the likely birth of the progenitor of the run-away HMXB system HD~153919 $\sim$6.4~Myr ago.

\item The distribution of the X-ray luminosities of the brightest pre--main-sequence stars shows signs of a moderate decrease in luminosity  relative to stars in younger regions like the Orion Nebula Cluster or Taurus Molecular Cloud (\S\ref{xrayir.sec}). This can be seen as a decrease in X-ray luminosity of 0.2--0.4~dex relative to the \citet{2007A&A...468..425T} relation for weak-line T-Tauri stars, an X-ray luminosity function with a steeper slope than for the Orion Nebula Cluster, and a lack of bright X-ray sources with $Age_{VI}>5$~Myr. However, for stars with $Age_{VI}$ younger than 5~Myr, the distribution of $L_X$ appears  not to vary with age. 

\item We use the observed mass function and X-ray luminosity function to estimate the total number of stars in the cluster (down to the hydrogen-burning limit at 0.08~$M_\odot$) and estimate the completeness limit for pre--main-sequence stars (\S\ref{imf.sec}). Assuming that the cluster follows a normal IMF, we estimate 5700--7500 stars projected within the {\it Chandra} field. The sample has a 50\%-completeness limit at 0.5~$M_\odot$ (meaning that a 0.5~$M_\odot$ star has a 50\% chance of detection). If we assume a universal X-ray luminosity function (neglecting effects of X-ray luminosity evolution), we find a consistent result of 6000 stars and a completeness limit at $L_X=10^{30.0}$~erg~s$^{-2}$. The cutoff of the sample in mass is not sharp due to the large amount of scatter in the $L_X$--$M$ relation. Some substellar-mass objects are identified, but the number of these objects (4--26) depends strongly on the assumption about stellar age. The typical X-ray luminosity of the detected substellar candidate is $\log L_X=29.9$~[erg~s$^{-1}$]. 

\item With low foreground extinction ($A_V\approx1.6$~mag corresponding to $N_H\approx3.5\times10^{21}$~cm$^{-2}$) X-ray median energy (spectral hardness) is most sensitive to plasma temperature and local absorption. For B-type stars, median energy ($ME$) and X-ray luminosity can be used to separate these objects into two groups -- one group with soft spectra ($ME<1.15$~keV), lower X-ray fluxes, and spectral types of B4 or earlier, and another group with harder spectra ($ME>1.15$~keV), intermediate X-ray luminosities, and spectral types ranging from B0 to B9 (\S\ref{ob.sec}). We argue that the first group is most likely stars with X-rays produced by stellar winds, while the second group is most likely stars with X-rays from coronal emission of pre--main-sequence binary companions. In two cases in the second group, X-ray flares are seen in the light curves providing additional evidence that the X-ray emission is from a pre--main-sequence companion. We estimate that 35-41\% of B stars have pre--main-sequence binary companions with masses $>$0.5~$M_\odot$. 

\item For low-mass stars, median energy can also be used to identify highly absorbed cluster members (\S\ref{hard.sec}). A small fraction of sources have X-ray median energies between 2~keV and 6~keV. While some of these may be contaminants, analysis on various optical, infrared, and X-ray color-magnitude diagrams suggests that many are bonefide cluster members. We hypothesize that the absorptions may be produced by circumstellar disks.

\end{enumerate}

\appendix
\twocolumngrid
\section{Gaia Distance Estimate \label{gaia.sec}}

\begin{figure}[h]
\centering
\includegraphics[width=0.35\textwidth]{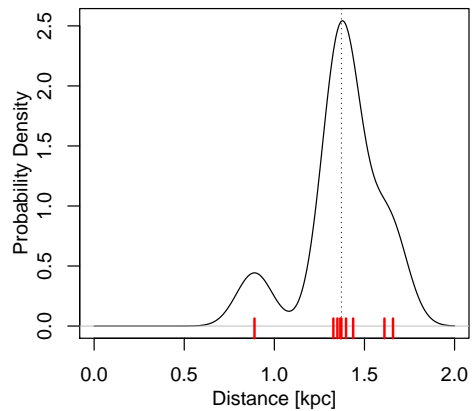} 
\caption{Distances based on parallaxes from the Gaia-Tycho catalog for 9 OB stars. A kernel-density estimate of the probability density function is shown by the black curve, red tic-marks show the measurements (the central tic-marks overlap), and the dotted line shows the median of the distribution at 1370~pc. \label{distance.fig}}
\end{figure}

Nine of the OB stars in NGC~6231 have parallax measurements in the Gaia-Tycho catalog \citep{2016arXiv160904172G} with uncertainties on parallax less than $\delta\pi<0.2$~mas. The distances indicated by the parallax measurements for these 9 stars are shown in Figure~\ref{distance.fig} as tic marks. The probability density function created by smoothing these data with a kernel with bandwidth $\sigma=0.1$~kpc is also shown, similar to the analysis performed for estimation of the distance to the Pleiades cluster using Gaia measurements by the \citet{2016arXiv160904172G}.  The median of this distribution is $1.37\pm0.42$~kpc. The uncertainty on the median is calculated by bootstrap resampling where uncertainties on individual measurements are simulated by adding a random variable drawn from the individual measurements' uncertainty distributions. 
Contributions of systematic uncertainties of the order $\pm0.3$~mas \citep{2016arXiv160904303L} yield a distance estimate $1.37\pm0.70$~kpc.

\newpage
\onecolumngrid
\acknowledgments MAK, NM, EDF, MG, JB, and RK acknowledge support from the Ministry of Economy, Development, and Tourism's Millennium Science Initiative through grant IC120009, awarded to The Millennium Institute of Astrophysics. MAK was also supported by a fellowship (FONDECYT Proyecto No.\ 3150319) from the Chilean Comisi\'on Nacional de Investigaci\'on Cient\'ifica y Tecnol\'ogica, and RK received support from FONDECYT Proyecto No.\ 1130140. The scientific results reported in this article are based on data obtained from the Chandra Data Archive, the Vista Variables in V\'ia Lact\'ea project, and the Two Micron All Sky Survey catalog.
This work make use of analysis methods developed at Penn State for the MYStIX project and {\it Chandra} data reduction procedures (including ACIS Extract) developed by Patrick Broos and Leisa Townsley. 
We thank the referee for many useful comments and suggestions. We also thank A.\ Sengupta for useful feedback on the article and P.\ Lucas for expert advice about the VVV survey. 

\facility{CXO(ACIS-I), ESO:VISTA(VVV)}

\software{
          ACIS Extract \citep{2010ApJ...714.1582B},
          astro \citep{CRANastro},
          astrolib \citep{1993ASPC...52..246L},  
          astrolibR \citep{CRANastrolibR},
          celestial \citep{2016ascl.soft02011R},
          SAOImage DS9 \citep{2003ASPC..295..489J}, 
          SIMBAD \citep{2000A&AS..143....9W},
          spatstat \citep{baddeley2005spatstat}, 
          TOPCAT \citep{2005ASPC..347...29T},
          XPHOT \citep{2010ApJ...708.1760G}
          }

\bibliography{mybib}



\end{document}